\documentclass{elsarticle}
\pdfoutput=1

\usepackage{amsmath}
\usepackage{amssymb}
\usepackage{multirow}
\usepackage{longtable}
\usepackage{lscape}
\usepackage{color}

\biboptions{sort&compress}

\newcommand{\capdef}{}
\newcommand{\mycaption}[2][\capdef]{\renewcommand{\capdef}{#2}%
        \caption[#1]{{\footnotesize #2}}}
\makeatletter
\renewcommand{\fnum@table}{\textbf{\tablename~\thetable}}
\renewcommand{\fnum@figure}{\textbf{\figurename~\thefigure}}
\makeatother

\newcommand{\ie}{i.e.}

\newcommand{\eg}{e.g.}

\newcommand{\cf}{cf.}

\newcommand{\eq}{eq.}

\newcommand{\eqs}{eqs.}

\newcommand{\fig}{figure}
\newcommand{\Fig}{Figure}
\newcommand{\figs}{figures}

\newcommand{\Ref}{Ref.}
\newcommand{\Refs}{Refs.}
\newcommand{\Sec}{Section}

\newcommand{\App}{Appendix}

\newcommand{\Tab}{table}

\newcommand{\equ}[1]{\eq~(\ref{equ:#1})}

\newcommand{\figu}[1]{\fig~\ref{fig:#1}}
\newcommand{\Figu}[1]{\Fig~\ref{fig:#1}}

\newcommand{\bi}{\begin{itemize}}
\newcommand{\ei}{\end{itemize}}

\newcommand{\be}{\begin{equation}}
\newcommand{\ee}{\end{equation}}
\newcommand{\bea}{\begin{eqnarray}}
\newcommand{\eea}{\end{eqnarray}}

\begin{document}


\begin{frontmatter}

\title{Are Gamma-Ray Bursts the Sources of Ultra-High Energy Cosmic Rays?}

\author[a,b]{Philipp Baerwald\corref{mycorrespondingauthor}}
\cortext[mycorrespondingauthor]{Corresponding author}
\ead{pxb46@psu.edu}

\author[a,c]{Mauricio Bustamante}
\ead{mauricio.bustamante@physik.uni-wuerzburg.de}

\author[a,c]{and Walter Winter}
\ead{winter@physik.uni-wuerzburg.de}

\address[a]{Institut f{\"u}r Theoretische Physik und Astrophysik, Universit{\"a}t W{\"u}rzburg, \\
D-97074 W{\"u}rzburg, Germany}
\address[b]{Department of Astronomy and Astrophysics; Department of Physics; Center for Particle and Gravitational Astrophysics; Institute for Gravitation and the Cosmos; \\ 
Pennsylvania State University, 525 Davey Lab, University Park, PA 16802, USA}
\address[c]{DESY, Platanenallee 6, 15738 Zeuthen, Germany}

\begin{abstract}
We reconsider the possibility that gamma-ray bursts (GRBs) are the sources of the ultra-high energy cosmic rays (UHECRs) within the internal shock model, assuming a pure proton composition of the UHECRs. For the first time, we combine the information from gamma-rays, cosmic rays, prompt neutrinos, and cosmogenic neutrinos quantitatively in a joint cosmic ray production and propagation model, and we show that the information on the cosmic energy budget can be obtained as a consequence. In addition to the neutron model, we consider alternative scenarios for the cosmic ray escape from the GRBs, \ie, that cosmic rays can leak from the sources. 
We find that the dip model, which describes the ankle in UHECR observations by the pair production dip, is strongly disfavored in combination with the internal shock model because a) unrealistically high baryonic loadings (energy in protons versus energy in electrons/gamma-rays) are needed for the individual GRBs and b) the prompt neutrino flux easily overshoots the corresponding neutrino bound. On the other hand, GRBs may account for the UHECRs in the ankle transition model if cosmic rays leak out from the source at the highest energies. In that case, we demonstrate that future neutrino observations can efficiently test most of the parameter space --- unless the baryonic loading is much larger than previously anticipated.
\end{abstract}



\end{frontmatter}

\section{Introduction}

One of the most interesting questions in astroparticle physics is that of the origin of the ultra-high energy cosmic rays (UHECRs). Possible source candidates are gamma-ray burst (GRBs) fireballs; see \Refs~\cite{Piran:2004ba,Meszaros:2006rc} for reviews. We focus on the internal shock model, where the prompt emission in gamma-rays is expected to come from internal collisions inside the ejected material~\cite{Paczynski:1994uv,Rees:1994nw}, accelerating particles to the highest energies. While recent observations point towards a heavier cosmic ray composition at the highest energies~\cite{Abraham:2010yv}, we focus on protons as candidates for the UHECRs in this study, for which plausible models for the particle acceleration and emission from GRBs exist\footnote{Note, however, that there are a number of papers discussing the effect of heavy nuclei, such as \Refs~\cite{Wang:2007xj,Murase:2008mr,Murase:2010gj,Metzger:2011xs}. We will comment on the case of heavy nuclei in the conclusions.

}. If GRBs produce the UHECRs, they may also dissipate a fraction of their energy into pion and, therefore, neutrino production.
The neutrino flux from GRB fireballs in the internal shock model was originally predicted in \Ref~\cite{Waxman:1997ti}, whereas alternative scenarios have been increasingly drawing attention; see \Refs~\cite{Murase:2008sp,Gao:2011xu,Gao:2012ay,He:2012tq,Zhang:2012qy,Murase:2013hh,Murase:2013ffa,Gao:2013fra,Asano:2013jea} for some recent works. Most importantly, the recent neutrino observations by the IceCube experiment have started to exert pressure on the conventional internal shock model~\cite{Abbasi:2012zw}. 

For the description of the UHECR observations, several transition models have been proposed in the literature~\cite{Hill:1983mk,Yoshida:1993pt,Wibig:2004ye,Hillas:2005cs}; see \Ref~\cite{Berezinsky:2013kfa} for a recent review. Since we only consider the highest cosmic ray energies and a pure proton composition, two are especially relevant for us: the \textbf{ankle model} assumes a transition between a (Galactic or different extragalactic) component below the ankle ($\sim 40 \, \mathrm{EeV}$) and an extragalactic component with an injection index $\alpha_p \simeq 2$ above the ankle. In this model, GRBs would only describe the extragalactic component, which means that we are not going to touch the contribution below the ankle, and we are not going to discuss some of the challenges for that. In the \textbf{dip model}, the extragalactic component extends to lower energies ($\sim 1 \, \mathrm{EeV}$), where the spectral shape is generated from a steeper injection spectrum with index $\alpha_p \simeq 2.5-2.7$ (depending on the source 
evolution) in combination with the dominant proton energy loss processes (pair production and interactions with cosmic microwave background (CMB) photons). The steep spectrum may be either an {\em ad hoc} assumption, or generated from a distribution of the maximal proton energies~\cite{Kachelriess:2005xh}. We will consider these two transition models directly in combination with the internal shock model for the source in terms of a combined source-propagation model, which will provide interesting hints on the required model parameters for the GRBs.

In order to predict the neutrino flux, several approaches have been followed in the literature: One may estimate the neutrino flux from the observed cosmic ray flux assuming that GRBs are the source of the UHECRs~\cite{Waxman:1997ti}, one may use the gamma-ray observations to predict the neutrino flux on a burst-by-burst basis~\cite{Guetta:2003wi,Becker:2005ej,Abbasi:2009ig}, or one may assume that cosmic rays are produced in the same processes as neutrinos, such as when only neutrons escape the source (``neutron model''), see, \eg, \Refs~\cite{Mannheim:1998wp,Ahlers:2011jj}. Let us discuss the connection between different pairs of messengers in greater detail in the following.

For the \textbf{gamma-ray--neutrino connection}, the predictions have been recently revised~\cite{Hummer:2011ms,Li:2011ah,He:2012tq}, yielding an order of magnitude lower neutrino flux than predicted earlier~\cite{Guetta:2003wi,Abbasi:2009ig,Abbasi:2012zw} (but not earlier numerical models as \Ref~\cite{Murase:2005hy}), and thus relaxing the tension partially. In addition, the quasi-diffuse flux normalization in these models is chosen somewhat {\em ad hoc}, such as that it relies on an externally provided baryonic loading (energy in protons versus energy in electrons/gamma-rays), chosen to be 10, and the number of observable (long) bursts per year, chosen to be 667 in \Refs~\cite{Abbasi:2009ig,Abbasi:2012zw}. 

The \textbf{cosmic ray--neutrino connection} is very stringent in the neutron model~\cite{Ahlers:2011jj}, in which protons are magnetically confined in the sources and only neutrons are able to escape. However, this connection is model-dependent. For example, alternative cosmic ray escape mechanisms have been recently studied in \Ref~\cite{Baerwald:2013pu}, which we consider in this paper as well. Note that, in fact, within a more general framework, the authors of \Ref~\cite{Kistler:2013my} conclude that the protons resulting from photopion processes are not sufficient to explain the cosmic-ray measurements. An additional puzzle in the neutron model is the magnitude of the predicted neutrino fluxes, which is significantly higher than the current bounds. This is already an indication that the pion production efficiency or the assumed value for baryonic loading in the gamma-ray--neutrino approach has been underestimated. On the other hand, the original computation~\cite{Waxman:1997ti} relies on the pion 
production 
efficiency, which implies that some of the corrections found in \Refs~\cite{Hummer:2011ms,Li:2011ah} 
apply, and 
on the energy injected into cosmic rays, which has to be reevaluated in view of more recent results such as HiRes~\cite{Abbasi:2007sv}, Telescope Array~\cite{AbuZayyad:2012ru} and Auger~\cite{Abraham:2010mj,Abreu:2011ph}.

The \textbf{connection between gamma-rays and UHECRs} has been heavily debated (see, \eg, \Refs~\cite{Eichler:2010ky,Waxman:2010fj}) and depends on a number of fudge factors, to be put in by hand as well. 
On the other side, it is clear that both the predicted neutrino and cosmic ray fluxes from gamma-rays will depend on common impact factors, such as the baryonic loading of the bursts. It has been pointed out that the definition of the baryonic loading depends on the energy range it is defined in, which means that it may carry a bolometric correction~\cite{Murase:2008mr,Eichler:2010ky}. 

In order to clarify these issues, it is therefore natural to choose a common normalization to draw a self-consistent picture, \ie, to normalize the predicted cosmic ray flux to the observation and to derive the baryonic loading, needed for the neutrino flux prediction, as a parameter. We follow this completely self-consistent strategy in this paper, which will allow us not only to constrain the parameters of common models, but also to obtain the information on the cosmic energy budget as a spin-off. Within this strategy, we identify the relevant impact factors, including the ones which may not be obvious from the beginning. 

This study is organized as follows:
We give the relevant relationships to describe the cosmic energy budget of UHECRs with GRBs in a model-independent way in \Sec~\ref{sec:modelindep}, where details are given in \ref{app:modelindep}.~Then, in \Sec~\ref{sec:models}, we discuss how the ankle and dip models can be accommodated within our combined source-propagation model. Details of the cosmic ray propagation are discussed in \ref{sec:crprop}, and of the statistical methods in \ref{sec:statistics}. We perform in \Sec~\ref{sec:scan} a more refined parameter space scan to support our findings for the ankle model. In addition, we discuss the impact of the choice of cosmic ray escape model and of local fluctuations of the GRB rate with respect to the star formation rate (SFR), since ensemble fluctuations could be relevant for the UHECR flux~\cite{Ahlers:2012az}. The impact of the maximal proton energy is discussed separately, in \ref{sec:maxe}. Finally, we summarize our results and present our conclusions in 
\Sec~\ref{sec:conclusions}. 

\section{Cosmic energy budget and observables for GRBs}
\label{sec:modelindep}

\begin{figure}[tp]
\begin{center}
\includegraphics[width=0.6\textwidth]{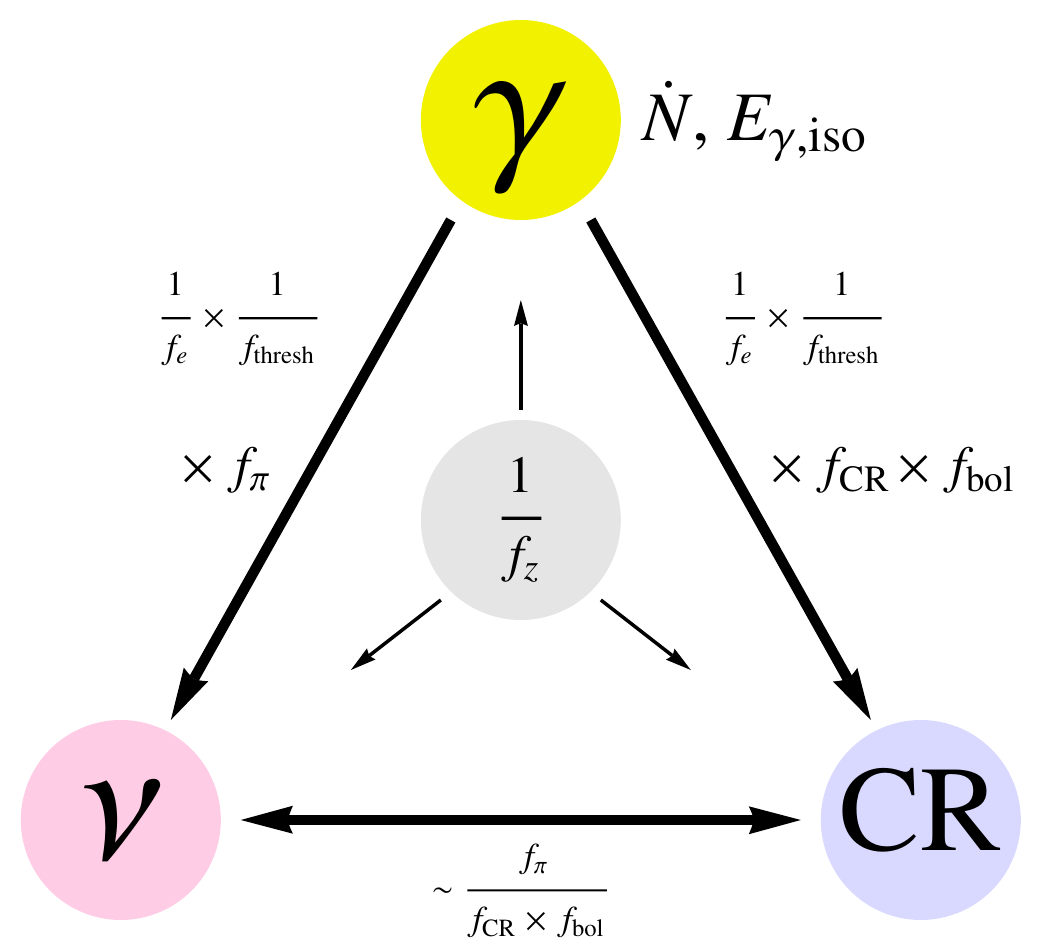}
\end{center}
\mycaption{\label{fig:triangle} Result for the multi-messenger connection (illustration). Here ``CR'' refers to UHECRs in the energy range between $10^{10} \, \mathrm{GeV}$ and $10^{12} \, \mathrm{GeV}$. The different labels refer to the number of observable GRBs per year ($\dot N$), the isotropic equivalent energy in gamma-rays ($E_{\gamma,\mathrm{iso}}$), the 
 cosmic evolution factor ($f_z > 1$),
 the baryonic loading ($f_e^{-1} \ge 10$), the instrument threshold correction ($f_{\mathrm{thresh}} \simeq 0.2-0.5$), the fraction of baryonic energy going into cosmic ray production ($f_{\mathrm{CR}}$), the fraction of baryonic energy going into pion production ($f_\pi$), and a bolometric correction factor ($f_{\mathrm{bol}} \ll 1$). }
\end{figure}

In this section, we review the multi-messenger picture among gamma-rays, neutrinos, and cosmic rays in a model-independent, analytical way. We also discuss different normalization methods of the fluxes, and how they are related to the observables. Note that we only present a short summary here, the detailed derivations can be found in \ref{app:modelindep}. Our findings are summarized in \figu{triangle}, which can be followed during this discussion.

The cosmic energy output of gamma-rays from GRBs can be characterized by the observables, such as the isotropic equivalent energy $E_{\gamma,\mathrm{iso}}$ per GRB, the number of observable GRBs per year $\dot N$, and the redshift distribution of the GRBs.\footnote{For the sake of simplicity and technical feasibility, we do not consider a luminosity distribution here. As detailed in \Ref~\cite{Kistler:2009mv}, it is possible to assume a threshold luminosity which is visible in the whole chosen redshift range. Hence, it is possible to calculate an average luminosity per burst which represents the distributed result well. Our results in this paper need to be interpreted as such appropriately averaged bursts.} As already suggested in \Ref~\cite{Le:2006pt}, we assume that the GRB rate in redshift does not exactly follow the measured star formation rate (SFR). We will use the factorization proposed by Kistler et al.~\cite{Kistler:2009mv}, which assumes that GRBs follow the SFR up to an additional 
evolution factor $(1+z)^\alpha$. For $\alpha>0$, this leads to an increased number of high $(z>2)$ redshift bursts compared to the pure SFR.
Although the exact number for $\dot N$ is not known, it must be of the order of 1000 bursts per year from observations, and is therefore a very robust measure for the normalization.

In order to estimate the quasi-diffuse neutrino flux from that, 
as it is done in \Refs~\cite{Abbasi:2009ig,Abbasi:2012zw}, one subtlety immediately arises: one needs the total number of bursts in the observable universe per year $\dot N_{\mathrm{tot}}$, which includes bursts below detection threshold in gamma-rays that nevertheless contribute to the neutrino flux. We therefore define a \textbf{threshold correction}
\begin{equation}
 f_{\mathrm{thresh}} \equiv \frac{\dot N}{\dot N_{\mathrm{tot}}} \le 1 \, ,
\end{equation}
which depends on instrument threshold and low-luminosity cutoff. For a simulation following \Refs~\cite{Kistler:2009mv,Wanderman:2009es}, we have found $f_{\mathrm{thresh}} \sim 0.3 - 0.5$; see \ref{app:modelindep}.
A very interesting recent study in this context is based on Swift data~\cite{Lien:2013qja}, from which one can estimate $f_{\mathrm{thresh}} \simeq 1000/4568 \simeq 0.22$, which is in the same ballpark. Hence, we will use $f_{\mathrm{thresh}} = 0.3$ in the following as a default value.

\begin{table}[t]
  \centering
  \begin{tabular}{|l|r|r|r|r|c|}
    \hline
     & &  & $\left. \dot{\tilde{n}}_{\mathrm{GRB}} \right|_{z=0}$ & $E_{\mathrm{CR}}^{[10^{10},10^{12}]}$  & \\
    SFR model & $\alpha$ & $f_z$ & $[\mathrm{Gpc}^{-3} \, \mathrm{yr}^{-1}]$ & $[10^{53} \, \mathrm{erg}]$ &  References \\
    \hline \hline
    Hopkins \& Beacom (2006)  & $1.2$ & 25.15 & 0.13 & 11.0 & \cite{Hopkins:2006bw,Kistler:2009mv} \\
                              & $0.0$ & 5.65  & 0.58 & 2.5  & \cite{Hopkins:2006bw} \\
    \hline
    Wanderman \& Piran (2010) & $0.0$ & 7.70  & 0.43 & 3.4  & \cite{Wanderman:2009es} \\
    \hline
    Madau \& Porciani (2000) & & & & & \cite{Porciani:2000ag} \\
    $\quad$ SF1              & $0.0$ & 9.89   & 0.35 & 4.3  & \\
    $\quad$ SF2              & $0.0$ & 14.42  & 0.23 & 6.3  & \\
    $\quad$ SF3              & $0.0$ & 14.36  & 0.23 & 6.2  & \\
    \hline
  \end{tabular}
  \mycaption{\label{tab:redshiftnorms} Cosmic evolution factor $f_z$, local GRB rate (without beaming correction), and required energy per GRB in UHECRs for different SFR histories and source evolution factors $\alpha$ (the star formation rate is corrected by a factor of $(1+z)^\alpha$). The results for $f_z$ are obtained using \equ{dotNcalc}, with the integration running from $z = 0$ to $z = 6$. The local GRB rate is obtained from \equ{obsmaster2} and the cosmic ray energy per bursts in the range $10^{10}$ to $10^{12} \, \mathrm{GeV}$ is obtained from \equ{cre2}, both by assuming $\dot N = 1000 \, \mathrm{yr}^{-1}$ and $f_{\mathrm{thresh}}=0.3$.}
\end{table}

A different quantity frequently used in the literature~\cite{Schmidt:1999iw,2001ApJ...559L..79S,Guetta:2003zp,Guetta:2006gq,Liang:2006ci,Pelangeon:2008qa,Wanderman:2009es,Lien:2013qja} for the normalization is the local GRB rate $\left. \dot{ \tilde{n}}_{\mathrm{GRB}} \right|_{z=0}$.\footnote{Here we use the local GRB rate $\left. \dot{ \tilde{n}}_{\mathrm{GRB}} \right|_{z=0}$ related to observations, the actual GRB rate $\dot n_{\mathrm{GRB}}$ is higher by correcting for the beaming factor of the GRBs; $\dot{\tilde{n}}_{\mathrm{GRB}} \equiv \dot n_{\mathrm{GRB}}/\langle f_{\mathrm{beam}} \rangle$. See \ref{app:modelindep} for a more detailed discussion.} It is related to the observable $\dot N$ by
\begin{equation}
\left. \dot{\tilde{n}}_{\mathrm{GRB}} \right|_{z=0} \simeq \frac{1}{\mathrm{Gpc}^3 \, \mathrm{yr}} \cdot \frac{\dot N \, [\mathrm{yr}^{-1}]}{968} \cdot f_{\mathrm{thresh}}^{-1} \cdot f_{z}^{-1} \, .
\label{equ:obsmaster2}
\end{equation}
Since the local GRB rate only represents the local environment at $z=0$, but $\dot N$ represents the whole GRB sample with the redshift distribution, the relationship includes a \textbf{cosmic evolution factor $f_z$} describing how representative the local GRB rate is for the whole sample. The stronger the evolution of the GRB rate in redshift (larger $\alpha$) is, the larger values of this correction factor are obtained. Note that our definition of $f_z$ in \equ{fz} of \ref{app:modelindep} is different from $\xi_Z$ in Waxman and Bahcall~\cite{Waxman:1998yy}, and includes the description of the $\Lambda$CDM cosmology. Typical values for $f_z$ range from $5$ to $25$, as listed in \Tab~\ref{tab:redshiftnorms} (third column) for different SFRs and evolution factors $\alpha$. In addition, we list the values for $\left. \dot{\tilde{n}}_{\mathrm{GRB}} \right|_{z=0} $ in the table (fourth column), which are obtained from \equ{obsmaster2} for $\dot N = 1000 \, \mathrm{yr}^{-1}$ and $f_{\mathrm{
thresh}}=0.3$. From the table, one can easily read off that the stronger the source evolution is, the smaller the local GRB rate will be. 
Note that if we extract $\dot N$ and $E_{\gamma,\mathrm{iso}}$ from the gamma-ray observations and use them for the normalization, the distribution of GRBs as a function of redshift, the neutrino, and cosmic ray fluxes will scale with $1/f_z$, as illustrated in \figu{triangle}.

In order to address the question of how much energy is needed per GRB, a frequently used approach is to derive the required local energy injection rate between $10^{10}$ and $10^{12} \, \mathrm{GeV}$ to reproduce the observations. 
The value given in \Ref~\cite{Waxman:1995dg} is $\dot{\varepsilon}_\mathrm{CR}^{\left[10^{10},10^{12}\right]} = 4.5 \cdot 10^{44} \, \mathrm{erg} \, \mathrm{Mpc}^{-3} \, \mathrm{yr}^{-1}$ which reproduced the observed UHECR flux at that time, when, however, data above $8 \cdot 10^{10} \, \mathrm{eV}$ were sparse (see also \Ref~\cite{Katz:2008xx} for an update using Auger data). Using the data from several experiments recently compiled by Gaisser, Stanev and Tilav~\cite{Gaisser:2013bla}, we obtain $1.5 \cdot 10^{44} \, \mathrm{erg} \, \mathrm{Mpc}^{-3} \, \mathrm{yr}^{-1}$, which is compatible with the original result. According to Waxman~\cite{Waxman:1995dg}, there is little sensitivity to the spectral injection index used for this calculation, as long as $dN_{\mathrm{CR}}/dE \propto E^{-\alpha_p}$ with $1.8 < \alpha_p < 2.8$. The required energy per GRB is then (see \ref{app:modelindep})
\begin{equation}
E_{\mathrm{CR}}^{[10^{10},10^{12}]} = 10^{53} \, \mathrm{erg} \cdot \frac{\dot \varepsilon_{\mathrm{CR}}^{[10^{10},10^{12}]}}{10^{44} \, \mathrm{erg} \, \mathrm{Mpc}^{-3} \, \mathrm{yr}^{-1}} \cdot \frac{968 \, \mathrm{yr}^{-1}}{\dot N} \cdot f_{\mathrm{thresh}} \cdot f_z \, .
\label{equ:cre2}
\end{equation}

We list required energies per GRB to achieve $\dot \varepsilon_{\mathrm{CR}}^{[10^{10},10^{12}]} = 1.5 \cdot 10^{44} \, \mathrm{erg} \, \mathrm{Mpc}^{-3} \, \mathrm{yr}^{-1}$, for different star formation rate and evolution models, in \Tab~\ref{tab:redshiftnorms}. The typical energy to be released in protons is between $10^{53}$ and $10^{54} \, \mathrm{erg}$, where, again, it is clear that the strong evolution case requires a rather large injected energy per burst, because there are so few bursts locally. Since this energy per burst only corresponds to the average burst, it is already an indication that the strong evolution model requires either a large average energy per burst or an extremely large baryonic loading.

Using a model for the source, energy partition arguments are typically used to relate the energy in protons and magnetic field to that of electrons and gamma-rays. In approaches to the neutrino production, such as in \Refs~\cite{Abbasi:2009ig,Abbasi:2012zw,Hummer:2011ms}, the \textbf{baryonic loading} $f_e^{-1} \simeq 10$ relates the total energy in protons (in the entire energy range) to the kinetic energy in electrons, which is assumed to be in equipartition with the energy in gamma-rays. We use this definition for the baryonic loading in the following, directly relating the total proton and gamma-ray energies for the sake of simplicity. However, for the relationship between gamma-ray observations and UHECRs, only the CR energy range between $10^{10}$ and $10^{12} \, \mathrm{GeV}$ is relevant, which implies that the baryonic loading is defined differently. We account for this by a \textbf{bolometric correction factor} $f_{\mathrm{bol}} < 1$ such that $f_{\mathrm{bol}}$ is the ratio of energy in protons 
between $10^{10}$ and $10^{12} \, \mathrm{GeV}$ to the one in the total energy range considered; see \equ{bol} in \ref{app:modelindep}. 
For a power-law without cutoff and the full proton energy range\footnote{Technically, the full energy range is defined in the SRF in our calculations, the comparison of the energies, however, needs to be done in the source frame. Hence, the limits need to be boosted to the source frame for the actual calculation, leading to a range from $\Gamma \cdot 1 \, \mathrm{GeV}$ (from the proton rest mass) to $\Gamma \cdot 10^{10} \, \mathrm{GeV}$.}, we find $f_{\mathrm{bol}}$ between about $0.2$ (for $\alpha_p = 2.0$) and $1.6 \cdot 10^{-4}$ (for $\alpha_p = 2.5$). Larger values are obtained for larger minimal proton energies, and somewhat smaller values for a (model-dependent) maximal proton energy significantly below $10^{12} \, \mathrm{GeV}$.

As a consequence, the required energy per burst in the UHECR range can be written in terms of GRB parameters as
\begin{equation}
E_{\mathrm{CR}}^{[10^{10},10^{12}]} = f_{\mathrm{CR}} \frac{f_{\mathrm{bol}}}{f_e} \, E_{\gamma, \mathrm{iso}} \, , \label{equ:ecrderiv2}
\end{equation}
where $f_{\mathrm{CR}}$ is the \textbf{fraction of baryonic energy going into cosmic ray production}, analogous to the fraction $f_\pi$ of \textbf{baryonic energy going into pion production} (pion production efficiency) as defined in \Refs~\cite{Waxman:1997ti,Guetta:2003wi}. One subtle point is that $f_\pi$ is the total amount of energy going into pions (neutral and charged) and not the average energy lost to pions in a single interaction. Hence, if cosmic rays escape as neutrons, typically $f_{\mathrm{CR}} \simeq 2 \cdot f_\pi \simeq 0.4$ for the pion production efficiency $f_\pi \sim 0.2$. This estimate is based on the consideration that the neutrons on average obtain about four times as much energy as the pions while there are roughly two times more (charged and neutral) pions produced in photohadronic interactions than neutrons. If the cosmic rays leaking from the source dominate, typically $f_\pi \ll 1$ and $f_{\mathrm{CR}} \gg f_\pi$.
In order to match the required energy injection per GRB in \equ{cre2}, we can easily see that $f_{\mathrm{CR}} \cdot f_{\mathrm{bol}} \cdot f_e^{-1} \simeq 2.5$ is required for, say, $E_{\gamma, \mathrm{iso}} \simeq 10^{53} \, \mathrm{erg}$,
Hopkins \& Beacom SFR without source evolution (conservative case, $\alpha = 0$), and $f_{\mathrm{thresh}}=0.3$. If cosmic rays efficiently escape as neutrons ($f_{\mathrm{CR}} \simeq 0.4$), then we obtain $f_e^{-1} \simeq 30$ for a proton spectral index of $\alpha_p=2.0$ (where $f_{\mathrm{bol}} \simeq 0.2$), and even larger values for $f_e^{-1}$ if a source evolution is included. A source of confusion in the literature seems to be the difference between baryonic loading $f_e^{-1}$ and the UHE baryonic loading $f_{\mathrm{bol}} \, f_e^{-1}$, which enters in \equ{ecrderiv2}, as pointed out in 
\Refs~\cite{Murase:2008mr,Eichler:2010ky}: it is not sufficient to have a large enough baryonic loading in total, one needs a large enough baryonic loading at the UHE. Therefore, if the baryonic loading is (implicitly) defined for the whole energy range (see \equ{loading}), as in most neutrino calculations, it has to be significantly larger than ten to describe the UHECR observations. This is also the reason why the predicted neutrino fluxes in \Ref~\cite{Ahlers:2011jj} are relatively high: the implied 
baryonic loading $\times$ pion production efficiency, which is not explicitly considered therein, is very high.

If one uses the observation in gamma-rays, one can predict the neutrino flux and cosmic ray injection in a particular model. With respect to the energy budget, the discussed correction factors will appear; see \ref{app:modelindep} for details. An illustration of the corrections affecting the different legs is shown in \figu{triangle}. One can clearly see that $f_z$ is an overall scaling factor because it is needed to obtain the local GRB rate from the observable $\dot N$. The factors baryonic loading and threshold correction affect neutrinos and cosmic rays in the same way. If, for instance, the UHECR flux is used for the normalization of the neutrino flux, such as in \Ref~\cite{Ahlers:2011jj}, these factors will automatically drop out in the calculation of the neutrino flux. On the other hand, the relative normalization between neutrino and cosmic ray flux scales $\propto f_\pi/(f_{\mathrm{CR}} \cdot f_{\mathrm{bol}})$. In the neutron model, $f_{\mathrm{CR}} \propto f_\pi$, 
and strong constraints on the model can be derived because the predicted neutrino fluxes are significantly above the current diffuse bounds~\cite{Ahlers:2011jj}. Since typically $f_{\mathrm{bol}} \ll 1$, a correspondingly large baryonic loading is implied (see discussion above). If, however, $f_{\mathrm{CR}} \gg f_\pi$, such as if the protons leak from the source~\cite{Baerwald:2013pu}, this constraint can be avoided.

In \figu{triangle}, it is clear that some of the scaling factors are dependent on the model and its input parameters ($f_\pi$, $f_{\mathrm{CR}}$, $E_{\gamma,\mathrm{iso}}$, $f_{\mathrm{bol}}$), whereas the remaining relevant parameter combination scaling the neutrino and cosmic ray fluxes is $\dot N \cdot f_e^{-1} \cdot f_{\mathrm{thresh}}^{-1}$. Assuming that $\dot{N} \simeq 1000 \, \mathrm{yr}^{-1}$ is known from observations and that $f_{\mathrm{thresh}} \simeq 0.3$, the UHECR observations can be directly used to measure the baryonic loading. In the following, we will fix these values and measure $f_e^{-1}$ for the sake of simplicity and readability. However, note that $f_e^{-1}$ is to be interpreted as the product of these quantities in the following, \ie, a higher instrument threshold correction or a larger number of observed bursts per year will reduce the required 
baryonic loading. The actual baryonic loading can be derived then from the values of $f_e^{-1}$ given in our figures as
\begin{equation}
 f_{e, \mathrm{actual}}^{-1} = \frac{1000 \, \mathrm{yr}^{-1}}{\dot N} \cdot \frac{f_{\mathrm{thresh}}}{0.3} \cdot f_e^{-1} \, 
\label{equ:rescale}
\end{equation}
for different choices of $\dot N$ or $f_{\mathrm{thresh}}$. The baryonic loading in the UHE range can be obtained as $f_{\mathrm{bol}} \, f_e^{-1}$, where $f_{\mathrm{bol}}$ depends on the spectral proton index and on the minimal and maximal proton energies.

Since the injected amount of cosmic rays always includes the product of number of sources as well as emission from a single source, the beaming factor cancels. Therefore, we do not include a beaming factor in our discussions (see \ref{app:modelindep}, where we keep track of it explicitly).

\section{Combined production and propagation model}
\label{sec:models}

Our GRB source model relies on NeuCosmA (Neutrinos from Cosmic Accelerators), as described in detail in \Ref~\cite{Baerwald:2011ee}. The photohadronic interactions are computed with the method in \Ref~\cite{Hummer:2010vx}, based on the physics of SOPHIA (Simulations Of Photo Hadronic Interactions in Astrophysics)~\cite{Mucke:1999yb}. Magnetic field effects on the secondaries, flavor mixing, and the helicity dependence of the muon decays are taken into account; see \Refs~\cite{Lipari:2007su,Baerwald:2010fk}. All known normalization corrections to the GRB neutrino flux predictions are taken into account~\cite{Hummer:2011ms}. For the description of the additional cosmic ray escape components, see \Ref~\cite{Baerwald:2013pu}. A more or less guaranteed component is the ``direct escape'' of protons for which the Larmor radius reaches the size of the acceleration region. Since the Larmor radius is proportional to energy, this component dominates at the highest energies if the proton acceleration is limited by the 
Larmor radius. In addition, protons may escape via diffusion, where the energy dependence could be weaker, depending on the diffusion coefficient; see next section for more details. For the cosmic ray propagation, we use a deterministic Boltzmann equation solver for the comoving cosmic ray density, following \Refs~\cite{Ahlers:2009rf,Ahlers:2010ty,Ahlers:2010fw,Anchordoqui:2011gy,Ahlers:2011jj,Ahlers:2011sd}. In comparison to \Refs~\cite{Ahlers:2009rf,Ahlers:2011jj}, we use \Ref~\cite{Hummer:2010vx} to compute the photohadronic energy losses due to interactions with the CMB photons and the resulting cosmogenic neutrino fluxes, which makes the code very efficient. See \ref{sec:crprop} for details and some phenomenological discussion relevant for this study.

Due to the large baselines involved, neutrino flavor oscillations are averaged; for the mixing parameters we have used the best-fit values from the global analysis in \Ref~\cite{Fogli:2012ua}, under the assumption of a normal mass hierarchy: $\sin^2 \theta_{12} = 3.07 \cdot 10^{-1}$, $\sin^2 \theta_{13} = 2.41 \cdot 10^{-2}$, $\sin^2 \theta_{23} = 3.86 \cdot 10^{-1}$, and $\delta_\mathrm{CP} = 1.08 \cdot \pi$.

For the sake of simplicity, we assume that all sources are alike in the cosmologically comoving frame, \ie, the source frame. By this choice, we imply that the cosmic ray injection factorizes into a redshift- and an energy-dependent part. If, for instance, the observables were fixed in the observer's frame, the maximal proton energy would depend on redshift and subtle spectral features would appear: for a fixed luminosity, high-redshift bursts would have a lower variability timescale (in the source frame) due to the redshift correction, and, consequently, the particle densities would be higher, photohadronic interactions would become more frequent, and this would introduce an artificial pull on the maximal proton energy. If, on the other hand, a luminosity distribution function were used, only few bursts would contribute to the maximal proton energies, which again would introduce spectral features and subtleties in the interpretation. For the ``standard'' GRB parameters, we use, unless noted otherwise, 
$\Gamma = 300$, $T_{90} = 10 \, \mathrm{s}$, $t_v = 10^{-2} \, \mathrm{s}$, and a luminosity $L_{\mathrm{iso}} = 10^{52} \, \mathrm{erg} \, \mathrm{s}^{-1}$. These parameters are given in the source frame in order to guarantee similar properties in that frame, which means that $L_{\mathrm{iso}} \simeq E_{\gamma,\mathrm{iso}}/T_{90}$. We define the acceleration efficiency $\eta$ (in the SRF) by $t'^{-1}_{\mathrm{acc}}= \eta c e B'/E'$, and use $\eta = 1.0$ unless noted otherwise.

The (target) photon spectrum is assumed to be a simple broken power-law with a lower spectral index $\alpha_\gamma = 1$, an upper spectral index $\beta_\gamma = 2$, and a spectral break at $\epsilon'_\gamma = 1 \, \mathrm{keV}$ (in the SRF).\footnote{Recent observations suggest somewhat larger $\beta_\gamma$, which however hardly affect the neutrino spectra (the main effect would be below the first break in the neutrino spectrum, where nonetheless the spectral index cannot be steeper than $E_\nu^{-1}$ from the kinematic of the weak decays).} The normalization of the photon spectrum is calculated based on the assumption that $E_{\gamma,\mathrm{iso}}$ is distributed over $N_\mathrm{sh} = T_{90}/t_v$ identical ejected matter shells of volume $V'_\mathrm{iso} = 16\pi \, \Gamma^5 \, c^3 t_v^3$. Additionally, it is assumed that the amount of emitted energy $E_{\gamma,\mathrm{iso}}$ is defined from the energy range from $0.2 \, \mathrm{keV}$ to $30 \, \mathrm{MeV}$ (in the source frame), which represents the energy 
range of the \textit{Fermi} GBM instrument.
For the proton spectrum, we assume a power-law with spectral index $\alpha_p$ which is exponentially suppressed above a maximal proton energy $E'_{p,\mathrm{max}}$. This maximal proton energy is derived by comparing the acceleration rate $t'^{-1}_{\mathrm{acc}}$ to the dominant loss rate. Our maximal proton energy is defined as the (lowest) energy at which the acceleration gains and the losses cancel. As loss rates we consider the dynamical loss rate, the synchrotron loss rate, and the photohadronic loss rate, so that the maximal proton energy is calculated as
\begin{equation}
 t^\prime_\mathrm{acc}\left(E_{p,\mathrm{max}}^\prime\right)
 = \min \left[ t_\mathrm{dyn}^\prime, t_\mathrm{syn}^\prime\left(E_{p,\mathrm{max}}^\prime\right), t_{p\gamma}^\prime\left(E_{p,\mathrm{max}}^\prime\right) \right] \;,
 \label{equ:EpmaxDetermination}
\end{equation}
where the different timescales are defined as in \Ref~\cite{Baerwald:2013pu}: $t_\mathrm{dyn}^\prime = \Delta d^\prime / c$ (with $\Delta d^\prime$ the shell width), $t_\mathrm{syn}^\prime\left(E^\prime\right) = 9 \, m^4/\left( 4 \, c \, e^4 B^{\prime2} E^\prime \right)$, and $t_{p\gamma}^\prime$ is calculated numerically. The normalization of the proton spectrum is normally derived by relating the energy in photons to the energy in protons via the baryonic loading factor $f_e^{-1}$. Here, we keep the baryonic loading as a free parameter and only fix it later by fitting the UHECR observation.

The cosmic ray injection function is given by \equ{crrew} in \ref{app:modelindep} (for details, see \equ{crmaster}), built up from the individual source spectrum. 
The sources are assumed to be distributed following the chosen SFR, corrected by an evolution factor $(1+z)^\alpha$, down to very small redshifts.\footnote{Choosing a different cutoff, such as $z_{\mathrm{min}}=0.02$, affects the spectral shape beyond $10^{11} \, \mathrm{GeV}$ somewhat.} In order to test the statistical significance of a model, we fit our UHECR flux prediction to the Telescope Array data~\cite{AbuZayyad:2012ru} (see \ref{sec:statistics} for details). For a comparison to data from the Pierre Auger and HiRes experiments, see \ref{sec:expdata}.
The best-fit normalization and energy calibration translate into the baryonic loading $f_e^{-1}$ of the model -- assuming that $\dot{N} \simeq 1000 \, \mathrm{yr}^{-1}$ and $f_{\mathrm{thresh}} \simeq 0.3$, as discussed above.

\subsection{Ankle model for cosmic ray transition}

\begin{figure}[ht]
	\centering
	\includegraphics[width=\textwidth]{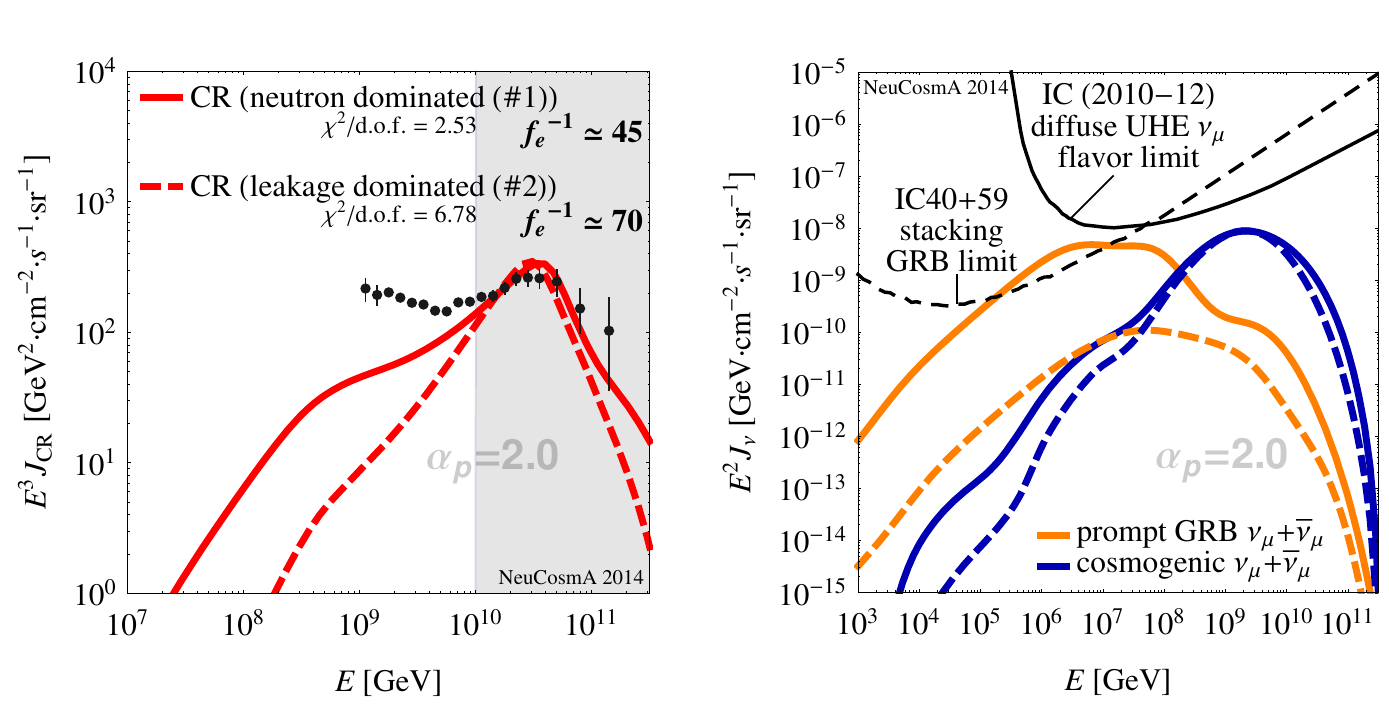}
	\mycaption{\label{fig:GRBCRprimer} Best-fit cosmic ray (left panel) and neutrino (right panel) fluxes as a function of energy for $\alpha_p=2$, the Hopkins \& Beacom star formation rate~\cite{Hopkins:2006bw}, and no cosmic evolution correction ($\alpha=0$). 
	In the left panel, the observed UHECR data from the Telescope Array~\cite{AbuZayyad:2012ru} is depicted as black circles
	together with our cosmic ray flux predictions (red curves). Additionally, the fit range is gray-shaded, and the $\chi^2$/d.o.f. and obtained $f_e^{-1}$ are given.
	In the right panel, the prompt (PeV) and cosmogenic (EeV) muon neutrino fluxes are given, together with the current bounds (see \ref{sec:statistics}). 
	The solid curves (neutron dominated (\#1)) correspond to our standard burst parameters with $\Gamma=300$ (see main text) and the neutron model; the dashed curves (leakage dominated (\#2)) use a higher $\Gamma=800$, leading to direct proton escape dominating at the highest energies.
	}
\end{figure}

In order to describe the extragalactic part of the ankle model, we use the energy range between $10^{10}$ and $10^{12} \, \mathrm{GeV}$ only; this energy range corresponds to the analytical discussion in \Sec~\ref{sec:modelindep}. Fits for two different model parameter sets, corresponding to the neutron model (\#1) and the direct escape model (\#2), are shown in \figu{GRBCRprimer}. The left panel depicts the UHECR fit, and the right panel the prompt and cosmogenic neutrino fluxes. Here a proton injection spectrum with $\alpha_p=2$, the Hopkins \& Beacom star formation rate, and no cosmic evolution correction ($\alpha=0$) are assumed. First of all, we note that the obtained baryonic loading from the fit lies between $45$ and $70$, in consistency with our analytical estimates from the previous section. The normalizations of prompt and cosmogenic neutrino spectra follow as a consequence of the UHECR fits; see right panel. In fact, the cosmogenic neutrino fluxes for both 
models are not very different because the cosmogenic neutrino production does not care how the protons escape from the source. Cosmogenic neutrinos can therefore be used as a model-independent test of the origin of the UHECRs up to higher redshifts, where the opacity for high-energy protons becomes large. The prompt neutrino fluxes are, on the other hand, very different: while the neutron model (\#1) is basically ruled out in consistency with \Ref~\cite{Ahlers:2011jj}, the direct escape model (\#2) flux is significantly below the current bounds (in that case, with a poorer $\chi^2/\mathrm{d.o.f.}$, though). The prompt neutrino flux prediction therefore strongly depends on the model for the UHECR escape.

\begin{figure}[tp]
	\centering
	\includegraphics[width=\textwidth]{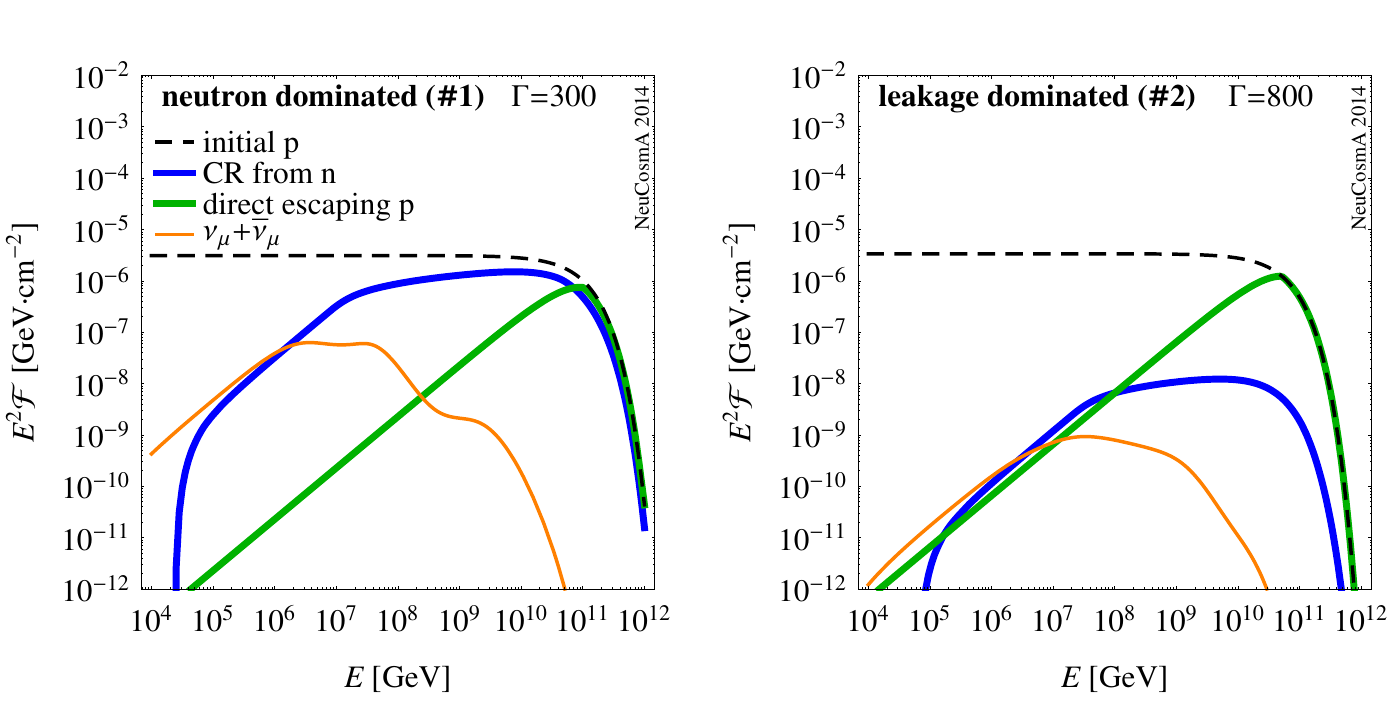} 
	\mycaption{\label{fig:GRBCRsourcemodels} Expected spectra from a single collision for our standard GRB parameters and $\Gamma=300$ (left panel) and $\Gamma=800$ (right panel), respectively ($z = 2$). The spectra are shown in the observer's frame, including only adiabatic losses due to the cosmic expansion, as in \Ref~\cite{Baerwald:2013pu}. Depicted are the input proton spectrum (in case all protons would just escape; thin dashed curve), the CR from neutron escape (thick blue/black curve), the contribution of directly escaping protons to the CR flux (thick green/gray curve), and the muon neutrino flux (after flavor mixing; thin orange/light gray curve).
	}
\end{figure}

It is illustrative to look into the single-collision source spectra for these two parameter sets, which are shown (protons without photohadronic and pair production losses) in \figu{GRBCRsourcemodels}. In this figure, the different components (initial injection, cosmic rays from neutrons, cosmic rays from direct proton escape, and neutrinos) are shown separately. It is important to note that the underlying theory in \Ref~\cite{Baerwald:2013pu} reduces to the conventional neutron model for high enough pion production efficiencies (left panel), whereas the direct escape component dominates if the pion production efficiency is low (right panel), \ie, different models are obtained for different sets of parameters.\footnote{Note that the neutron component is slightly harder than the input proton spectrum as a result of the high-energy (multi-$\pi$) processes.} Since the direct escape component is harder than the neutron escape component, the 
corresponding cosmic ray spectrum in \figu{GRBCRprimer}, left panel, dashed curve, becomes harder as well. Therefore, in principle, larger $\alpha_p$ can produce a better fit of the shape (the $\chi^2$/d.o.f. is significantly smaller), at the 
expense of a larger $f_e^{-1}$ (see \equ{ecrderiv2}, where $f_{\mathrm{bol}}$ is smaller then). Furthermore, a diffusive escape component could look closer to the neutron model, especially if Kolmogorov-like diffusion is assumed.
Therefore, we anticipate that both options are, in principle, possible.

Now one can argue how much the results depend on the chosen parameters. If the strong evolution case ($\alpha=1.2$) with more high-redshift bursts is used, which describes the observations better, the fit works equally well with somewhat larger predicted cosmogenic neutrino fluxes and a clearly excluded prompt neutrino flux in the neutron model (\#1). However, the main qualitative difference is the larger required baryonic loading of around $200$, which comes from the larger value of $f_z$; see, \eg, \equ{crrew2}. In a sense, the SFR evolution is therefore the most modest assumption one can make in order to obtain a baryonic loading which has been anticipated to be realistic so far. As far as the dependence on the assumed GRB parameters is concerned, we perform a more detailed parameter space study for the ankle model in \Sec~\ref{sec:scan}.

\subsection{Dip model for cosmic ray transition}

For the dip model, we extend the fit energy range to between $10^{9}$ and $10^{12} \, \mathrm{GeV}$. This energy range is large enough to cover the pair production dip, but does not extend to lower energies where the diffusion of cosmic rays on the intergalactic magnetic fields is expected to become important for spectral effects~\cite{Berezinsky:2002nc,Lemoine:2004uw,Kotera:2007ca}.

\begin{figure}[t!]
	\centering
	\includegraphics[width=\textwidth]{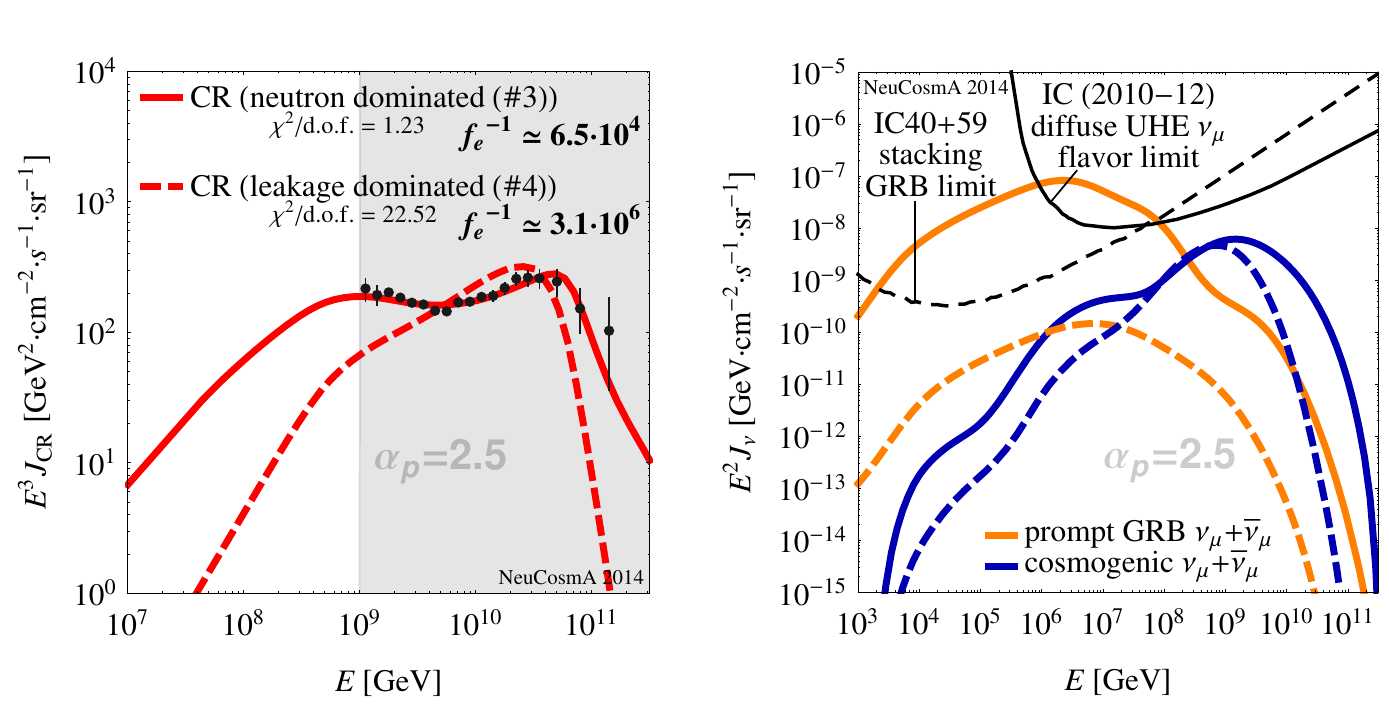} \\
	\mycaption{\label{fig:GRBCRprimerDIP} Best-fit cosmic ray (left panel) and neutrino (right panel) fluxes as a function of energy for $\alpha_p=2.5$, the Hopkins \& Beacom star formation rate~\cite{Hopkins:2006bw}, and no cosmic evolution correction ($\alpha=0$). In the left panel, the observed UHECR data from the Telescope Array~\cite{AbuZayyad:2012ru} is depicted as black circles
	together with our cosmic ray flux predictions (red curves). Additionally, the fit range is gray-shaded, and the $\chi^2$/d.o.f. and obtained $f_e^{-1}$ are given.
	In the right panel, the prompt (PeV) and cosmogenic (EeV) muon neutrino fluxes are given, together with the current bounds (see \ref{sec:statistics}). 
	The solid curves (neutron dominated (\#3)) correspond to our standard burst parameters with $\Gamma=300$ (see main text) and the neutron model; the dashed curves (leakage dominated (\#4)) use a higher $\Gamma=600$ and lower $L_{\mathrm{iso}} = 10^{50.5} \, \mathrm{erg} \, \mathrm{s}^{-1}$, leading to direct proton escape dominating at the highest energies.
}
\end{figure}

\begin{figure}[tp]
	\centering
	\includegraphics[width=\textwidth]{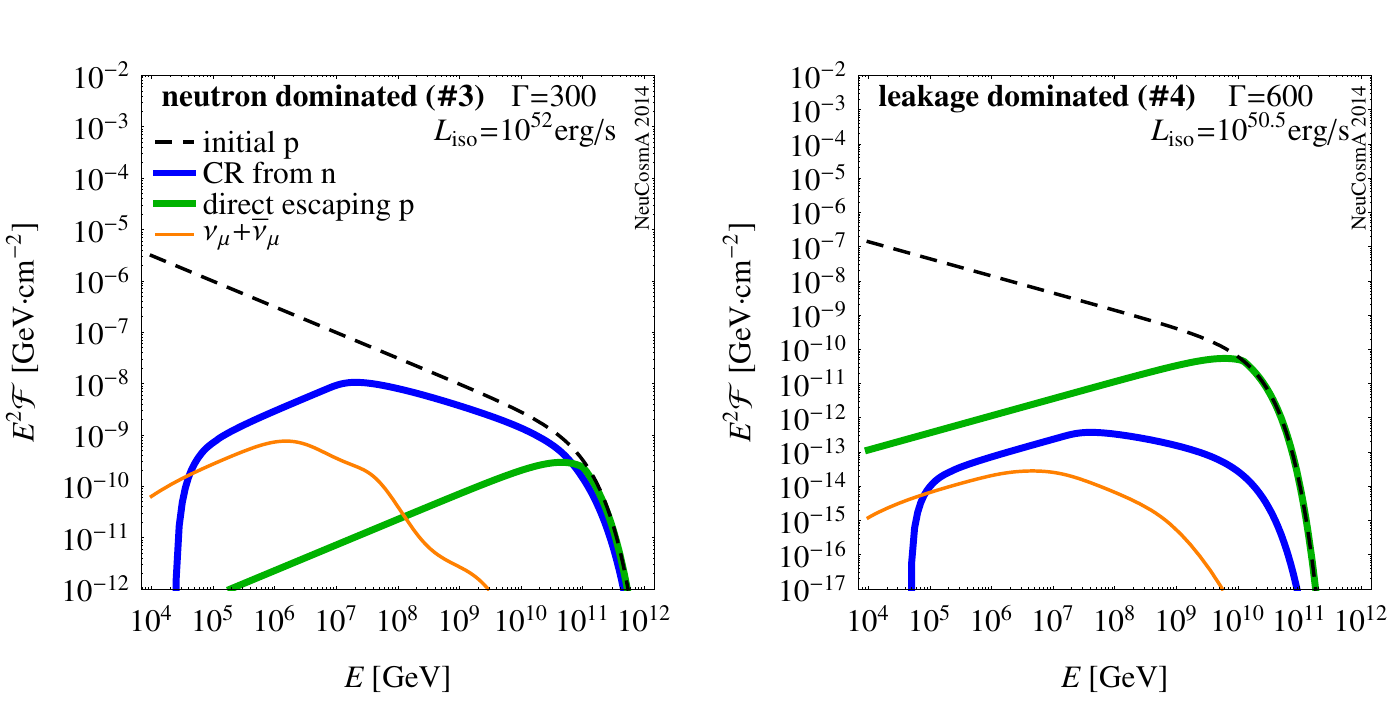}
	\mycaption{\label{fig:GRBCRsourcemodelsDIP} Expected spectra from a single collision for our standard GRB parameters and $\Gamma=300$, $L_{\mathrm{iso}} = 10^{52} \, \mathrm{erg} \, \mathrm{s}^{-1}$ (left panel) and $\Gamma=600$, $L_{\mathrm{iso}} = 10^{50.5} \, \mathrm{erg} \, \mathrm{s}^{-1}$ (right panel), respectively ($\alpha_p=2.5$, $z = 2$). The spectra are shown in the observer's frame, including only adiabatic losses due to the cosmic expansion, as in \Ref~\cite{Baerwald:2013pu}. Depicted are the input proton spectrum (in case all protons would just escape; thin dashed curve), the CR from neutron escape (thick blue/black curve), the contribution of directly escaping protons to the CR flux (thick green/gray curve), and the muon neutrino flux (after flavor mixing; thin orange/light gray curve).
	}
\end{figure}

We show two possible scenarios in \figu{GRBCRprimerDIP}, one for the neutron model (\#3) and one for the direct escape-dominated case (\#4). The corresponding single-collision spectra without photohadronic and pair production energy losses are shown in \figu{GRBCRsourcemodelsDIP}. Compared to the ankle model, the difference between the two examples is even more extreme in terms of the prompt neutrino flux: while the neutron model is clearly excluded, the direct escape model is significantly below the current bounds (right panel of \figu{GRBCRprimerDIP}). The reason is that the neutrino production at the peak of the prompt neutrino spectrum (say $10 \, \text{PeV}$) follows the cosmic ray spectrum at about a factor of 20 higher energy (say $200 \, \text{PeV}$). At this energy, the dip model reproduces the observed cosmic ray spectrum much better than the ankle model (left panel of \figu{GRBCRprimerDIP}), which however implies that the cosmic ray flux has to be larger there than for the ankle model. As a 
consequence, the neutrino 
flux 
overshoots the prompt flux bounds. 
This can be 
avoided in the direct escape case, where however the spectral fit is not as good as for the neutron model because of the harder spectrum.\footnote{One could change the spectral index to improve the fit here, but at the expense of the baryonic loading. Changing the spectral index from $\alpha_p=2.5$ to $\alpha_p=2.7$ decreases the bolometric correction $f_{\mathrm{bol}}$ by a factor of 30, which leads to an increase of the needed baryonic loading $f_e^{-1}$ by the same factor.} 

The challenge for the dip model in the context of GRBs is actually the required baryonic loading $f_e^{-1} \simeq 10^5$, needed to describe the observation. This value is significantly larger than for the ankle model and comes from the correction factor $f_{\mathrm{bol}}$, which is smaller the steeper the spectrum is. Of course, $f_{\mathrm{bol}}$ depends on the minimal proton energy as well, especially for $\alpha_p>2$, but it is nevertheless clear that the conventional assumptions for the baryonic loading of GRBs are challenged in that model. Note that the UHECR baryonic loading $f_{\mathrm{bol}} \, f_e^{-1} \sim 10$, as for the ankle model.

Another interesting issue is how to obtain the spectral injection indices $\alpha_p \simeq 2.5 - 2.7$, required for the dip model, from the normally anticipated injection indices for Fermi shock acceleration $\alpha_p \simeq 2.0 - 2.2$. One possibility could be the effect of turbulences on Fermi shock acceleration, which reduces the effectivity of the acceleration and changes the expected spectral index to $\alpha_p \geq 2.3$~\cite{Lemoine:2006gg}. On the other hand, it has been proposed for AGNs to use a distribution function on the maximal proton energy in combination with an injection index $\alpha_p \simeq 2$ to generate this steep spectrum~\cite{Kachelriess:2005xh}. We anticipate that this works for GRBs as well if an appropriate luminosity distribution function were chosen (which translates into a distribution of the maximal proton energy). However, we do not expect that this affects the required baryonic loading qualitatively, since most bursts 
will then not have high enough maximal proton energies to match the ultra-high energy part of the observed spectrum (\ie, $\dot N$ is effectively smaller because it only captures the bursts with high enough proton energies), and this needs to be compensated by the baryonic loading.

Based on these observations and on parameter space scans (not shown here), we therefore conclude that the combined production-propagation GRB internal shock model is challenged in the context of the dip cosmic ray transition model, because a) very high baryonic loadings are required, and b) the prompt neutrino flux easily overshoots the neutrino flux bound in the neutron model case. We therefore focus on the ankle model in the following.

\section{Statistical analysis for cosmic ray ankle model}
\label{sec:scan}

In the previous section, we illustrated that the ankle model for the cosmic ray transition is compatible with the current neutrino bounds for certain parameter sets if the cosmic rays can escape other than by neutron escape. While this observation is quite generic, we now discuss a) for which parts of the parameter space this observation holds and b) how it depends on the escape model. Note that the fit of the cosmic ray observation depends on the shape of the cosmic ray escape spectrum and, possibly, on the transition of the cosmic ray contribution at lower energies. Therefore, the results presented in this section are less generic and somewhat more model-dependent than the general observations in the previous section. We first discuss the dependence on the cosmic ray escape model, then on the redshift evolution of the sources.

\subsection{Impact of cosmic ray escape model}

\begin{figure}[tp]
	\centering
	\includegraphics[width=0.5\textwidth]{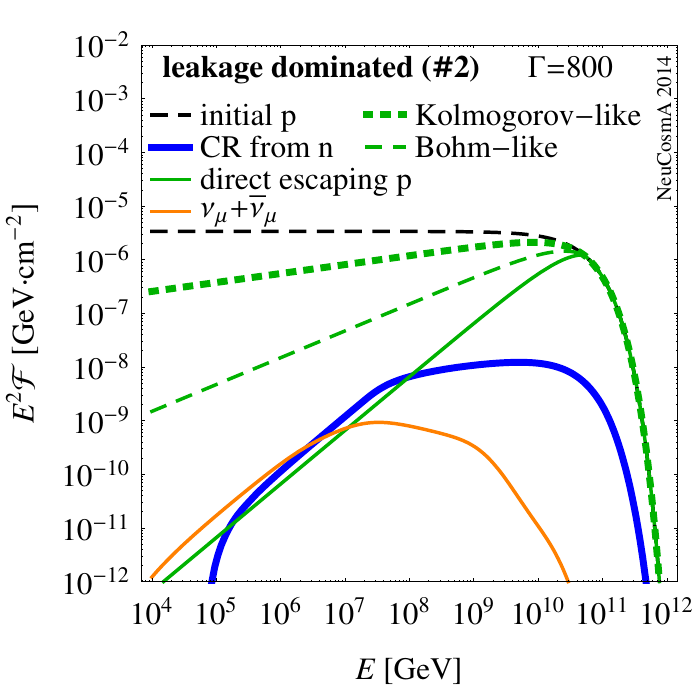}
	\mycaption{\label{fig:diff} Different assumptions for additional escape components (in a single collision), together with neutrino and neutron escape spectra. Figure corresponds to the example in the right panel of \figu{GRBCRsourcemodels}.}
\end{figure}

The particle escape depends on details of acceleration, turbulence, and dynamics, and is therefore very much model dependent. We consider three different possibilities (for details, see \Ref~\cite{Baerwald:2013pu}), illustrated for one set of parameters in the single-collision spectra in \figu{diff}:
\begin{description}
	\item[Neutron model.] The cosmic rays escape as neutrons and the protons are magnetically confined. This is the assumption frequently used in the literature. In \figu{diff}, it corresponds to the thick dark/blue curve.
	\item[Direct escape.] All protons from the edges of the shells will escape over a width of $R_L'$. That is, although the protons are magnetically confined, they can still leave the source if the distance to the edge is comparable to the Larmor radius. The fraction of directly escaping protons is $\propto R_L'/\Delta d' \propto E'$, where $\Delta d'$ is the shell width. This fraction is more or less guaranteed, and corresponds to the fraction of particles escaping without scattering. However, depending on the burst parameters, it can be sub-dominant compared to the neutrons if the pion production efficiency is high enough, or the maximal proton energy is limited by synchrotron or photohadronic losses. The direct escape spectrum is very hard, as illustrated in \figu{diff}. Note that the shell width increases (after the collision) in the same way as the Larmor radius for an adiabatic index $4/3$ (for a relativistic gas/plasma)~\cite{Baerwald:2013pu}, which means that direct escape from one shell does not 
depend on subtleties of the time evolution.
	\item[Diffusive escape.] A less conservative estimate for the escape of the protons is that a fraction $\lambda'/\Delta d'$ can escape, where $\lambda'=\sqrt{D' \, t'_{\mathrm{dyn}}}$ is the diffusion length over the dynamical timescale $t'_{\mathrm{dyn}}$, and $D'$ is the (spatial) diffusion coefficient. For Bohm diffusion, $D' \propto R_L' \propto E'$; for Kolmogorov diffusion, $D' \propto (E')^{1/3}$; and the fraction of escaping particles is proportional to the square root of that. Assuming that at the highest energies all particles can escape (efficient diffusion), as for direct escape, different possible diffusion components are illustrated in \figu{diff}. In the following, we will use the (more conservative) Bohm case for illustration. In that case, the diffusion coefficient is inversely proportional to the magnetic field, $D' \propto B'^{-1}$. Note, however, that in this case diffusion length and shell width do not scale in exactly the same way, which means that this scenario can only be used for a 
rough quantitative, more optimistic estimate for particle escape. 
\end{description}

\begin{figure}[tp]
	\centering
	\vspace*{-0.18\textheight}
	\includegraphics[width=\textwidth]{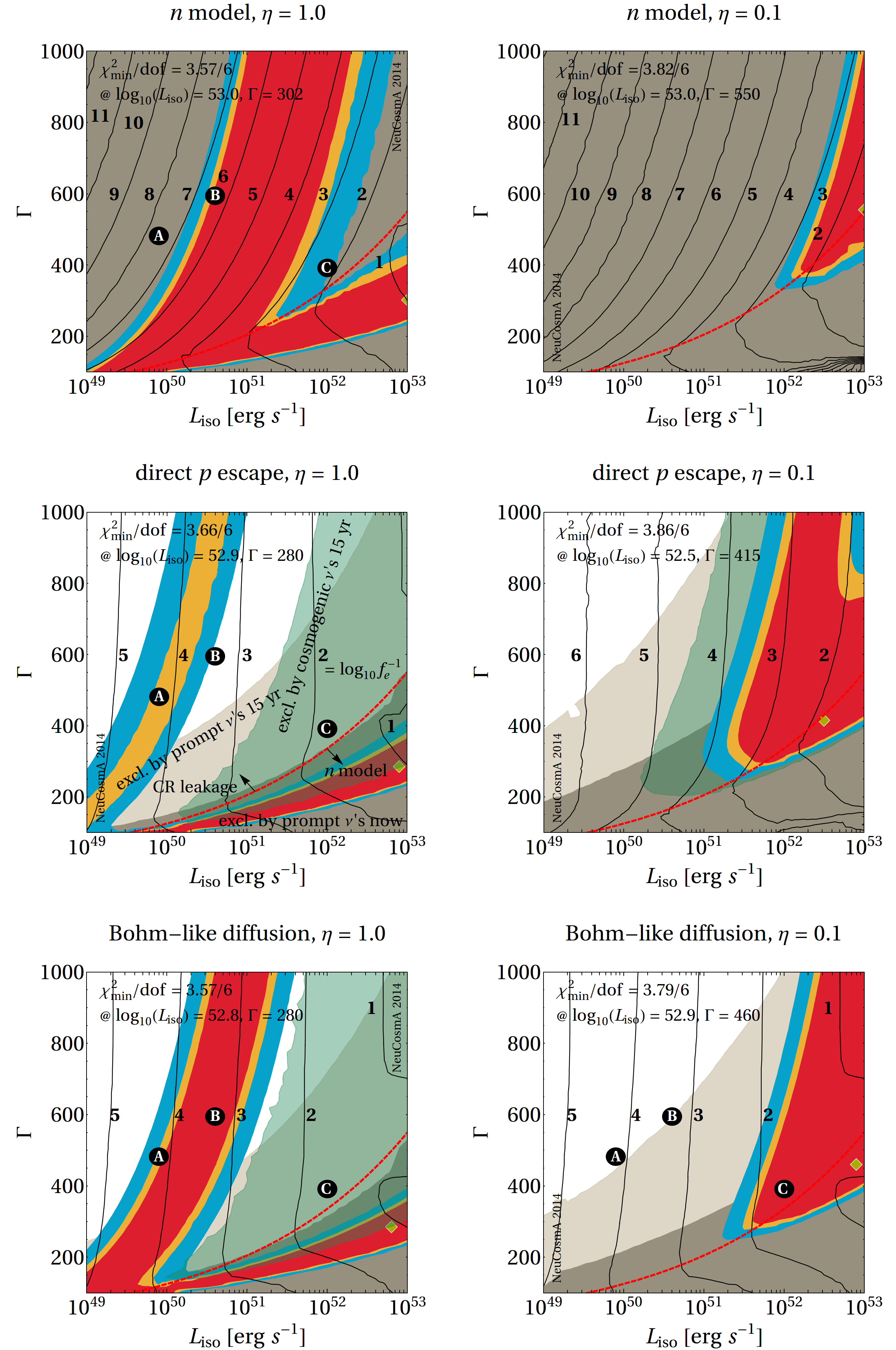}
	\mycaption{\label{fig:fits} Filled contours: allowed regions (red/darkest gray: 90\% C.L., yellow/light gray: 95\% C.L., blue/darker gray: 99\% C.L.) as a function of $L_{\mathrm{iso}}$ and $\Gamma$ for the fit to cosmic ray data from the Telescope Array~\cite{AbuZayyad:2012ru}, in the energy range between $10^{10}$ and $10^{12} \, \mathrm{GeV}$, assuming that all GRBs are alike. In the left column, $\eta=1.0$, in the right column, $\eta=0.1$. The different rows correspond to the neutron escape (first row), direct proton escape (second row), and Bohm-diffusive escape (third row) UHECR escape models. The red-dashed curve separates the ``direct escape dominated'' region (above curve) and the ``neutron model'' region. The dark gray shading marks the current IceCube-excluded region; the light gray shading, the expected exclusion from the GRB analysis after 15 years; and the green shading, the expected exclusion from the cosmogenic neutrino analysis after 15 years (note that in the top left panel this region, while present, is totally contained within the exclusion region from prompt neutrinos and is therefore not visible). The iso-baryonic loading contours (numbers are 
	$\mathrm{log}_{10} f_e^{-1}$) are shown also, where the baryonic loading is obtained as a result of the fit. Here $\alpha_p = 2.0$, $t_v = 0.01$ s (in the source frame), and SFR evolution of the sources by Hopkins \& Beacom ($\alpha = 0$) have been chosen. 
}
\end{figure}

\begin{figure}[tp]
	\centering
	\vspace*{-0.15\textheight}
	\includegraphics[trim=.5cm 0.1cm 2cm 5cm,width=\textwidth]{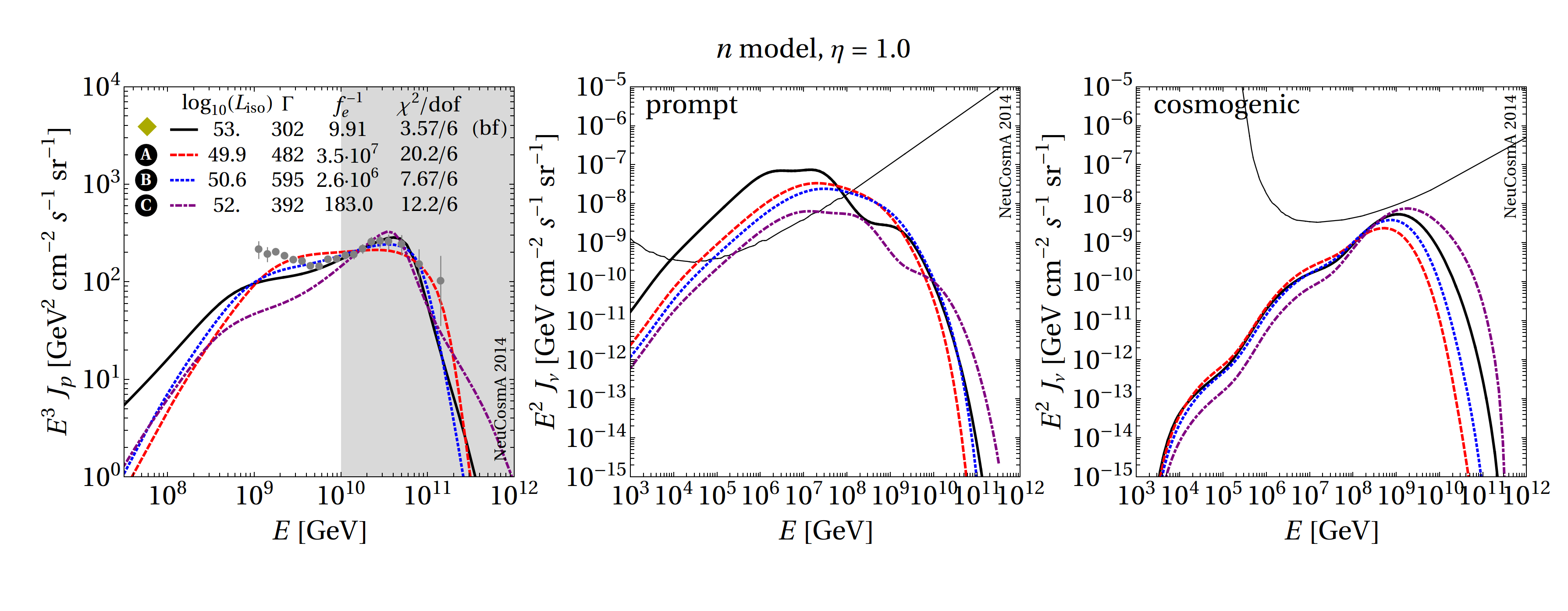} \\
	\includegraphics[trim=.5cm 0.1cm 2cm 5cm,width=\textwidth]{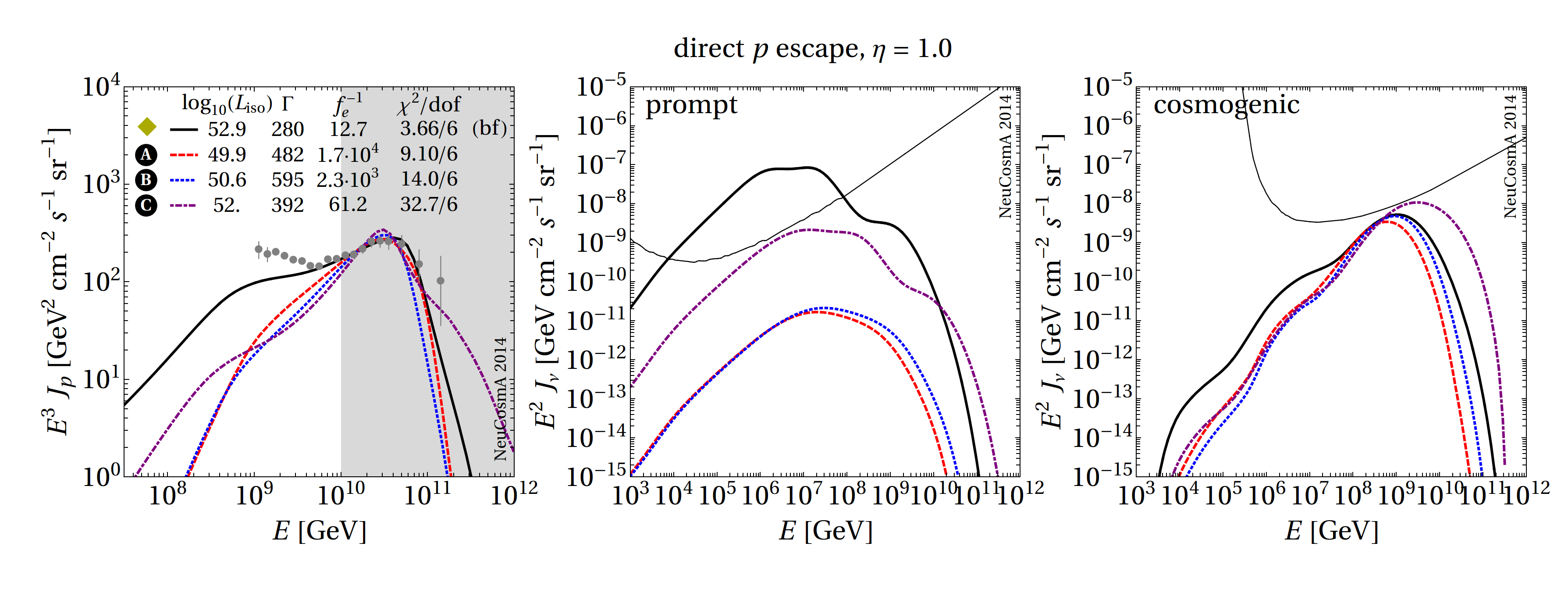} \\
	\includegraphics[trim=.5cm 0.1cm 2cm 5cm,width=\textwidth]{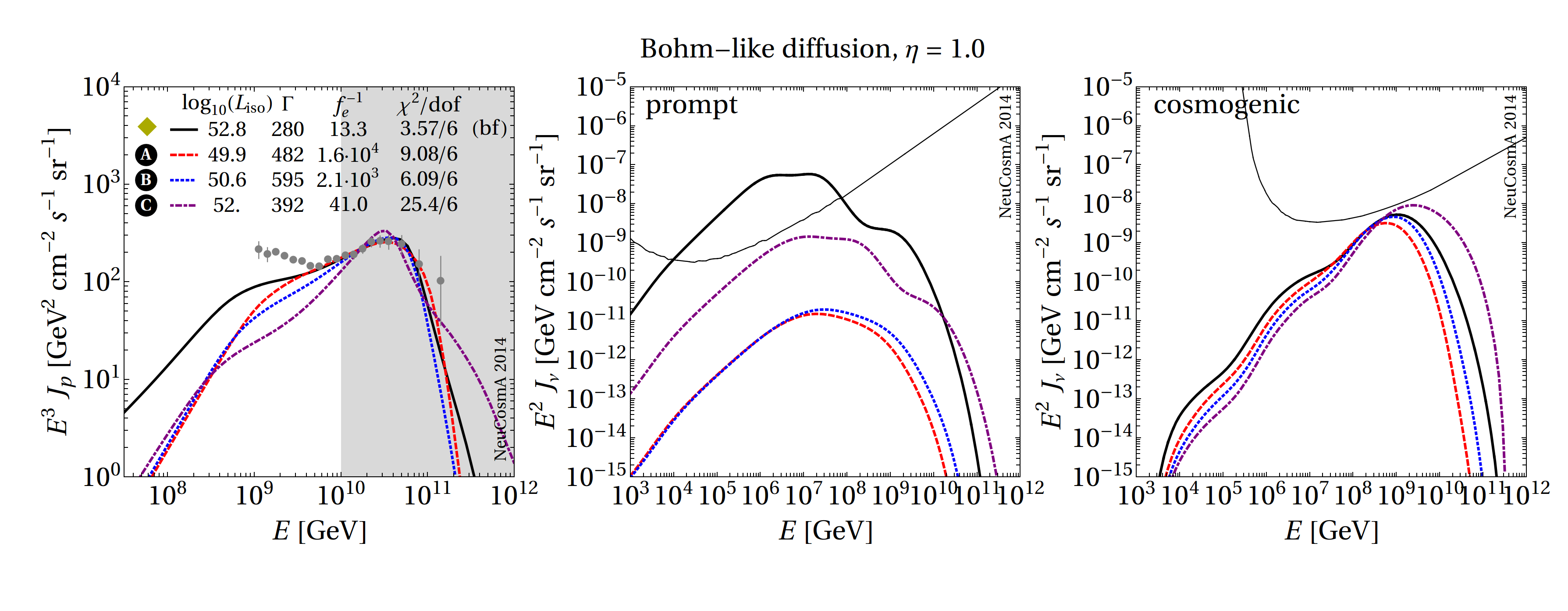} \\
	\includegraphics[trim=.5cm 3cm 2cm 5cm,width=\textwidth]{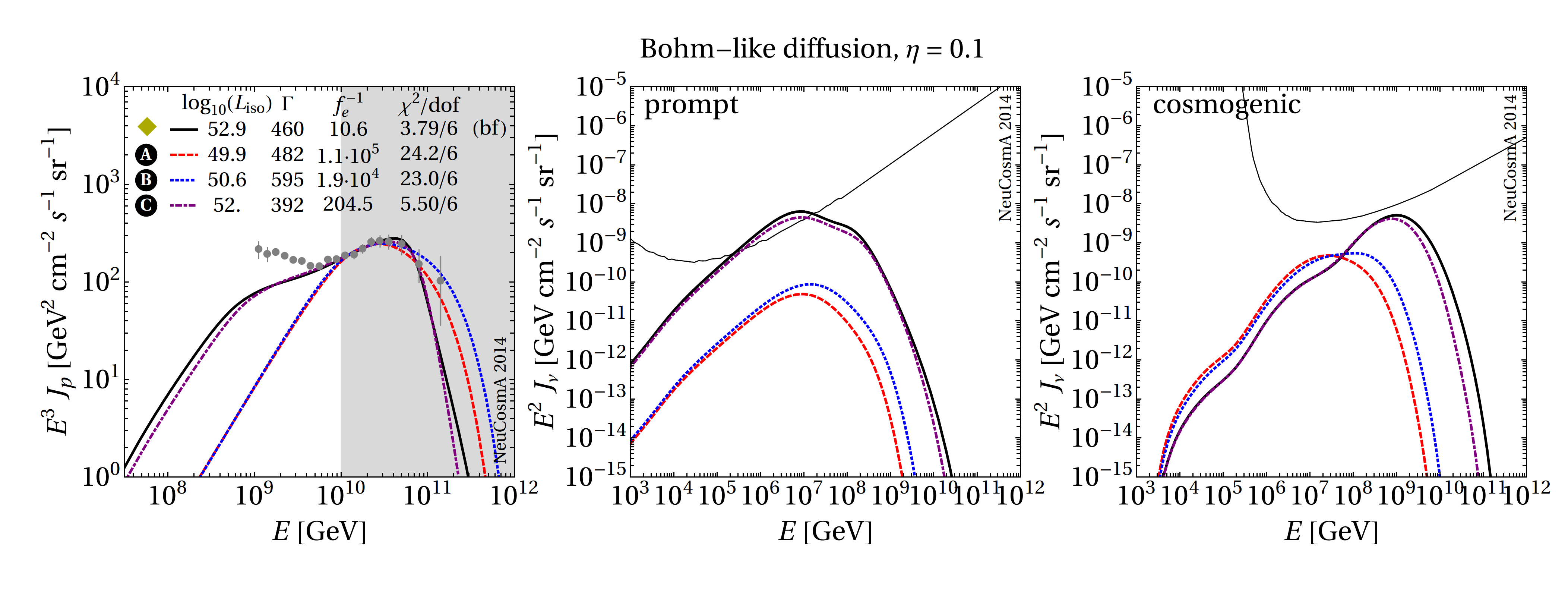}
	\mycaption{\label{fig:spectra}Cosmic ray, prompt neutrino, and cosmogenic neutrino spectra (in columns) from GRBs in the range $z=0$ to $6$, for selected points in the parameter space plane $\Gamma$ vs.~$L_\mathrm{iso}$, corresponding to the markers in \figu{fits}. The different rows correspond to the upper left, middle left, lower left, and lower right panels in \figu{fits}. The fit range is gray-shaded.}
\end{figure}

Our main results are presented in \figu{fits} as parameter space scans as functions of $L_{\mathrm{iso}}$ and $\Gamma$, and in \figu{spectra}, where the cosmic ray, prompt, and cosmogenic neutrino spectra for several points marked in \figu{fits} are shown -- similar to our earlier figures. The different rows in \figu{fits} correspond to the three different assumptions for the UHECR escape discussed above; the different columns, to two different acceleration efficiencies determining the maximum proton energies. The different rows in \figu{spectra} correspond to different panels in \figu{fits}, as indicated. 

Before we come to the details, let us briefly summarize the procedure. We compute the neutrino and cosmic ray spectra for an individual GRB with certain parameters using the GRB source model, we distribute that over redshift based on a choice of the source evolution, assuming that all bursts are alike in the cosmologically comoving frame, and we propagate the cosmic rays down to redshift zero. Then we fit the predicted cosmic ray spectrum to the Telescope Array surface detection data \cite{AbuZayyad:2012ru} between $10^{10}$ and $10^{12} \, \mathrm{GeV}$, \ie, we determine the normalization and the energy calibration within the systematic energy uncertainty of the experiment. This fit is shown as filled contours in the panels of \figu{fits}, where the best-fit point is marked by a diamond, and the best-fit parameters are given in the upper left corners. Thus, the fit regions represent the cosmic ray observation only. From the normalization, we can then derive the baryonic loading required within each GRB for 
this source model, which 
is overlaid as solid (unfilled) contours in terms of 
$\mathrm{log}_{10} f_e^{-1}$. In addition, we compute the excluded regions from neutrino observations by comparing the prompt and cosmogenic neutrino fluxes with the current GRB stacking and ultra-high energy analysis bounds, respectively. These excluded regions are for current and future (15 years) data shown as shaded regions in \figu{fits}, as labeled in the plots. Details of the statistical analysis are described in \ref{sec:statistics}. Note that the GRB source model determines the pion production efficiency and the shapes of the prompt neutrino and ejected cosmic ray spectra, while the cosmic ray propagation model determines the shape of the cosmogenic neutrino spectrum and observed cosmic ray spectrum. The normalization of the neutrino spectra is just a consequence of the requirement that GRBs ought to describe the UHECR observation, \ie, of the normalization of the observed cosmic ray spectrum to data.

Let us first of all discuss the neutron model (first row in \figu{fits}). Considering the fit contours, it is clear that the neutron model provides an excellent fit to cosmic ray observations for reasonable parameter sets if only the cosmic rays are considered (and not also the neutrinos, as we will see). For instance, the best fit for $\eta=1$ (left panel) is at $\Gamma \approx 302$ and $L_{\mathrm{iso}} \approx 10^{53} \, \mathrm{erg \, s^{-1}}$, and corresponds to $f_e^{-1} \approx 10$, while the best fit for $\eta=0.1$ (right panel) is at a higher $\Gamma$. This shift is a consequence of the less effective particle acceleration leading to lower maximal proton energies, which reduces the quality of the fits. For a more detailed discussion of the dependence on $\eta$ and its impact on the conclusions, see \ref{sec:maxe}. In short, a lower acceleration efficiency shifts the range in parameter space in which a certain maximal proton energy is reached to the right, 
roughly along the red dashed curve. This effect can be seen in the plots of \figu{epmax} (in \ref{sec:maxe}), where each panel shows the contours for $E_{p,\mathrm{max}} = 10^{10.25} \, \mathrm{GeV}$ (black solid curve) and $10^{10.75} \, \mathrm{GeV}$ (black dotted curve) for different $\eta$-values, including intermediate cases.
Because these best-fit parameter values are in a plausible range, it was widely accepted that the neutron model for cosmic ray escape describes the UHECR production in GRBs. This hypothesis was however rejected in \Refs~\cite{Ahlers:2011jj,Abbasi:2012zw} by relating the cosmic ray and neutrino spectra and by using the current neutrino bounds. Indeed, the shaded regions in the first row of \figu{fits}, corresponding to the current IceCube-excluded regions, span the whole parameter space, and the neutron model can be excluded everywhere already. Compared to earlier references, we have the source model prediction for the baryonic loading as well, which increases tremendously above the red (dashed) curve. The reason is that the pion production efficiency drops there, and in fact it turns out that above this curve other escape mechanisms become important, whereas below this curve neutron escape dominates in all of our models.
Note that we allow for arbitrarily high baryonic loadings here derived from the normalization, and the required baryonic loadings for this model have to be extremely high in the upper left corner. In practice, depending on the minimal proton energy at injection, it can be estimated that baryonic loadings $\gtrsim 10^4$ could mean that $pp$ self-interactions among the protons become important, an effect which we do not consider. In addition, gamma-rays from $\pi^0$ decays produced by $p \gamma$ interactions or proton synchrotron radiation could violate the Fermi bounds, if they can escape.\footnote{The restrictions will be less severe in the other cosmic ray escape models, since the neutral and charged pion photoproduction are directly related and the pion production efficiency is low. In addition, proton synchrotron losses typically do not constrain the maximal proton energy. For a more detailed discussion on the impact of the gamma-ray bounds for efficient pion production, see \Ref~\cite{Ahlers:2011jj}.}
Several exemplary spectra for the upper left panel in \figu{fits} are shown in the first row of \figu{spectra}, marked by the diamond (best-fit) and dots. Good fits are obtained for the best-fit point and point~B, bad fits for points~A (requires an upscaling of the energy) and~C (too high proton energy, leading to a strong spectral peak). However, one can clearly see that all prompt neutrino fluxes overshoot the current bounds. At point~C, the cosmogenic neutrino flux will eventually become larger because of the larger $E'_{p,\mathrm{max}}$. 

As already indicated, the situation changes completely if direct (middle row of \figu{fits}) or diffusive escape (lower row of \figu{fits}) of protons is included at the highest energies. Whereas all models are similar below the red dashed curve, where neutron escape dominates, there are two important differences above that curve: first of all, the required baryonic loadings are significantly smaller because direct or diffusive cosmic ray escape dominates, and second, the neutrino bounds can only reach as far as the pion production efficiency allows for significant neutrino production. In fact, in the left panels, a part of the parameter space is even probed better by cosmogenic neutrinos in the future, which do not care about the cosmic ray escape mechanism. From these plots (taking into account intermediate values of $\eta$, as discussed in \ref{sec:maxe}), it is clear that 
\begin{enumerate}
 \item There are parts of the parameter space with moderate $\Gamma \gtrsim 400$ and baryonic loadings $10 \lesssim f_e^{-1} \lesssim 100$ which are still allowed if cosmic rays can efficiently escape by diffusion (see lower right panel), which however can be tested by future IceCube data.
\item There are parts of the parameter space with either extremely large $\Gamma$ or extremely large $f_e^{-1}$ which are inaccessible by future IceCube bounds; see also \ref{sec:maxe} for more details. Since these parts require extreme parameters (on average), they will be challenged elsewhere. For example, the branch in the middle left panel will disappear if the energy calibration of the cosmic ray measurements can be improved (see \ref{sec:maxe}).
\end{enumerate}
Therefore, it is clear that if IceCube does not find high-energy neutrinos from GRBs, it will be very difficult to maintain the paradigm that GRBs are the sources of the UHECRs in the internal shock model. However, current bounds are not yet strong enough if cosmic rays can escape by mechanisms other than neutron production. 

We also show the spectra for several points in \figu{spectra}. Comparing the middle two rows (direct escape versus diffusion), it is clear that the cosmic ray spectra for direct escape are harder, and therefore provide worse fits. While the cosmic ray spectra in the second row (direct escape) appear to be similar for points~A and~B, the fit for point~B is still much worse because the energy recalibration is penalized. In all cases, the best-fit prompt neutrino fluxes overshoot the current bound for the prompt neutrino flux. For $\eta=1$ (upper three rows), the maximal proton energy for point~C is high, and therefore the cosmogenic neutrino flux is high, too. Comparing the lower two rows ($\eta=1$ and $\eta=0.1$ for diffusive escape), one can easily see that lower acceleration efficiencies help for the shape, but too low proton energies (points~A and~B) are penalized because of the energy calibration error.

\begin{figure}[tp]
	\centering
	\includegraphics[width=\textwidth]{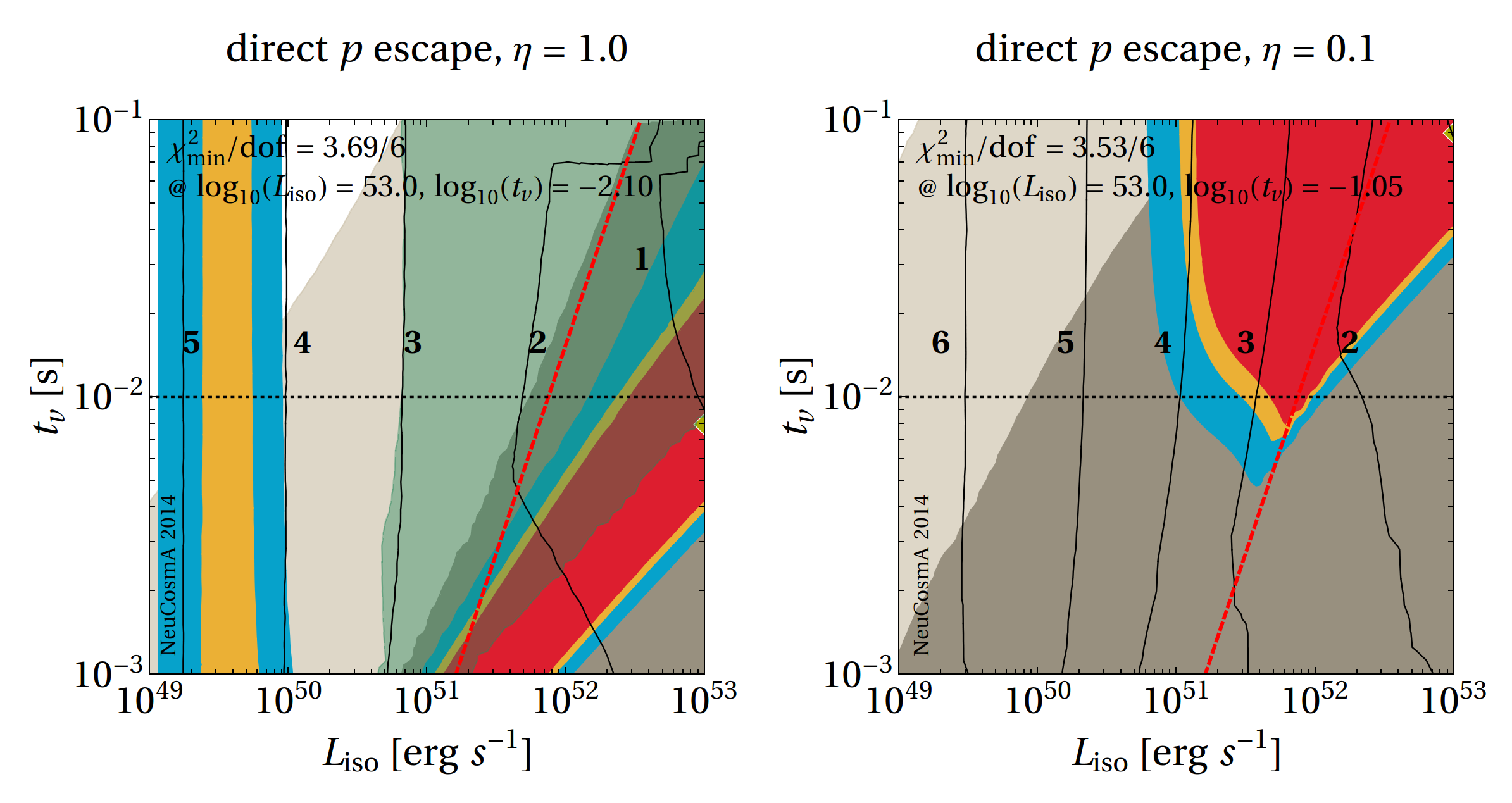}
	\mycaption{\label{fig:tvdep} Same as \figu{fits}, middle panels (direct escape), as a function of $L_{\mathrm{iso}}$ and $t_v$ for a fixed $\Gamma = 300$. The left (right) panel corresponds to an acceleration efficiency of $\eta = 1.0$ ($0.1$). The horizontal dotted lines mark the standard value of $t_v = 10^{-2}$ s.}
\end{figure}

So far, we have fixed several of the parameters and have shown the dependence as a function of $L_{\mathrm{iso}}$ and $\Gamma$. In \figu{tvdep}, we instead fix $\Gamma$ and vary $t_v$. While the result looks qualitatively different, it does not reveal any new allowed regions. This is expected, since the pion production efficiency roughly scales $\propto L_{\mathrm{iso}}/(\Gamma^4 \, t_v \, \epsilon'_\gamma )$~\cite{Waxman:1997ti,Guetta:2003wi}, with $\epsilon'_\gamma$ chosen to be the photon break energy in the SRF, and the neutron model and prompt neutrino production follow the pion production efficiency. That is, there is a degeneracy among $L_{\mathrm{iso}}$, $\Gamma^4$, $t_v$, and $\epsilon'_\gamma$, which means that the relevant features will be visible in any of the relevant parameter combinations. The other regions follow, more or less, the maximal proton energy; see Fig.~5 in \Ref~\cite{Baerwald:2013pu}.

\subsection{Effect of cosmological source evolution}
\label{sec:fluct}

\begin{figure}[tp]
	\centering
	\includegraphics[width=\textwidth]{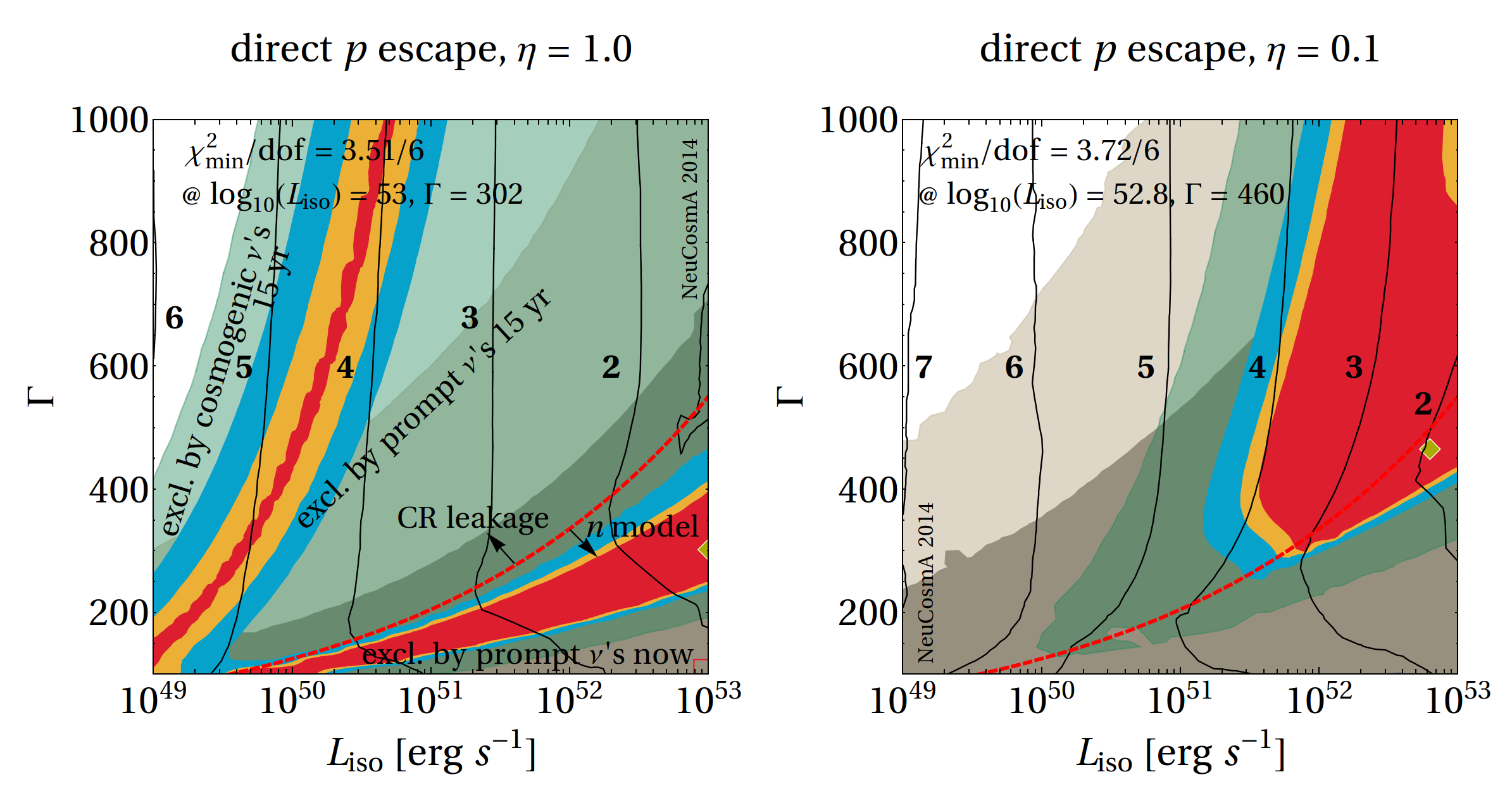}
	\mycaption{\label{fig:grbevol} Same as \figu{fits}, middle panels (direct escape), but assuming GRB evolution of the sources.}
\end{figure}

So far, we have assumed SFR evolution of the sources. It is, however, plausible that GRBs evolve more strongly, which we will test here. In addition, because of the very limited statistics, it is not possible yet to exclude that there are local deviations from the SFR and the GRB rate, which impact the cosmic ray flux at the highest energies.

We show in \figu{grbevol} the fits as in \figu{fits}, middle panels (direct escape), but for GRB evolution of the sources (star formation rate times $(1+z)^{1.2}$~\cite{Kistler:2009mv}). At a first glance, from comparing to \figu{fits}, there is qualitative impact on the fit regions. There are, however, two important differences. First of all, the required baryonic loading for a specific parameter set is higher everywhere, which is a consequence of the higher factor $f_z$ in \equ{cre2} for the strong evolution case. This is also the reason why we have not taken the GRB evolution as our baseline case. Second, the cosmogenic neutrino bound will have a much larger impact than in the SFR case, which is expected because it is well known that the cosmogenic neutrino flux is higher for the strong evolution case.

\begin{figure}[tp]
 \centering
 \includegraphics[width=\textwidth]{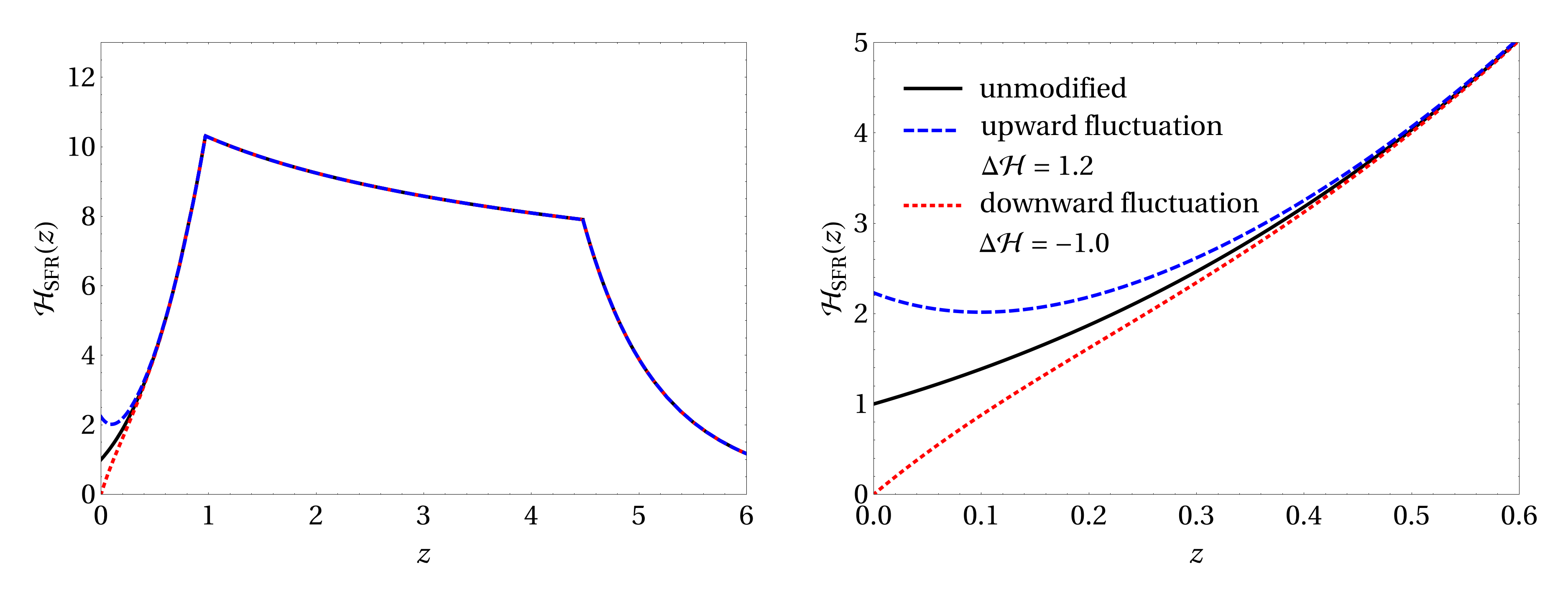}
 \mycaption{\label{fig:hmod}{\it Left:} Unmodified SFR evolution with redshift (black, solid line), and SFR with a local upwards (blue, dotted line) and downwards fluctuation (red, dashed line). {\it Right:} Magnification for low redshifts.}
\end{figure}

\begin{figure}[tp]
  \centering
  \includegraphics[width=\textwidth]{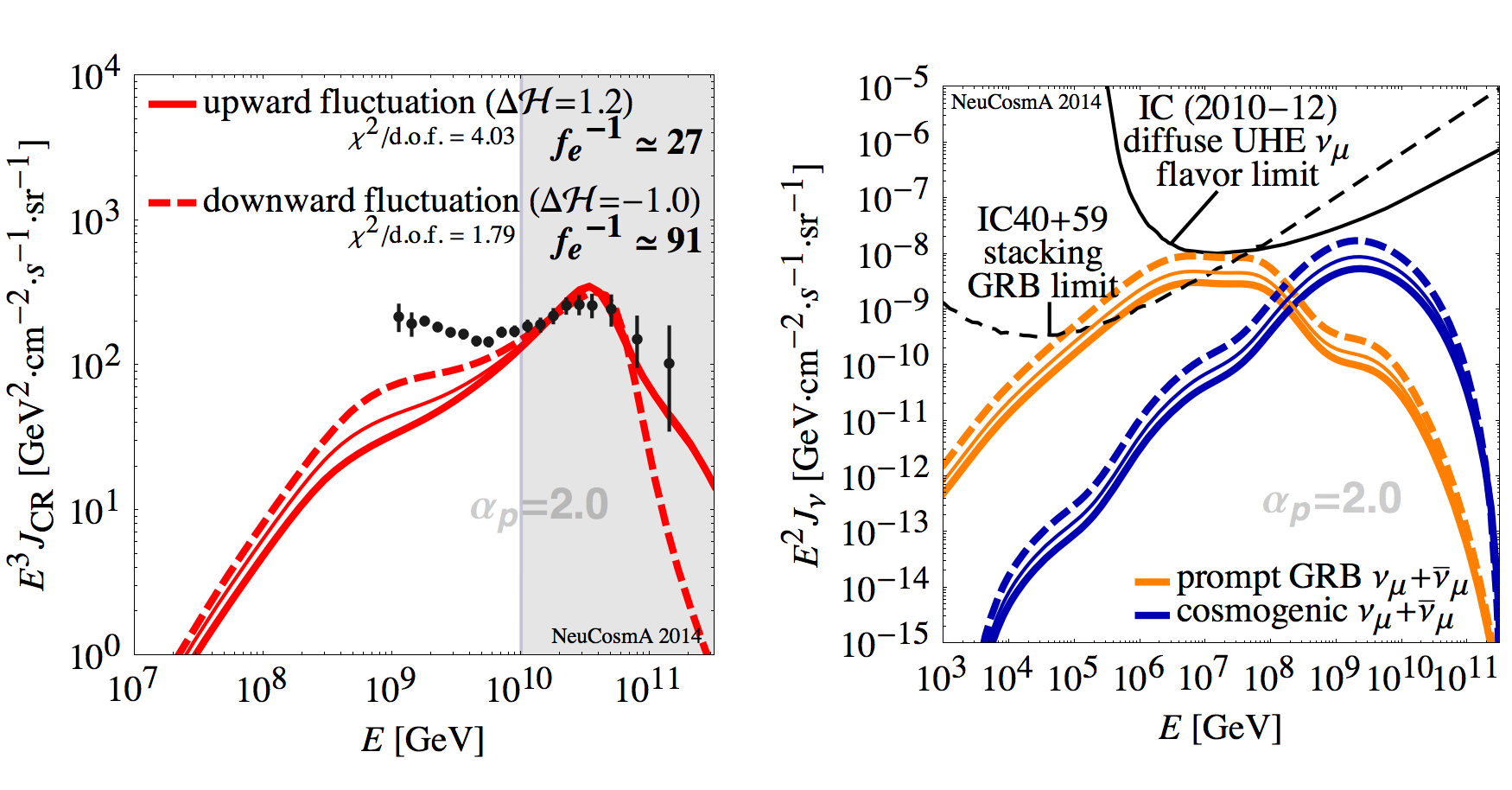}
  \mycaption{\label{fig:fluc} The effect of ensemble fluctuations on the CR (left column) and neutrino spectra (right column), for model~1 in \figu{GRBCRprimer}. The original results obtained in the absence of fluctuations are included as thin lines for comparison.}
\end{figure}

\begin{figure}[tp]
	\centering
	\includegraphics[width=\textwidth]
	{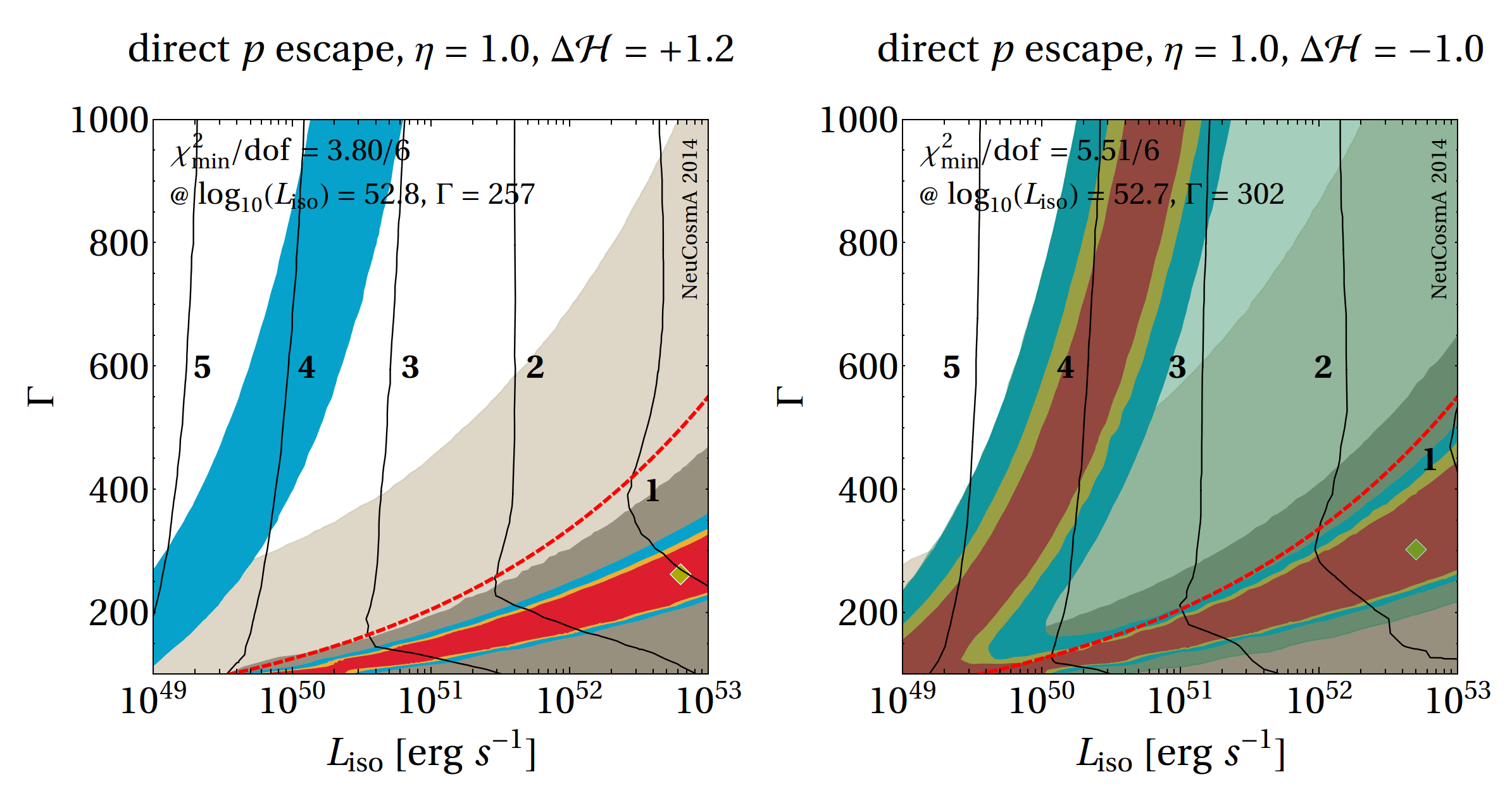}
	\mycaption{\label{fig:flucscan} Same as \figu{fits}, middle left panel (direct escape with $\eta = 1.0$), but assuming a local upward (left panel) and downward (right panel) fluctuation of the SFR of $\Delta\mathcal{H} = 1.2$ and $-1.0$, respectively. Note that, in the right panel, the left border of the exclusion region from cosmogenic neutrinos (in green) coincides with the left border of the 99\% C.L.~allowed region (in blue).}
\end{figure}

\enlargethispage{\baselineskip}

Another potential issue, which can affect our conclusions, is the fact that the GRB rate could deviate from the SFR in our local environment. That is especially relevant for UHECRs, for which the mean free path is only about $1 \, \mathrm{Gpc}$ ($z \simeq 0.25$) at $10^{10} \, \mathrm{GeV}$, and $100 \, \mathrm{Mpc}$ ($z \simeq 0.024$) at $10^{11} \, \mathrm{GeV}$; see \figu{fig_interaction_length} in \ref{sec:crprop}. Therefore, local fluctuations in the GRB rate will affect the UHECR spectrum and hence the required baryonic loading, whereas the GRB observations, which are dominated by higher redshifts, are hardly affected. We illustrate a local deviation (below redshift 0.25) in \figu{hmod}; here the unchanged version of the dimensionless distribution of sources in redshift $\mathcal{H}(z)$ is defined in \equ{defH} (see \ref{app:modelindep}). We have estimated the local deviation from statistics: distributing 1000 bursts over redshift, only about six will be in the range $z \lesssim 0.25$.\footnote{One can 
also estimate this from the local GRB rate of $\sim 1 \, \mathrm{Gpc^{-3} \, yr^{-1}}$, since $z \simeq 0.25$ 
corresponds to a mean free path of the protons of $R \simeq 1 \, \mathrm{Gpc}$ at $10^{10} \, \mathrm{GeV}$. The average GRB rate between $z=0$ and $z=0.25$ is roughly $1.5$ (see \figu{hmod}), which leads to $4 \pi R^3/3 \cdot 1.5 \cdot 1 \, \mathrm{Gpc^{-3} \, yr} \simeq 6 \, \mathrm{yr}^{-1}$.} The $1 \sigma$ relative (Gaussian) error is therefore roughly $1/\sqrt{6} \simeq 0.41$, and the $3 \sigma$ range can be 
estimated as $-1 \lesssim \Delta \mathcal{H} \lesssim +1.2$.

The effect of such local deviations is illustrated in \figu{fluc} for model~\#1 of \figu{GRBCRprimer}. A local upward fluctuation of the GRB rate clearly reduces the required baryonic loading, and, at the same time, the prompt and cosmogenic neutrino fluxes. This is different from earlier figures, where the cosmogenic neutrino flux remained almost unchanged. A local upward fluctuation of the GRB rate may therefore be a plausible explanation for the non-observation of neutrinos from GRBs, and at the same time allow for reasonable $f_e^{-1}$. However, the goodness of fit is slightly reduced, because the local enhancement relatively increases the high-energy part of the cosmic ray spectrum. A local downward fluctuation of the GRB rate causes the opposite: a better fit, at the expense of higher neutrino fluxes and baryonic loadings.

These results should be taken into consideration in the interpretation of our parameter scans: in \figu{flucscan} we show the parameter scans for a direct proton escape scenario with a local upward (left panel) and downward (right panel) fluctuation of $\Delta \mathcal{H} = 1.2$ and $-1.0$, respectively. The upward fluctuation (left panel) is capable of reducing the required values of baryonic loadings enough to make the cosmogenic neutrino exclusion (after 15 years of exposure) disappear, at the expense of the spectral shape change, which makes the left fit branch vanish.
The corresponding downward fluctuation increases the cosmogenic neutrino flux, which means that the future cosmogenic bound can even partially exclude the left branch of the fit.

\section{Summary and conclusions}
\label{sec:conclusions}

In this study, 
we have considered the standard (internal shock) fireball scenario, currently used for the state-of-the-art GRB stacking neutrino analyses, and combined it with UHECR propagation assuming a pure proton composition. This combined source-propagation model yields the cosmic ray, prompt neutrino, and cosmogenic neutrino fluxes, using the prompt gamma-ray observations as input. The source model predicts the shape of the UHECR injection spectrum in the cosmologically comoving frame as a function of the GRB parameters (such as $\Gamma$ and $L_{\mathrm{iso}}$), and the pion production efficiency. By requiring that GRBs are the sources of the UHECRs, the baryonic loading of the sources, which is the main free parameter which has so far been assumed {\em ad hoc}, can be derived in a completely self-consistent way. As a consequence, a self-consistent picture of the cosmic energy budget from GRBs can be drawn, connecting cosmic ray, gamma-ray, and neutrino observations; see \figu{triangle}.

Earlier studies have indicated that neutron production as the cosmic ray escape mechanism from the sources is already strongly constrained by neutrino observations, since cosmic rays and neutrinos are produced in the same (photohadronic) processes. We have confirmed this conclusion, but we have demonstrated that the neutrino and cosmic ray fluxes predicted by this assumption can only be justified for baryonic loadings $f_e^{-1} \gg 10$ of the bursts for typical GRB parameter values -- where we are using the IceCube definition for the baryonic loading, \ie, total energy in protons versus total energy in electrons/photons. We have therefore studied alternative escape scenarios for the cosmic ray protons, such as direct escape (without scattering) as the minimally guaranteed component, and diffusion. In fact, all of our cosmic ray escape models are identical in the parameter space where the neutron production is efficient, while other escape mechanisms dominate in the parameter space where the pion production 
efficiency is small.

As far as the transition model to a different (Galactic or extragalactic) component at lower energies is concerned, we have considered the ankle model and the dip model. In the context of the GRB source model, the dip model is strongly disfavored because of the constraints from the neutrino bounds and the extremely large values required for the baryonic loading $f_e^{-1} \gtrsim 10^5$. On the other hand, we have demonstrated that the ankle model is still plausible for $10 \lesssim f_e^{-1} \lesssim 100$, $\Gamma \gtrsim 400$, and $L_{\mathrm{iso}} \gtrsim 10^{52.5} \, \mathrm{erg \, s^{-1}}$, if the cosmic-ray protons can efficiently escape by diffusion (assuming SFR evolution of the sources). Note that while these parameters sound reasonable, they are probably already quite far-fetched for the average GRB. This part of the parameter space can, however, be tested by the IceCube experiment in the next 10--15 years. There is nonetheless a part of the parameter space which will survive these tests and which 
requires significantly 
lower isotropic luminosities, 
but which also requires $f_e^{-1} \gtrsim 1000$ to compensate for that. This part of the parameter space is especially favored by Pierre Auger data, compared to HiRes and Telescope Array data. Improved energy resolution for the UHECR measurements will limit this possibility in the future. For the case of strong source redshift evolution, the cosmogenic neutrino flux bounds will give more efficient exclusions than the stacking bounds from the prompt fluxes in parts of the parameter space.

In order to support our findings by analytical arguments, we have rederived the main relationships for the cosmic energy budget using GRBs, with a number of interesting observations. First of all, we have used the isotropic equivalent luminosity per GRB and the number of observable GRBs per year in the universe $\dot N \simeq 1000 \, \mathrm{yr}^{-1}$ for the normalization, which are, compared to the local ($z=0$) GRB rate, directly measurable quantities. As a consequence, the local GRB rate can be derived from $\dot N$, and we have shown that it must be $\mathcal{O}(0.1) \, \mathrm{Gpc}^{-3} \, \mathrm{yr}^{-1}$ (to be corrected by the beaming factor) if GRBs evolve more strongly than the SFR (not including a possible population of low-luminosity GRBs). The injected energy into cosmic rays per burst in the energy range between $10^{10}$ and $10^{12} \, \mathrm{GeV}$ must then be larger than $10^{54} \, \mathrm{erg}$. We have therefore chosen the SFR evolution case as baseline, for which 
these requirements (and therefore the required baryonic loading) are somewhat less severe. 
We could also identify the reason why the required baryonic loading is larger than previously anticipated: it matters if it is defined with respect to the UHECR range or the total energy range, which implies that these two are related by a bolometric correction. If the IceCube definition is used (energy in protons in the total energy range), baryonic loadings as low as $10$ are typically too low to describe the UHECR observations. Finally, note that the beaming factor drops out of our framework, \ie, our analysis is not sensitive to the beaming factor. Although we only observe a fraction of the bursts beamed in our direction and all GRBs will contribute to the cosmic ray flux, the beaming is automatically corrected for by using the isotropic equivalent energy.

There is one caveat in the interpretation of our baryonic loadings: our results actually scale with $\dot N \cdot f_e^{-1} \cdot f_{\mathrm{thresh}}^{-1}$, where $f_{\mathrm{thresh}}$ corrects for bursts below the instrument threshold. We have chosen $f_{\mathrm{thresh}}= 0.3$ for our computations, whereas it is possible that this number is somewhat smaller, \eg, a recent study based on Swift data~\cite{Lien:2013qja} can be used to estimate $f_{\mathrm{thresh}} \simeq 1000/4568 \simeq 0.22$, which would slightly lower the required baryonic loading with respect to the numbers we give in this study. A rescaling of our results for arbitrary $f_{\mathrm{thresh}}$ is, however, trivial (see \equ{rescale}). Another possibility to reduce the required baryonic loading is a local upward fluctuation of the GRB rate compared to the SFR, which is plausible within current statistics. Such a fluctuation would also reduce both the prompt and cosmogenic neutrino flux predictions, but it would increase the tension with the 
UHECR spectral shape.

It should also be noted as a limitation of our present analysis that the used GRB model assumes that the proton, neutrino, and photon emissions all originate in internal collisionless shocks occurring at a single, representative, average radius $R_C = 2 \Gamma^2 c t_v$ from the central emitter, where $\Gamma$ is the average Lorentz factor of the burst and $t_v$ is its variability timescale, a global property of the burst's light-curve. In reality, however, it is expected that the emission of the different species occurs at different radii: in the collisions that occur at low radii, close to the emitter, matter densities are higher and so photon-photon and nucleon-photon interactions prevent photons and nucleons from escaping. On the other hand, most of the neutrinos created in photohadronic interactions come from these low radii, since neutrinos are able to free-stream out of the dense matter ejecta. At larger collision radii, which imply lower matter densities, protons can escape directly from 
the edges of the matter shells. 
Similarly, as already discussed in \Refs~\cite{Wang:2007xj,Murase:2008mr}, the gamma-rays and heavier nuclei with the highest energies may come from these large collision radii. 
A detailed treatment of collisions occurring at different radii will be presented elsewhere~\cite{FBprep}.

Including heavier nuclei into the assumed UHECR component may allow for higher CR energies, as nuclei can be accelerated more efficiently due to their higher charges. However, the nuclei can escape without being photodisintegrated only for parameter sets for which the photon densities are low enough; see \Refs~\cite{Wang:2007xj,Murase:2008mr}. In fact, the disintegration and pion production efficiencies are proportional to each other~\cite{Murase:2010gj}, which means that the UHECR nuclei escape without disintegration in our direct/diffusive escape regimes, and that we anticipate that our results do not qualitatively change there. For higher photon densities, however, photodisintegration can become efficient and the neutrino and neutron production will be reduced, which has to be compensated by higher baryonic loadings. 

We conclude that it is still possible to draw a self-consistent picture for the cosmic energy budget of the UHECRs if GRBs are their sources in the internal shock model. This picture requires a cosmic ray escape mechanism other than neutron escape {\em and} baryonic loadings that are significantly larger than the commonly assumed value of ten, while the neutron model is already ruled out by current neutrino bounds. Future IceCube bounds will however severely constrain the parameter space and, especially, the baryonic loading. Note that we have not included luminosity distributions yet, which will complicate the interpretation and the predicted UHECR spectrum. Especially if there are many low-luminosity GRBs close by, they will affect the shape of the predicted UHECR injection spectrum because lower maximal proton energies are implied.

\appendix

\section{The UHECR energy budget from GRBs}
\label{app:modelindep}

Here we show the detailed derivations leading to the conclusions in \Sec~\ref{sec:modelindep}.

\subsection{Observation of prompt gamma-rays, and local GRB rate}

For the description of the redshift distribution, we follow \Ref~\cite{Kistler:2009mv}. The comoving GRB rate $[\mathrm{Mpc}^{-3} \, \mathrm{yr}^{-1}]$ is given by
\begin{equation}
\dot n_{\mathrm{GRB}} = \mathcal{E}(z) \cdot \dot \rho_*(z) \, , \label{equ:comGRB}
\end{equation}
where 
\begin{equation}
\mathcal{E}(z) = \mathcal{E}_0 (1+z)^\alpha \, \label{equ:defE}
\end{equation}
 describes the evolution of the fraction of stars resulting in GRBs, $\alpha \simeq 1.2$, and $\rho_*(z)$ is the (comoving) star formation density $[M_\odot \, \mathrm{Mpc}^{-3} \, \mathrm{yr}^{-1}]$. 
The \textbf{observed} redshift distribution of GRBs $\mathrm{d} \dot N/\mathrm{d}z$ $[\mathrm{yr}^{-1}]$ can be written as~\cite{Kistler:2009mv}
\begin{equation}
\frac{\mathrm{d} \dot N}{\mathrm{d}z} = F(z) \frac{ \dot n_{\mathrm{GRB}} }{\langle f_{\mathrm{beam}} \rangle} \, \frac{\mathrm{d}V/\mathrm{d}z}{1+z} = F(z) \frac{ \mathcal{E}(z) \, \dot \rho_*(z)}{\langle f_{\mathrm{beam}} \rangle} \, \frac{\mathrm{d}V/\mathrm{d}z}{1+z} \, , \label{equ:kistler}
\end{equation}
where the last factor is the comoving volume correction.\footnote{This correction is defined as
\begin{equation}
 \frac{\mathrm{d}V}{\mathrm{d}z} = 4\pi D_H \frac{1}{h\left(z\right)} d_c^2\left(z\right) \, , \quad \mathrm{with} \quad 
 h\left(z\right) = \sqrt{\Omega_m\left(1+z\right)^3 + \Omega_\Lambda} \quad \mathrm{and} \quad
 d_c\left(z\right) = D_H \int_0^z \frac{\mathrm{d}z^\prime}{h\left(z^\prime\right)} \, .
\end{equation}
Here $D_H = 4.255 \, \mathrm{Gpc}$, $\Omega_m = 0.27$, $\Omega_\Lambda = 0.73$, taken from \Ref~\cite{Komatsu:2010fb}.}
Note that here the beaming factor is needed to correct for the invisible GRBs beamed into different directions ($0 < \langle f_{\mathrm{beam}} \rangle^{-1} < 1$) and $F(z)$ accounts for the ability to observe the GRB, such as the detector threshold ($0 < F(z) < 1$). In the main text, we use $\dot{\tilde{n}}_{\mathrm{GRB}} \equiv \dot n_{\mathrm{GRB}}/\langle f_{\mathrm{beam}} \rangle$ for the sake of simplicity, which is lower than the actual GRB rate by the beaming factor.

It is useful to define the adimensional redshift evolution $\mathcal{H}(z)$ of the GRBs by normalizing the comoving GRB rate to the local rate leading to
\begin{equation}
\mathcal{H}(z) = \frac{ \dot n_{\mathrm{GRB}} }{\left. \dot n_{\mathrm{GRB}} \right|_0} = (1+z)^\alpha \, \frac{\dot \rho_*(z)}{\dot \rho_*(0)} \, ,
\label{equ:defH}
\end{equation}
such that $\mathcal{H}(z=0)=1$. In addition, we distinguish the \textbf{total number of bursts per year in the observable universe} $\dot N_{\mathrm{tot}}$ and the \textbf{number of observable bursts per year} $\dot N$, where, from \equ{kistler},
\begin{equation}
\frac{\mathrm{d} \dot N}{\mathrm{d}z} = F(z) \frac{\mathrm{d} \dot N_{\mathrm{tot}}}{\mathrm{d}z} \, , \quad
\dot N = \int\limits_0^\infty \frac{\mathrm{d} \dot N}{\mathrm{d}z} \, \mathrm{d}z \, , \quad \dot N_{\mathrm{tot}} = \int\limits_0^\infty  \frac{\mathrm{d} \dot N}{\mathrm{d}z} \frac{1}{F(z)} \, \mathrm{d}z \, .
\label{equ:dotn}
\end{equation}
As a consequence, one can compute $\dot N_{\mathrm{tot}}$ from $\dot N$ if the threshold function of the instrument is known. We define the ratio $f_{\mathrm{thresh}} \equiv \dot N/\dot N_\mathrm{tot}$
as the fraction of observable bursts because of the instrument threshold. For instance, using a power-law luminosity distribution proposed by Wanderman \& Piran \cite{Wanderman:2009es} with the redshift distribution from Kistler {\it et al.}~\cite{Kistler:2009mv}, we obtain a ratio $f_{\mathrm{thresh}} \simeq 0.5$ for a threshold of $1.75 \cdot 10^{-8} \, \mathrm{erg} \, \mathrm{s^{-1}} \mathrm{cm^{-2}}$ if we assume that bursts can only have a luminosity in the range $10^{50}$ to $10^{54} \, \mathrm{erg}\,\mathrm{s}^{-1}$, as implied in \Ref~\cite{Wanderman:2009es}. This result is of course dependent on the chosen distributions and cutoffs, \eg, when we extend the distribution to lower luminosities, say, $10^{49} \, \mathrm{erg}\,\mathrm{s}^{-1}$, the ratio goes down to 0.3, which is the value that we have adopted in this study. In any case, one should keep in mind that there is a factor of two to three difference between $\dot N_{\mathrm{tot}}$ and $\dot N$.

It is now useful to relate the total redshift distribution of GRBs to \equ{kistler} by using \equ{defH} and \equ{dotn} as
\begin{equation}
 \frac{\mathrm{d} \dot N_{\mathrm{tot}}}{\mathrm{d}z} = \frac{\left. \dot n_{\mathrm{GRB}} \right|_{z=0} }{\langle f_{\mathrm{beam}} \rangle} \, \mathcal{H}(z) \, \frac{\mathrm{d}V/\mathrm{d}z}{1+z} \quad , \label{equ:kistlertot}
\end{equation}
where $\left. \dot{\tilde{n}}_{\mathrm{GRB}} \right|_{z=0} = \left. \dot n_{\mathrm{GRB}} \right|_{z=0}/\langle f_{\mathrm{beam}} \rangle$ is the often-quoted ``local GRB rate'', which is of the order of one burst per $\mathrm{Gpc}^3$ and year; see, \eg, \Ref~\cite{Wanderman:2009es}. It is reduced with respect to the actual GRB rate $\left. \dot n_{\mathrm{GRB}} \right|_{z=0}$, which includes the GRBs beamed in different directions which are not directly observable, by the beaming factor. We can now derive $\dot{N}_{\mathrm{tot}}$ as
\begin{equation}
	\dot N_{\mathrm{tot}} = \frac{\left. \dot n_{\mathrm{GRB}} \right|_{z=0}}{\langle f_{\mathrm{beam}} \rangle} \cdot \int\limits_0^\infty \mathcal{H}(z) \, \frac{\mathrm{d}V/\mathrm{d}z}{1+z} \, \mathrm{d}z \equiv \frac{\left. \dot n_{\mathrm{GRB}} \right|_{z=0}}{\langle f_{\mathrm{beam}} \rangle} \cdot 4\pi \, D^3_H \cdot f_z \quad ,
	\label{equ:dotNcalc}
\end{equation}
where we have defined
\begin{equation}
 f_z \equiv \frac{1}{4\pi \, D^3_H} \int\limits_0^\infty \mathcal{H}(z) \, \frac{\mathrm{d}V/\mathrm{d}z}{1+z} \, \mathrm{d}z \, .
\label{equ:fz}
\end{equation}
 This cosmic evolution factor describes how representative the local GRB rate is for the whole distribution, and it therefore depends on the SFR and GRB evolution. We then can derive the local GRB rate as a function of the observable $\dot N$ as
\begin{equation}
\frac{\left. \dot n_{\mathrm{GRB}} \right|_{z=0}}{\langle f_{\mathrm{beam}} \rangle} = \frac{\dot N}{f_{\mathrm{thresh}}} \frac{1}{4 \pi \, D_H^3 f_z} \simeq \frac{1}{\mathrm{Gpc}^3 \, \mathrm{yr}} \cdot \frac{\dot N \, [\mathrm{yr}^{-1}]}{968} \cdot f_{\mathrm{thresh}}^{-1} \cdot f_{z}^{-1} \, ,
\label{equ:obsmaster}
\end{equation}
which contains \equ{obsmaster2}.
Note that the value 968 comes from the volume term $4 \pi \, D_H^3 \simeq 968 \, \mathrm{Gpc}^3$. Typical values for $f_z$ can be read off for different star formation and evolution models in \Tab~\ref{tab:redshiftnorms}.

\subsection{Cosmic ray injection and observation}

The cosmic-ray injection rate $\mathcal{L}_{\mathrm{CR}}(E,z)$ [$\mathrm{GeV^{-1} \, Mpc^{-3}\, s^{-1}}$] can be extrapolated from a single-source isotropic emission spectrum $dN_{\mathrm{CR}}^\mathrm{iso}/dE$ in the source frame [$\mathrm{GeV}^{-1}$] as
\begin{equation}
\mathcal{L}_{\mathrm{CR}}(E,z) = \frac{\mathrm{d}N_{\mathrm{CR}}^\mathrm{iso}}{\mathrm{d}E}\cdot \frac{1}{ \langle f_{\mathrm{beam}} \rangle}  \cdot \dot n_{\mathrm{GRB}}(z) \quad . \label{equ:crmaster0}
\end{equation}
In order to see the origin of the beaming factor, consider one GRB which ejects cosmic rays at a rate per volume $Q'_{\mathrm{CR}}(E)$ [$\mathrm{GeV^{-1} \, cm^{-3} \, s^{-1}}$] in the shock rest frame. Then \equ{crmaster0} changes to
\begin{equation}
\mathcal{L}_{\mathrm{CR}}(E,z) = \underbrace{ \frac{Q'_{\mathrm{CR}}(E) \cdot V'_{\mathrm{iso}}  \cdot T'_{90} }{\Gamma} }_{dN_{\mathrm{CR}}^\mathrm{iso}/dE\mathrm{ source frame}} \cdot \underbrace{ \frac{1}{ \langle f_{\mathrm{beam}} \rangle} }_{\mathrm{beaming corr.}} \cdot \underbrace{ \dot n_{\mathrm{GRB}}(z) }_{\mathrm{comoving GRB rate}} \quad . \label{equ:crmaster}
\end{equation}
Here the first term corresponds to the total spectrum [$\mathrm{GeV}^{-1}$] released from the single GRB over the duration $T'_{90}$, and the next-to-last factor multiplies that times the number of GRBs per Mpc$^{3}$ and year. The actual source volume is expressed by the ``isotropic volume'' of the burst~\cite{Baerwald:2011ee}
\begin{equation}
	V'_{\mathrm{iso}} = 4\pi \, R_C^2 \cdot \Delta d' = 4\pi \, R_C^2 \cdot \Gamma \cdot c \cdot t_v \, ,
	\label{equ:visoISM}
\end{equation}
where $t_v$ is the variability timescale in the source frame; for details see \Refs~\cite{Baerwald:2011ee,Baerwald:2013pu}. Thus the beaming factor in \equ{crmaster} enters because energy and volume $V'_{\mathrm{iso}}$ are computed by assuming isotropic emission, whereas only a fraction $\langle f_{\mathrm{beam}} \rangle^{-1}$ of that energy is actually emitted by the GRB. 
In addition, note that \equ{crmaster} factorizes in an energy-dependent part and a redshift-dependent part, as it is often assumed in the literature.

In order to address the UHECR connection, a frequently used approach is to use the local energy injection rate between $10^{10}$ and $10^{12} \, \mathrm{GeV}$~\cite{Waxman:1995dg}, which can be obtained from \equ{crmaster0} as
\begin{equation}
\dot \varepsilon_{\mathrm{CR}}^{[10^{10},10^{12}]} = \int\limits_{10^{10} \, \mathrm{GeV}}^{10^{12} \, \mathrm{GeV}} \mathcal{L}_{\mathrm{CR}}(E,0) \, E \, \mathrm{d}E = \frac{ \left. \dot n_{\mathrm{GRB}} \right|_{z=0} }{\langle f_{\mathrm{beam}} \rangle }  \, \underbrace{ \int\limits_{10^{10} \, \mathrm{GeV}}^{10^{12} \, \mathrm{GeV}}    \frac{\mathrm{d}N_{\mathrm{CR}}^\mathrm{iso}}{\mathrm{d}E} \, E \, \mathrm{d}E }_{\equiv E_{\mathrm{CR}}^{[10^{10},10^{12}]}} \, ,
\label{equ:crinj}
\end{equation}
which is again proportional to the local GRB rate. From \equ{crinj}, using \equ{obsmaster}, we can then derive \equ{cre2}.

\subsection{Neutrinos and multi-messenger physics with GRBs}

In order to include the neutrinos in the discussion, we have to take into account that they come from photohadronic interactions, for which the proton and photon densities in the source are the required input; see, \eg, \Ref~\cite{Hummer:2010vx}.
Typically, the energy in protons is related to the energy in electrons/photons by partition arguments, similar to the approach used in \Refs~\cite{Abbasi:2009ig,Abbasi:2012zw}, \ie,
\begin{equation}
 \int\limits_{0}^{\infty}   \frac{\mathrm{d}N_{p}}{\mathrm{d}E} \, E \, \mathrm{d} E = \frac{1}{f_e}  \, \int\limits_{0}^{\infty}  \frac{\mathrm{d} N_{\gamma}}{\mathrm{d} \varepsilon} \, \varepsilon \, \mathrm{d}\varepsilon  =  \frac{1}{f_e} E_{\gamma,\mathrm{iso}}\, ,
\label{equ:loading}
\end{equation}
where we compute it in the source frame for this discussion. Note that minimal and maximal proton and photon energies are to be defined within $\mathrm{d}N_p/\mathrm{d}E$ and $\mathrm{d} N_\gamma/\mathrm{d} \varepsilon$, respectively.
The energy range of the gamma-rays is typically given by the instrument, such as {\em Fermi} GBM. The energy range of the protons spans the whole proton spectrum, from the set minimal to the set maximal energy, at least covering the energy range relevant for the neutrino production which is different from that of the UHECR.\footnote{The neutrino spectra peak at about $\sim 1 \, \mathrm{PeV}$ due to secondary cooling~\cite{Baerwald:2010fk}, and come from $\sim 10 \, \mathrm{PeV}$ protons.} The factor $f_e^{-1}$ is commonly known as ``baryonic loading''.
Let us assume that a fraction $f_{\mathrm{CR}} \le 1$ of the protons can escape from the source as cosmic rays and define
\begin{equation}
f_{\mathrm{bol}} \equiv \left(
\int\limits_{10^{10} \, \mathrm{GeV}}^{10^{12} \, \mathrm{GeV}}    \frac{\mathrm{d}N_{p}}{dE} \, E \, \mathrm{d}E
\right) \left/ \left(
\int\limits_{0}^{\infty}   \frac{\mathrm{d}N_{p}}{\mathrm{d}E} \, E \, \mathrm{d}E
\right) \right.
\label{equ:bol} 
\end{equation}
as the bolometric correction factor describing how much of the proton energy sits in the UHE range.\footnote{Note that we take the UHE range in the source frame (in principle, the energy has to be higher in the source frame to match the observed $10^{10} \, \mathrm{GeV}$), which is however similar to the observed UHE range because the mean free path of the protons at $10^{10} \, \mathrm{GeV}$ is only $1 \, \mathrm{Gpc}$, and therefore $z \le 0.25$. }
This bolometric correction depends on the energy range of the proton spectrum, the proton spectral index, and the maximal proton energy, and is for all practical applications $\le 1$. 
We can then derive from \equ{loading} and \equ{bol} the energy injected into UHECR from the individual burst, \equ{ecrderiv2}.

In order to see the connection between gamma-rays, cosmic rays, and neutrinos including the processes in the source, we rewrite \equ{crmaster0} using \equ{obsmaster}:
\begin{equation}
\mathcal{L}_{\mathrm{CR}}(E,z) = \frac{\mathrm{d}N_{\mathrm{CR}}^\mathrm{iso}}{\mathrm{d}E}\cdot  \frac{\dot N}{f_{\mathrm{thresh}}}  \cdot \frac{1}{ 4 \pi \, D_H^3 \, f_z}    \cdot  \mathcal{H}(z) \, . \label{equ:crrew}
\end{equation}
We can then write the injected energy in the UHECR range with \equ{ecrderiv2} as
\begin{equation}
\int\limits_{10^{10} \, \mathrm{GeV}}^{10^{12} \, \mathrm{GeV}} \mathcal{L}_{\mathrm{CR}}(E,z) \, E \, \mathrm{d}E = f_{\mathrm{CR}} \cdot \frac{f_{\mathrm{bol}}}{f_e} \cdot  E_{\gamma, \mathrm{iso}}  \cdot  \frac{\dot N}{f_{\mathrm{thresh}}}  \cdot \frac{1}{ 4 \pi \, D_H^3 \, f_z}    \cdot  \mathcal{H}(z)  \, , \label{equ:crrew2}
\end{equation}
which shows the relationship to the gamma-ray observations.

Since neutrinos do not interact, it is straightforward to define an injection function similar to that of the cosmic rays as (\cf, \equ{crmaster0})
 \begin{equation}
 \mathcal{L}_{\nu}(E,z)  =  \frac{\mathrm{d}N_\nu}{\mathrm{d} E_\nu}\cdot  \frac{1}{ \langle f_{\mathrm{beam}} \rangle}    \cdot  \dot n_{\mathrm{GRB}}(z)   \, , \label{equ:injnu}
 \end{equation}
where the neutrinos only suffer from energy losses due to the adiabatic expansion of the universe. Note that the beaming factor can be interpreted in different ways here: either only the bursts beamed in our direction can be seen (read in combination with first factor), or only a fraction of the isotropic energy is actually injected (read in combination with last factor). In order to connect with the physics of the sources, we assume here, for the sake of simplicity, that a fraction $f_\pi$ of the proton energy goes into pion production (pion production efficiency~\cite{Waxman:1997ti,Guetta:2003wi}), that 50\% of the pions produced in photohadronic interactions are charged pions, and each lepton in the pion decay obtains about 25\% of the pion energy. Then we have from \equ{injnu} with \equ{obsmaster} and \equ{loading} the injected energy into neutrinos
\begin{equation}
\int\limits_{E_{\nu,\mathrm{min}}}^{E_{\nu,\mathrm{max}}} \mathcal{L}_{\nu}(E,z) \, E \, \mathrm{d}E \simeq \frac{f_{\pi}}{8}  \cdot \frac{1}{f_e} \cdot  E_{\gamma, \mathrm{iso}}  \cdot  \frac{\dot N}{f_{\mathrm{thresh}}}  \cdot \frac{1}{ 4 \pi \, D_H^3 \, f_z}    \cdot  \mathcal{H}(z)  \, . \label{equ:nurew}
\end{equation}
Of course, this simple estimate does not take into account the energy dependence of the proton interaction length and the normalization change from the cooling of secondaries~\cite{Hummer:2011ms,Li:2011ah}, which we fully take into account in our numerical simulations, but it can serve as a first estimate. For the gamma-rays, similar considerations can be made. However, it is straightforward to identify the common scaling factors from \equ{kistlertot} using \equ{obsmaster}. These considerations, including \eqs~(\ref{equ:crrew2}) and~(\ref{equ:nurew}), are pictorially represented in \figu{triangle}.

\section{Cosmic ray propagation}
\label{sec:crprop}

As a fundamental part of our study, we have computed the propagation of cosmic rays (CRs) from their origin, at a cosmological source with redshift $z$, to Earth, taking into account the effects of energy losses en route, due to the adiabatic cosmological expansion and to the interaction with the photons of the cosmic microwave background (CMB) and the cosmic infrared background (CIB). For a selection of literature on the subject, see \Refs~\cite{Greisen:1966jv,Zatsepin:1966jv,Blumenthal:1970nn,Berezinsky:1988wi,Ahlers:2009rf,Ahlers:2011sd,Stanev:2004kk,Aloisio:2006wv,Kotera:2011cp}, with which our results agree. We have assumed that CRs are composed solely of protons.

The interactions that we have considered between protons and background photons are a) $e^+e^-$ pair production, \ie, $p + \gamma \rightarrow p + e^+ + e^-$, and b) photohadronic processes. In a first approximation, the latter are described by the resonant process $p + \gamma \rightarrow \Delta^+\left(1232\right) \rightarrow n + \pi^+$; we have, however, used the NeuCosmA photohadronics code to include many more processes (see section \ref{SecProblemLossPGamma}). An accompanying ``guaranteed'' flux of cosmogenic neutrinos is predicted from the decays of the secondary neutrons and pions \cite{Beresinsky:1969qj,Stecker:1978ah}, \eg, $n \rightarrow p + e^- + \bar{\nu}_e$ and $\pi^+ \rightarrow \mu^+ \nu_\mu \rightarrow \bar{\nu}_\mu e^+ \nu_e + \nu_\mu$ (see \Refs~\cite{Beresinsky:1969qj,Stecker:1978ah,Ahlers:2010fw,Allard:2006mv,Kotera:2010yn,Engel:2001hd,DeMarco:2005kt})

The CR propagation is performed by solving the Boltzmann transport equation for the comoving number density of protons [$\mathrm{GeV}^{-1} \, \mathrm{Mpc}^{-3}$],
\begin{equation}
 Y_p\left(E,z\right) = a^3\left(z\right) n_p\left(E,z\right) = n_p\left(E,z\right)/\left(1+z\right)^3 \; ,  
\end{equation}
with $n_p$ the real number density and $a\left(z\right) = \left(1+z\right)^{-1}$ the scale factor. The transport equation is (see, \eg, \Refs~\cite{Ahlers:2011jj,Anchordoqui:2011gy}):
\begin{equation}\label{equ:EqYiEvol}
 \dot{Y}_p = \partial_E\left(HEY_p\right) + \partial_E\left(b_{e^+e^-} Y_p\right) + \partial_E\left(b_{p\gamma} Y_p\right) + \mathcal{L}_\mathrm{CR} \; ,
\end{equation}
with $E$ the proton energy in the source frame (see, \eg, \Ref~\cite{Ahlers:2009rf}). In the r.h.s.~of \equ{EqYiEvol}, the first term accounts for continuous energy losses due to the adiabatic cosmological expansion, with $H\left(z\right) = H_0 \sqrt{\Omega_m\left(1+z\right)^3 + \Omega_\Lambda}$ the Hubble parameter. We have used the local value $H_0 = 70.5 \, \mathrm{km} \, \mathrm{s}^{-1} \, \mathrm{Mpc}^{-1} = 2.28475 \cdot 10^{-18} \, \mathrm{s}^{-1}$, and the energy densities of matter and cosmological constant given by $\Omega_m = 0.27$ and $\Omega_\Lambda = 0.73$, respectively \cite{Komatsu:2010fb}. The second and third terms, respectively, account for continuous energy losses due to $e^+e^-$ pair production and photohadronic ($p\gamma$) production on the photon backgrounds, with the corresponding energy-loss rates $b_{e^+e^-}$ and $b_{p\gamma}$, where $b \equiv dE/dt$ [$\mathrm{GeV} \, \mathrm{s}^{-1}$]. The fourth term describes the CR injection rate per comoving volume; it is factorized as $\
mathcal{L}_\mathrm{CR}\left(E,z\right) = \mathcal{H}_\mathrm{CR}\left(z\right) Q_\mathrm{CR}\left(E,z\right)$, where $Q_\mathrm{CR}$ [$\mathrm{GeV}^{-1} \, \mathrm{Mpc}^{-3} \, \mathrm{s}^{-1}$] is the injection spectrum at the source and $\mathcal{H}_\mathrm{CR}$ is the adimensional comoving redshift evolution, defined in \equ{defH}.

By means of the relation $\mathrm{d}z = -\mathrm{d}t\left(1+z\right)H\left(z\right)$, \equ{EqYiEvol} can be recast as an equation in redshift. We have written original computer code to numerically solve \equ{EqYiEvol}, propagating the proton flux from $z=6$ down to $z=0$. The diffuse proton flux $J_p$ at Earth [$\mathrm{GeV}^{-1} \, \mathrm{cm}^{-2} \, \mathrm{s}^{-1} \, \mathrm{sr}^{-1}$] is obtained at the last step of the calculation, from the local density, through $J_p\left(E\right) = \left[ c/\left(4\pi\right) \right] n_p\left(E,0\right)$.

In the present analysis we treat protons and neutrons as the same particles, which is a good approximation because neutrons decay into protons with a rate larger than the pair production rate, and the photohadronic interactions are symmetric between protons and neutrons to a first approximation. However, we introduce a cutoff in the cosmogenic neutrino spectrum from neutron decays where the photohadronic interaction rate exceeds the neutron decay rate.


Notice that that the implementation of a solution for \equ{EqYiEvol} that we have used is strictly valid only for proton energies above $\sim 10^{9}\, \mathrm{GeV}$. Below this energy, diffusion effects in the intergalactic magnetic fields may affect he spectral shape~\cite{Aloisio:2006wv} -- depending on the magnetic fields, of course.

\subsection{Energy loss rates due to $p\gamma$ interactions}\label{SecProblemLossPGamma}

In general, the interaction rate\footnote{Note that, with our choice of units for $n_\gamma$ and $\sigma_{p\gamma}^\mathrm{tot}$, the rate $\Gamma_{p \gamma \rightarrow p^\prime b}$ is output in cm$^{-1}$; multiplication by $c$ gives it the appropriate units, s$^{-1}$.} (probability of interaction per unit time per particle) between protons and an isotropic background photon field $n_\gamma$ [$\mathrm{GeV}^{-1} \, \mathrm{cm}^{-3}$], at proton energy $E$, is calculated at each redshift step as in \eq~(4) of \Ref~\cite{Hummer:2010vx} (see also \Refs~\cite{Ahlers:2009rf,Ahlers:2011sd}). For our results, the calculation is performed by NeuCosmA and considers interactions of the type $p+\gamma \rightarrow p^\prime+ b$, with the daughter particles ($b$), typically $\pi^+$, $\pi^-$, $\pi^0$, or $K$ \cite{Hummer:2010vx}. As explained in \App~B of \Ref~\cite{Hummer:2010vx}, for a given photon background (CMB, CIB), NeuCosmA calculates the energy loss rate $b_{p\gamma}\left(E,z\right)$ [$\mathrm{GeV} \, \mathrm{s}^{-
1}$] that enters \equ{EqYiEvol} in a fast and efficient way based on a parametrization of SOPHIA results \cite{Mucke:1999yb}.

The other relevant energy loss rate, $b_{e^+e^-}$, due to $e^+e^-$ pair production in the interaction $p + \gamma \rightarrow A + e^+ + e^-$ on an isotropic photon background $n_\gamma\left(\epsilon,z\right)$, is calculated following Blumenthal, using Eqs.~(13) and (14) of \Ref~\cite{Blumenthal:1970nn}.

\subsection{Scaling of the photon backgrounds and of the energy loss rates}

The isotropic CMB photon number density [$\mathrm{GeV}^{-1} \, \mathrm{cm}^{-3}$] is a blackbody spectrum, with present-day temperature $T = 2.725 \, \mathrm{K}$, that scales adiabatically with redshift as $ n_\gamma^\mathrm{CMB}\left(\epsilon,z\right) = \left(1+z\right)^2 n_\gamma^\mathrm{CMB}\left(\epsilon/\left(1+z\right),z=0\right)$ (see, \eg, \cite{Anchordoqui:2011gy}). On the other hand, since the CIB receives contributions from sources at different redshifts, its scaling is more complicated, and depends on the choice of evolution of the sources of the CIB photons and of the local ($z=0$) photon number density. We follow the CIB scaling presented in \App~C of \Ref~\cite{Ahlers:2009rf}: for the source evolution we have chosen the SFR by Hopkins \& Beacom \cite{Hopkins:2006bw} with no high-redshift correction, \ie, $\alpha = 0$ (see \Tab~\ref{tab:redshiftnorms}, third row), and for the local CIB spectrum, the popular model by Franceschini {\it et al.}~\cite{Franceschini:2008tp}. The energy loss rates on 
the CMB and CIB due to photohadronic interactions and pair production, $b_{p\gamma}^\mathrm{CMB}$, $b_{e^+e^-}^\mathrm{CMB}$, $b_{p\gamma}^\mathrm{CIB}$, and $b_{e^+e^-}^\mathrm{CIB}$, are calculated following the procedure outlined in the previous subsection, with either $n_\gamma = n_\gamma^\mathrm{CMB}$ or $n_\gamma = n_\gamma^\mathrm{CIB}$. The total interaction rate which enters \equ{EqYiEvol} receives contributions from interactions on the CMB and CIB, \ie, $b^\mathrm{tot}\left(E,z\right) = b^\mathrm{CMB}_{p\gamma}\left(E,z\right) + b^\mathrm{CMB}_{e^+e^-}\left(E,z\right) + b^\mathrm{CIB}_{p\gamma}\left(E,z\right) + b^\mathrm{CIB}_{e^+e^-}\left(E,z\right)$.

\begin{figure}[tp!]
 \centering
 \includegraphics[width=0.5\textwidth]{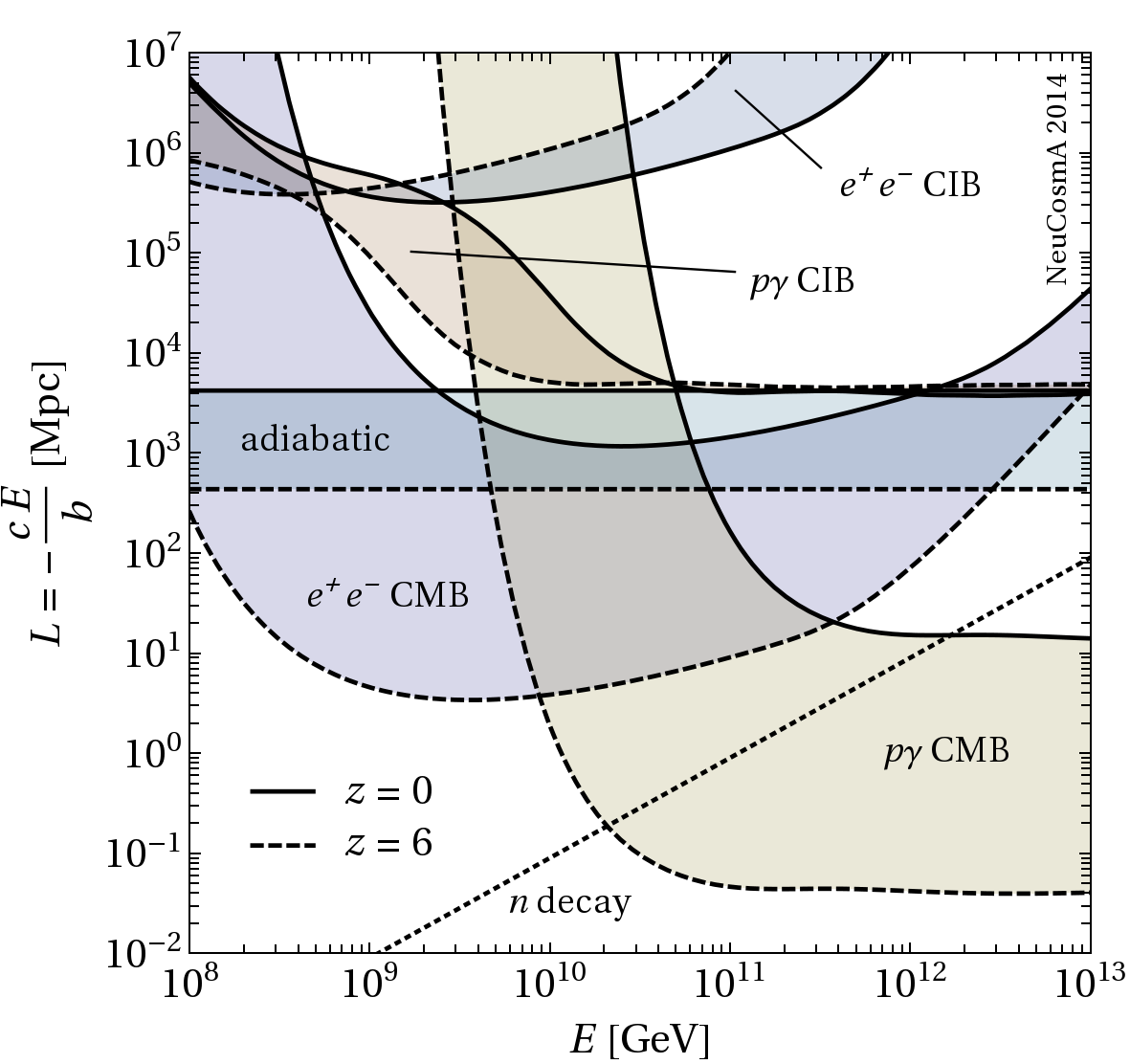}
 \caption{Cosmic-ray proton interaction length $L = -cE/b$ for the adiabatic losses due to the cosmological expansion, and for photohadronic and pair-production losses on the CMB and the CIB. Two values of redshifts where used: $z = 0$ (solid lines) and $z = 6$ (dashed lines). Compare to Fig.~3 in \Ref~\cite{Allard:2006mv}, Fig.~1 in \Ref~\cite{Takami:2007pp}, and Fig.~7 in \Ref~\cite{Kotera:2011cp}.}
 \label{fig:fig_interaction_length}
\end{figure}


The interaction length can be calculated from the energy loss rate as $L = -c E / b$. \Figu{fig_interaction_length} shows the attenuation length corresponding to adiabatic losses only (in this case, $b_\mathrm{adiabatic} = -c H\left(z\right)$), and the interaction lengths due to photohadronic and pair-production losses on the CMB and the CIB. At low redshifts, note that the low-energy interaction length is dominated by the CIB. From around $E \sim 10^{8.5}\, \mathrm{GeV}$, the CMB interactions become dominant, and the total interaction length decreases due to the total energy loss rate $b^\mathrm{tot}$ becoming larger. At higher redshifts, the interaction length is dominated by the CMB throughout the whole energy range: since the CMB photon 
density grows with redshift, the interaction length is shorter for $z=6$ than for $z=0$.

\begin{figure}[tp!]
 \centering
 \includegraphics[width=\textwidth]{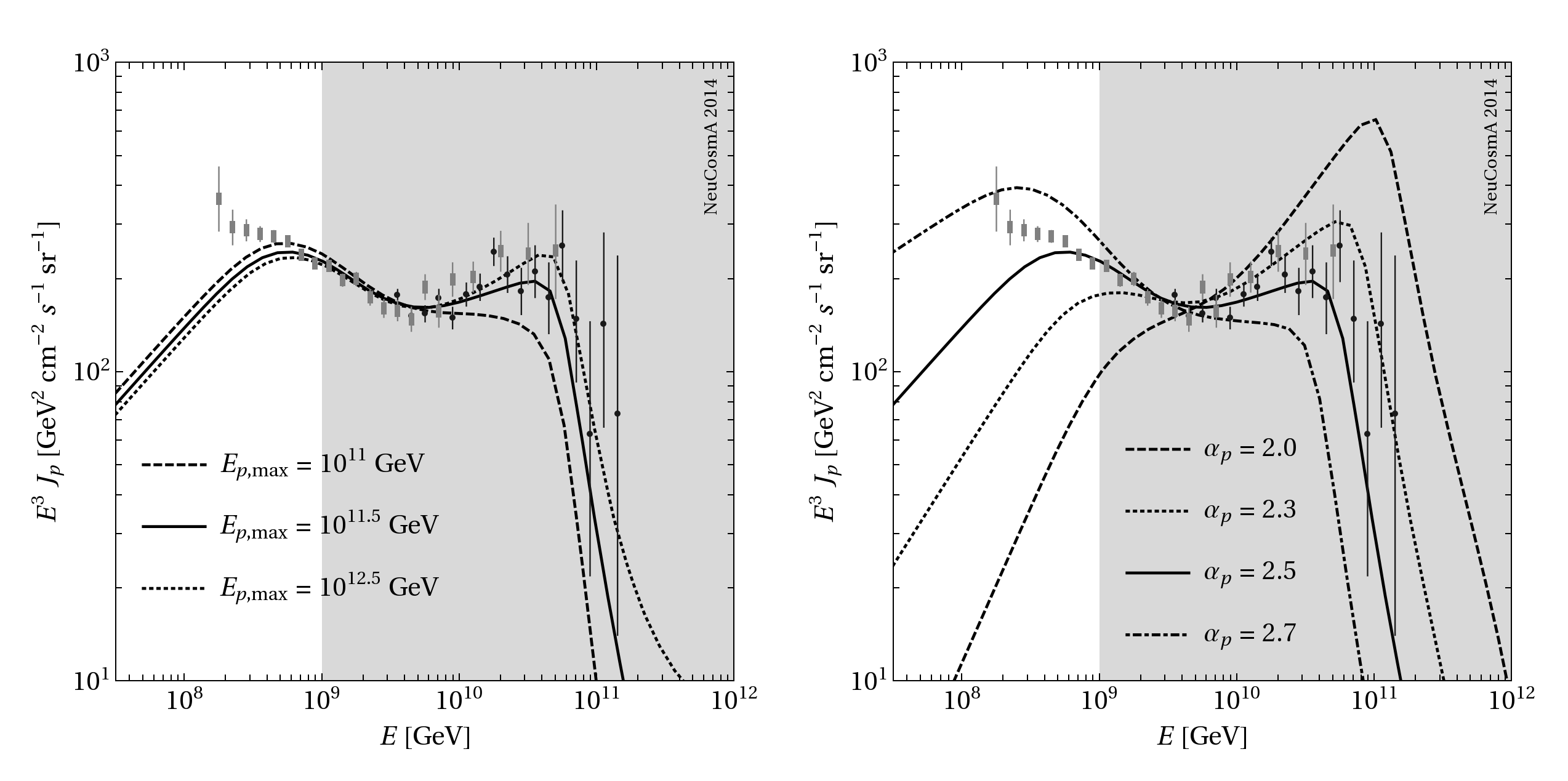}
 \caption{UHECR proton spectrum at Earth, fitted to the HiRes data \cite{Abbasi:2007sv} in the range $10^9\--10^{12}\, \mathrm{GeV}$ (gray band). The CR injection function has been assumed to be a simple exponentially-damped power law $Q_\mathrm{CR} \propto E^{-\alpha_p} e^{-E/E_{p,\mathrm{max}}}$. The GRBs have been assumed to follow the SFR by Hopkins \& Beacom without any high-redshift correction \cite{Hopkins:2006bw}. {\it Left:} variation with $E_{p,\mathrm{max}}$, for a fixed $\alpha_p = 2.5$. {\it Right:} variation with $\alpha_p$, for a fixed $E_{p,\mathrm{max}} = 10^{11.5}\, \mathrm{GeV}$.}
 \label{fig:cr_spectra_generic_injection}
\end{figure}

As an example, \figu{cr_spectra_generic_injection} shows the CR proton flux obtained by using a generic redshift-independent CR injection function $Q_\mathrm{CR}\left(E\right) \propto E^{-\alpha_p} e^{-E/E_{p,\mathrm{max}}}$, normalized by fitting the resulting local CR flux to the HiRes monocular UHECR data \cite{Abbasi:2007sv} in the range $E \in \left[10^9,10^{12}\right]\, \mathrm{GeV}$, marked by the gray filled region. The fitting procedure is described in \ref{sec:statistics}. For the redshift evolution of the CR sources, we have assumed the SFR by Hopkins \& Beacom (with $\alpha = 0$) \cite{Hopkins:2006bw}. In the left panel, we fixed $\alpha_p = 2.5$ and varied the maximum proton energy $E_{p,\mathrm{max}} = 10^{11}$, $10^{11.5}$, and $10^{12.5}\, \mathrm{GeV}$. While all three curves are clearly able to fit the lower-energy data points, which have smaller uncertainties, too low a value of the maximum proton energy will fail to fit the highest-energy points. In the right panel we fixed $E_{p,\mathrm{
max}} = 10^{11.5}\, \mathrm{GeV}$ 
and instead varied the spectral index of the CR injection 
function, $\alpha_p = 2.0$, $2.3$, $2.5$, and $2.7$. A value of $\alpha_p = 2.5$ provides the best fit, since in essence it corresponds to a dip model which is able to reproduce both the low- and high-energy features of the data. On the other hand, a value of $\alpha_p = 2.0$ results in the worst fit in the range $E \in \left[10^9,10^{12}\right]\, \mathrm{GeV}$, but would yield a very good fit if instead the range $E \in \left[10^{10},10^{12}\right]\, \mathrm{GeV}$ was used, corresponding to an ankle model. 

\subsection{Cosmogenic neutrinos}\label{SecCosmoNuSubInjection}

\begin{figure}[tp!]
 \centering
 \includegraphics[width=0.5\textwidth]{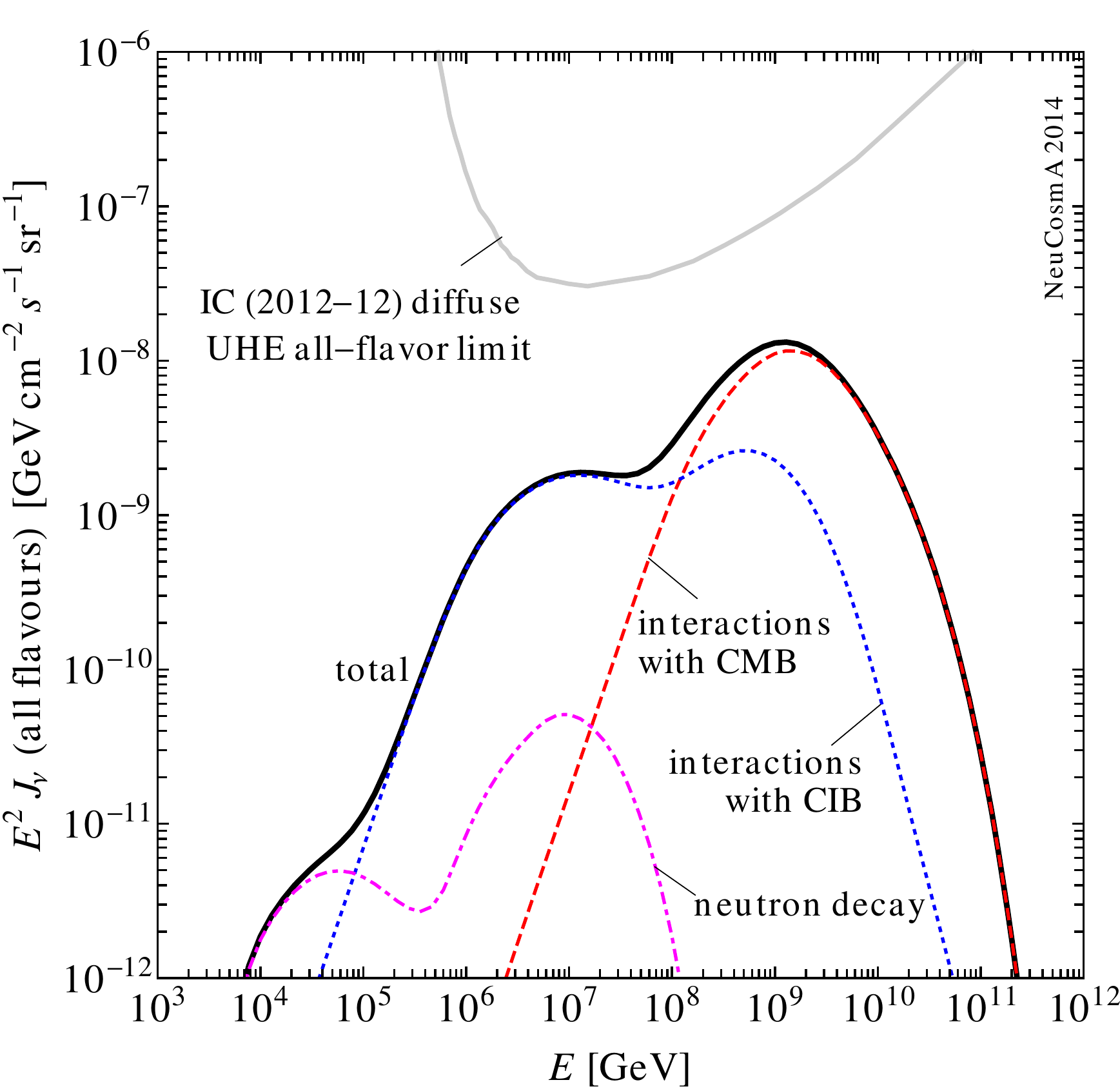}
 \caption{Cosmogenic neutrino spectrum associated to a CR flux with $\alpha_p = 2.5$ and $E_{p,\mathrm{max}} = 10^{11.5}\, \mathrm{GeV}$, fitted to HiRes UHECR data. The different contributions are shown individually, and the relevant IceCube upper bound is included (see \ref{sec:statistics}). The results are consistent with those by Kotera {\it et al.} \cite{Kotera:2010yn}.}
 \label{fig:cosmneutrino_by_contribution}
\end{figure}

Cosmogenic neutrinos are created in interactions of CR protons with the cosmological photon backgrounds and come from two different decay chains: the decay of pions/muons/kaons and the decay of neutrons. From NeuCosmA, we obtain the neutrino injection spectra per energy, volume, and time; after taking into account flavor mixing, the flux of $\nu_\alpha$ ($\alpha = e, \mu, \tau$) at Earth is $J_{\nu_\alpha}\left(E\right) = \left[ c / \left(4\pi\right) \right] n_{\nu_\alpha}\left(E,0\right)$. \Fig~\ref{fig:cosmneutrino_by_contribution} shows the different contributions that make up the all-flavor cosmogenic neutrino flux. Note that our results are compatible with existing calculations, \eg, by Kotera {\it et al.} \cite{Kotera:2010yn}.




\section{Details of the statistical analysis}
\label{sec:statistics}

We fit the UHECR proton flux generated by our propagation code (see \ref{sec:crprop}) to the surface observation data recorded by the Telescope Array (TA)~\cite{AbuZayyad:2012ru}. These consist of pairs $(E_i,\left( E^3 J_\mathrm{CR} \right)_i^\mathrm{TA})$, with $\sigma_i$ the uncertainty on $\left( E^3 J_\mathrm{CR} \right)_i^\mathrm{TA}$. We define a simple two-parameter $\chi^2$ function as
\begin{equation}\label{equ:Chi2Definition}
 \chi^2\left(f_e^{-1}, \delta_E\right) =
 \sum_i\left( \frac{ E^{\prime3}_i J_\mathrm{CR} \left( E^\prime_i, f_e^{-1} \right) - \left( E^3 J_\mathrm{CR} \right)_i^\mathrm{TA} } { \sigma_i } \right)^2 
 + \left( \frac{\delta_E}{\sigma_E} \right)^2 \;,
\end{equation}
with the sum performed over the data points that have energies $E_i \geq 10^{10} \, \mathrm{GeV}$ for the ankle model and $E_i \geq 10^{9} \, \mathrm{GeV}$ for the dip model. When minimized, this function yields the value of the normalization $f_e^{-1}$, and the energy-scale displacement $\delta_E$, defined so that $E \to E^\prime \equiv \left( 1 + \delta_E \right) E$. Since the (Gaussian) error bars on each TA flux data point $i$ are asymmetric, we have chosen the uncertainty $\sigma_i$ for each to be the size of the upper bar, if, for given values of $f_e^{-1}$ and $\delta_E$, the calculated flux $E^{\prime3}_i J_\mathrm{CR} \left( E^\prime_i, f_e^{-1} \right)$ lies above the central value of the data point; otherwise, we have equaled it to the size of the lower bar. For the systematic energy uncertainty of the TA experiment we have used $\sigma_E = 0.21$, following \Ref~\cite{Barcikowski:2013wsa}.

For the estimation of the number of expected neutrino events and the calculation of the limits, we use a simple approach which folds the neutrino flux prediction with the parameters of the measurement. The number of neutrino events $\#\nu$ is calculated as
\begin{equation}
	\#\nu = \int \mathrm{d}E \, J_\nu(E) \cdot A_\mathrm{eff}(E) \cdot t_\mathrm{exp} \cdot 4\pi \quad ,
	\label{equ:numberofneutrinoevents}
\end{equation}
where $J_\nu(E)$ is the neutrino flux as function of energy (in $\left[\mathrm{GeV}^{-1} \, \mathrm{cm}^{-2} \, \mathrm{s}^{-1} \, \mathrm{sr}^{-1} \right]$), $A_\mathrm{eff}(E)$ the energy-dependent effective area including Earth attenuation effects (in $\left[\mathrm{cm}^2 \right]$), $t_\mathrm{exp}$ the exposure (in $\left[\mathrm{s}\right]$), and $4\pi$ is the factor for the full solid angle (in $\left[ \mathrm{sr}\right]$). The standard 90\% C.L.~exclusion limit by Feldman and Cousins for an arbitrary flux is obtained by choosing the normalization in \equ{numberofneutrinoevents} to obtain 2.44 events~\cite{Feldman:1997qc} (background-free case). 

The differential limits in this study are given by
\begin{equation}
	E^2 J_{\nu,\mathrm{limit}} (E) = \frac{E}{2.3 \cdot A_\mathrm{eff} \cdot t_\mathrm{exp} \cdot 4\pi} \quad ,
	\label{equ:neutrinolimit}
\end{equation}
that is, a neutrino flux exactly following the differential limit over one order of magnitude in energy will yield one event.
The current limit for the prompt neutrino flux is based on the model-independent solid angle-averaged effective area from the combined IC40+59 GRB stacking analysis~\cite{Abbasi:2012zw} with the exposure being estimated from comparing the 215 bursts of the combined sample to the assumed 667 (long) bursts per year, \ie, $t_\mathrm{exp} = 215/667$ yr. For the cosmogenic neutrinos, we calculate the current limit from the average effective area for a $4\pi$-isotropic $\nu_\mu$ flux during 615.9 days lifetime with the IC79 and IC86 configurations, given in \Ref~\cite{Aartsen:2013bka}. 
Both of these analyses are considered to be background-free, the stacking analysis because of timing and directional information, the UHE analysis because of the cut in energy.

To calculate the extrapolated neutrino upper bounds after $t_\mathrm{exp} = 15 \, \mathrm{yr}$ of full detector exposure, for the prompt neutrinos we simply rescale the current bound by the factor $215/\left(15 \cdot 667 \right)$, while for the cosmogenic neutrinos we rescale the corresponding current bound by $615.9/\left( 15 \cdot 365 \right)$.

\section{Impact of acceleration efficiency, maximal proton energy, and energy calibration}
\label{sec:maxe}

The acceleration efficiency is one of the factors determining the maximal proton energy: the maximal proton energy is obtained by equating the acceleration rate (depending on the acceleration efficiency) with the synchrotron, adiabatic, or photohadronic cooling rate in the model, whichever is larger; see \equ{EpmaxDetermination}. On the other hand, the maximal proton energy can be fit to the UHECR observation within the energy calibration uncertainty, which means that some uncertainty is acceptable. Here we discuss the interplay between acceleration efficiency/maximal proton energy and our fits, and the impact of the energy calibration uncertainty.

\begin{figure}[tp]
	\centering
	\includegraphics[width=\textwidth]{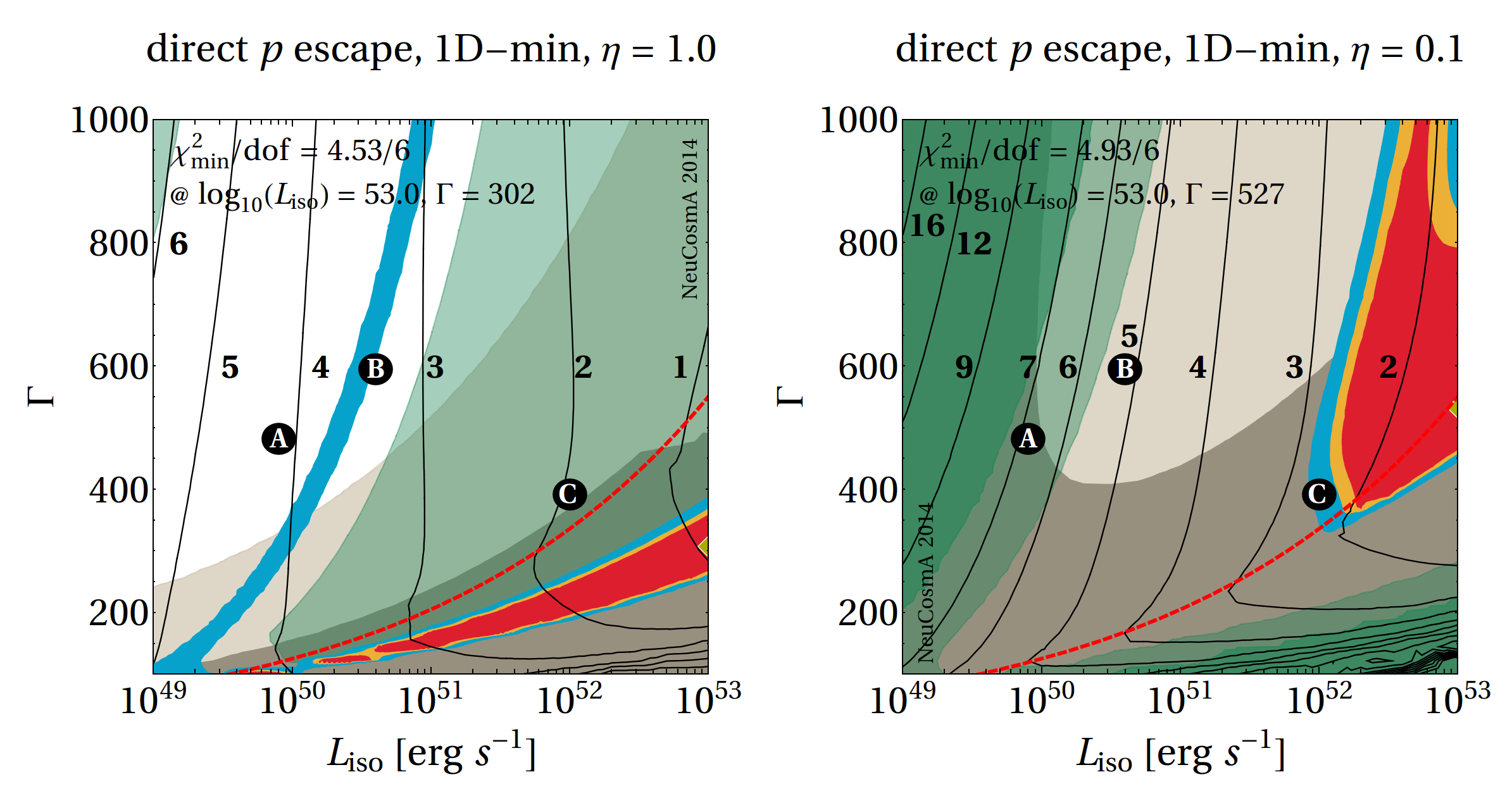}
	\mycaption{\label{fig:1dmin} Same as \figu{fits}, middle panels (direct escape), but with \textbf{only the normalization constant left as a free parameter} in the minimization process. In the right panel, the area colored in dark green corresponds to the parameter space region excluded by cosmogenic neutrinos with the current exposure.}
\end{figure}

\begin{figure}[tp]
	\centering
	\includegraphics[width=\textwidth]{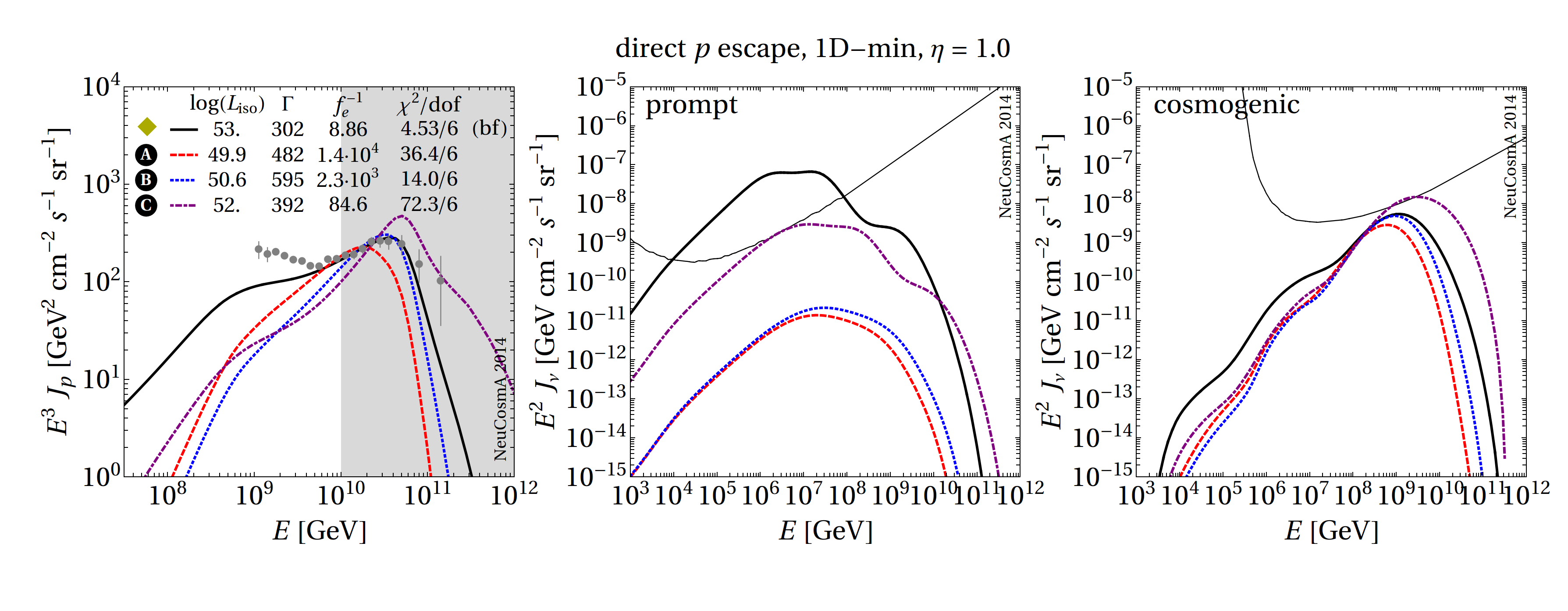}
	\vspace*{-0.6cm}\;
	\includegraphics[width=\textwidth]{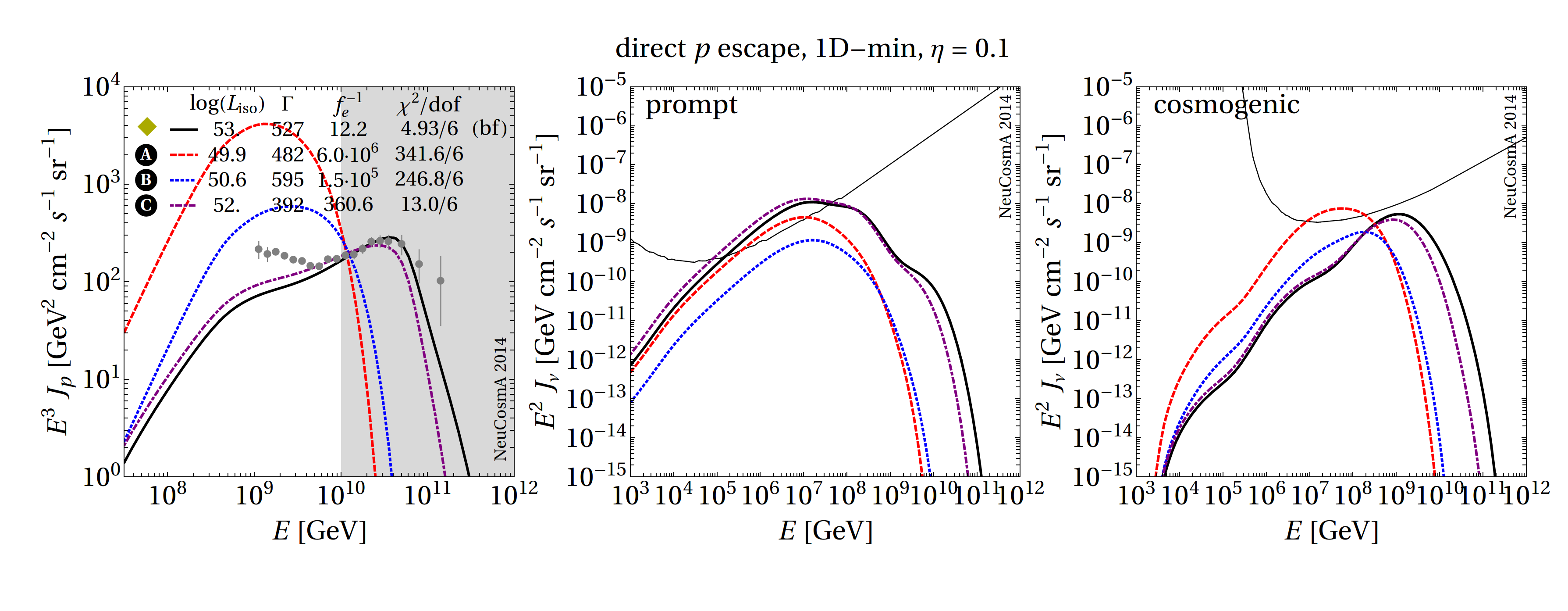}
	\vspace{-1cm}
	\mycaption{\label{fig:1dminspectra}Cosmic ray, prompt neutrino, and cosmogenic neutrino spectra (in columns) for selected points in the parameter space plane $\Gamma$ vs.~$L_\mathrm{iso}$, corresponding to the markers in \figu{1dmin}.}
\end{figure}

Let us focus on the energy calibration first. For that purpose, we keep the $\delta E=0$ in \equ{Chi2Definition} fixed, and only vary the normalization. The resulting parameter scan plots in the $\Gamma$ vs.~$L_\mathrm{iso}$ plane are shown in \figu{1dmin}. The required values of the baryonic loading in this case are larger, since the proton spectrum curves can only be shifted vertically when attempting to fit them to the TA data. The larger values of $f_e^{-1}$ lead 
to higher neutrino event yields and, therefore, to larger exclusion regions. In fact, for $\eta = 0.1$ (right panel), the cosmogenic neutrino event yield is high enough to result in an exclusion region even at the current detector exposures (dark green areas), something which had not been evidenced for any other parameter scan plot so far in this study. In \figu{1dminspectra}, the proton and neutrino spectra are shown for the best-fit points in \figu{1dmin} and for the selected points A-C. Clearly, as indicated by the values of the reduced $\chi^2/\mathrm{d.o.f.}$, the fits are worse than the ones we had found when we allowed both the normalization constant and the energy-scale displacement to be free parameters in the fitting procedure, \cf~\figu{fits} (middle row) and \figu{spectra} (second row from the top). Particularly for the points A and B for an efficiency of $\eta = 0.1$, the fit to TA data is notably worse compared to our earlier results, since the proton spectra for these points peak at much too low 
energies. 
Similarly, for $\eta=1$, point~C overshoots the maximal proton energy (although corrected for by the GZK cutoff), which may be partially compensated by the energy calibration. These adjustments come at the expense of a penalty $\chi^2$ (last term in \equ{Chi2Definition}), which means that the maximal proton energy will nevertheless have an impact on the fit quality (see below). Note that the qualitative shape of the fit contours in \figu{1dmin} is however similar to the middle row in \figu{fits}, though the left fit branch for $\eta=1.0$ is noticeably wider when the energy calibration error is included in the fit. Therefore, we expect that improved energy reconstruction in the UHECR measurements can constrain this part of the parameter space.

\begin{figure}[tp]
	\centering
	\includegraphics[width=\textwidth]{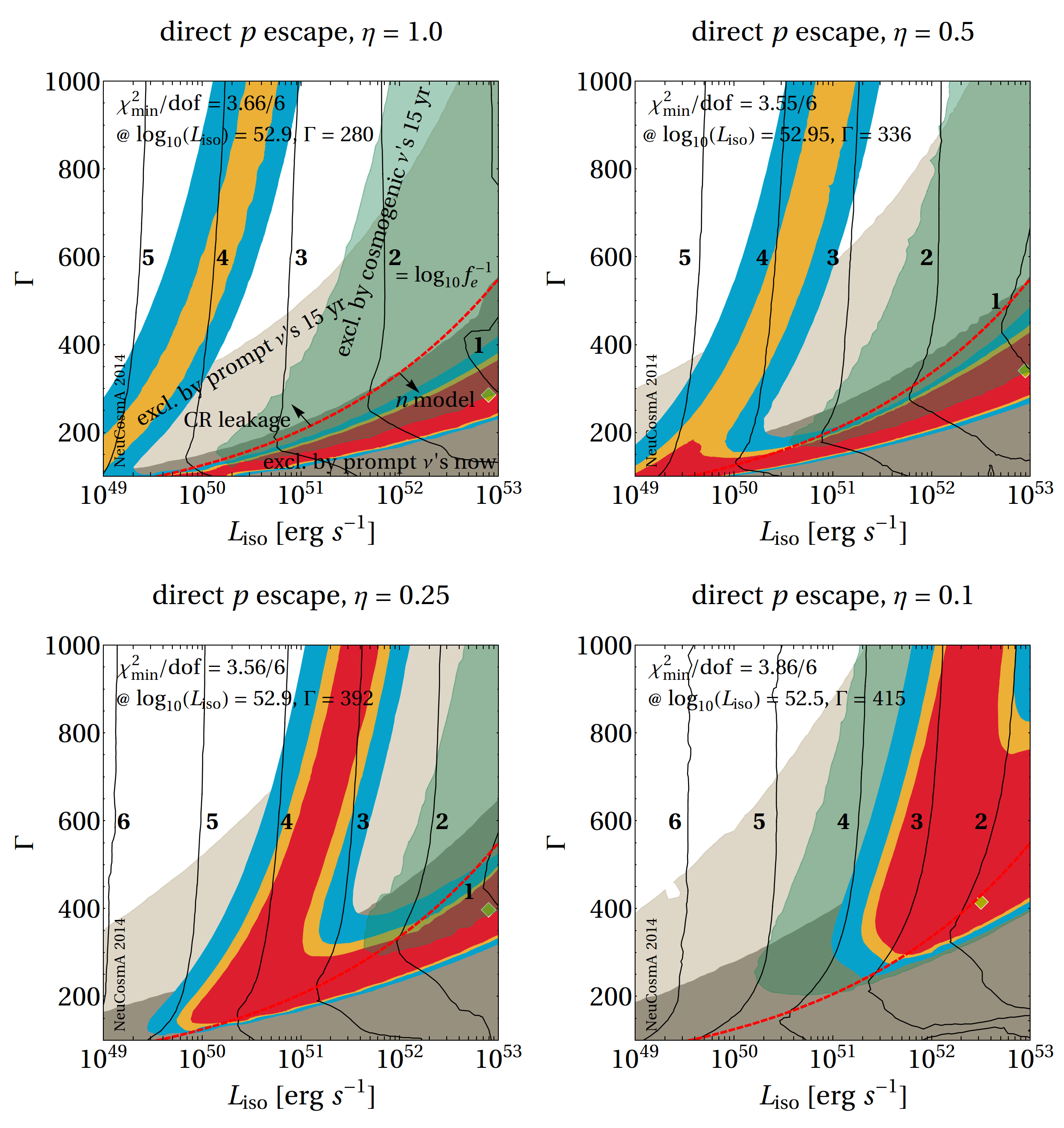}
	\mycaption{\label{fig:etadep} Figure similar to \figu{fits} (middle row) for acceleration efficiencies $\eta = 1.0$, $0.5$, $0.25$, and $0.1$.}
\end{figure}


\begin{figure}[tp]
	\centering
	\includegraphics[width=\textwidth,clip=true]{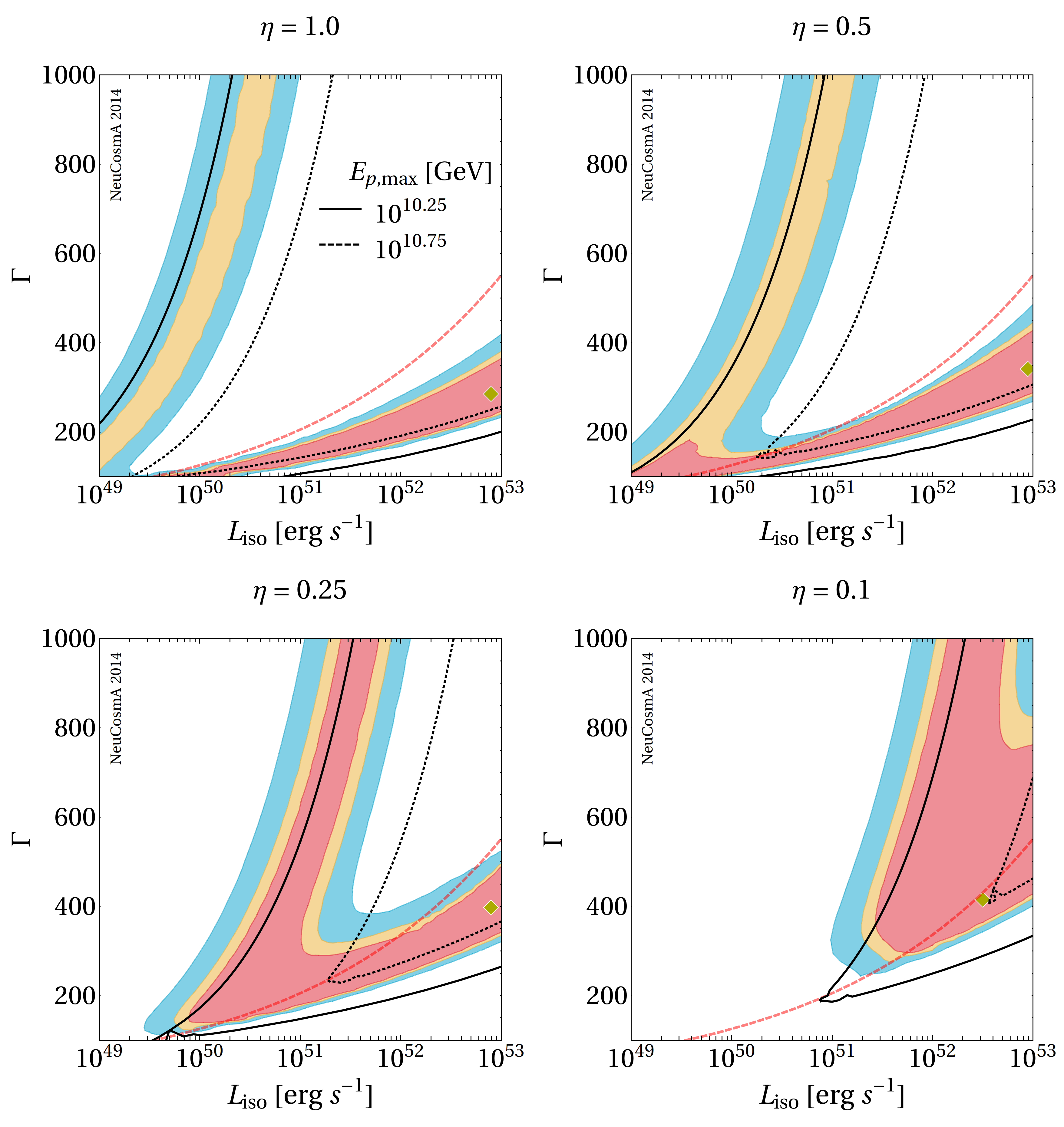}
	\mycaption{\label{fig:epmax}Contours of $E_{p,\max} = 10^{10.25}\, \mathrm{GeV}$ (solid black lines) and $10^{10.75}\, \mathrm{GeV}$ (dotted black lines) in the $\Gamma$ vs.~$L_\mathrm{iso}$ plane for acceleration efficiencies of $\eta = 1.0, 0.5, 0.25,$ and $0.1$. The energy $E_{p,\max}$ is in the source frame. The $\chi^2$ regions corresponding to 90\% (red), 95\% (yellow), and 99\% C.L. (blue), and the dashed red curve below which neutron escape dominates, are included for reference.}
\end{figure}

A related issue is the impact of the acceleration efficiency and therefore of the maximal proton energy on the fit.
First of all, one may ask if the transition between the left and right columns in \figu{fits} is continuous.
We therefore show in \figu{etadep} the dependence on the acceleration efficiency in the different panels. One can clearly see that the transition between the lower and higher acceleration efficiency is indeed continuous if the two intermediate values are taken into account. If the acceleration efficiency is allowed as another free parameter, one therefore can cover, in principle, most of the plane with the left fit branch. This will however not affect our qualitative conclusions. 

Note that the IceCube bounds depend on the acceleration efficiency as well. Lower acceleration efficiencies require an upscaling of the proton flux by energy calibration and normalization, which in turn increases the prompt neutrino flux. Higher acceleration efficiencies increase the contribution of high-energy protons, and therefore the flux cosmogenic neutrinos.
Comparing to the future IceCube bounds, it is clear that regions with moderately small baryonic loadings $f_e^{-1} \lesssim 100$ can be excluded, and that $f_e^{-1} \sim 1000$ requires an average $\Gamma \gtrsim 800$ (left branch in lower left panel), which seems unrealistic. On the other hand, $\Gamma \simeq 300$ needs $f_e^{-1} \gtrsim 10^{4.5}$ (left branch in upper left panel). Therefore, reasonable parameter ranges can be excluded, unless the baryonic loading is extremely large.

Finally, to illustrate the relationship between acceleration efficiency and maximal proton energy, we show the maximal proton energy as a function of the acceleration efficiency in \figu{epmax}. We note from comparing \figu{etadep} with \figu{epmax} that the fit contours follow the maximal proton energy predicted by the model. 
This is not surprising, and has been also seen above: if the maximal proton energy is too small, the UHECR prediction cannot be fit. This may be partially compensated by the energy calibration, but at the expense of a penalty $\chi^2$ (see above). If the maximal proton energy is too high, the relatively hard spectra of the cosmic ray escape components (even for the neutron model it is harder than $E^{-2}$ because of multi-pion production) will lead to a strong peak before the GZK cutoff (\cf, curve~C in upper left panels of \figu{1dminspectra}). In the statistical analysis, such a peak is disfavored because of its shape. This feature is certainly somewhat model-dependent, but it can be easily taken into account by considering different acceleration efficiencies as in this appendix.

\section{Impact of the choice of the experimental UHECR data}
\label{sec:expdata}

Our results so far have been obtained by fitting our simulated UHECR spectra to the Telescope Array (TA) surface detection data \cite{AbuZayyad:2012ru}. However, TA is not the only giant air shower experiment that has measured the UHECR spectrum: the Pierre Auger Observatory (PAO) and the High Resolution Fly's Eye (HiRes) experiment, for instance, have also published UHECR spectra; the former using hybrid fluorescence and surface detector observations \cite{Abreu:2011ph,Abraham:2010mj}, and the latter using its monocular mode during two runs (HiRes-I and HiRes-II) \cite{Abbasi:2007sv}. There are only a few differences among the different data sets, but the PAO spectrum exhibits particular features: it is approximately a factor of 2 lower than the HiRes and TA spectra, has a more pronounced dip at $\sim 10^{9.3} \, \mathrm{GeV}$, and is measured to higher energies than the other two, where the GZK cutoff becomes evident \cite{Abraham:2008ru}.

To test the effect that using different experimental UHECR data sets has on our conclusions, we used the proton spectral index $\alpha_p = 2.0$ and repeated the minimization procedure that we carried out using the TA data (see \ref{sec:statistics}), but now using the PAO and HiRes data instead. As in the main text, we have focused on a transition model, so that the minimization was carried out within the energy range $10^{10}$ \--- $10^{12}\, \mathrm{GeV}$.

In the present Appendix, when performing the minimization we have added an extra $\chi^2$ penalty to ensure that the fitted spectra do not exceed the upper ends of the error bars of the data points that lie below $10^{10}\, \mathrm{GeV}$. That is because one can always explain a missing contribution by an additional (such as Galactic) component at lower energies, but an excess clearly contradicts data. For the purpose of calculating the penalty, the lower ends of the error bars of the data points below $10^{10}\, \mathrm{GeV}$ are extended to minus infinity, since we only want to make sure that the data points are not exceeded by the spectra. The penalty is calculated as
\begin{eqnarray}
 \chi^2_\text{pen}\left(f_e^{-1}, \delta_E\right) &=&
 \sum_i
 \Theta \left( E^{\prime3}_i J_\mathrm{CR} \left( E^\prime_i, f_e^{-1} \right) - \left( E^3 J_\mathrm{CR} \right)_{i,\text{u.b.}}^\mathrm{exp} \right) \nonumber \\
 && \qquad \cdot
 \left( \frac{ E^{\prime3}_i J_\mathrm{CR} \left( E^\prime_i, f_e^{-1} \right) - \left( E^3 J_\mathrm{CR} \right)_{i,\text{u.b.}}^\mathrm{exp} } { \sigma_i } \right)^2 \;,
\end{eqnarray}
where $\Theta$ is the Heaviside step function, $\left( E^3 J_\mathrm{CR} \right)_{i,\text{u.b.}}^\mathrm{exp}$ is the $1\sigma$ upper bound on the $i$-th experimental data point (with $\mathrm{exp}$ = HiRes, PAO, or TA), and the sum is performed over all data points below $10^{10} \, \mathrm{GeV}$. While the effect of including $\chi^2_\text{pen}$ in the fitting procedure is minimal when using the TA and HiRes UHECR data, it becomes more significant when fitting to the PAO data.

\begin{figure}[tp]
	\centering
	\includegraphics[width=\textwidth,clip=true]{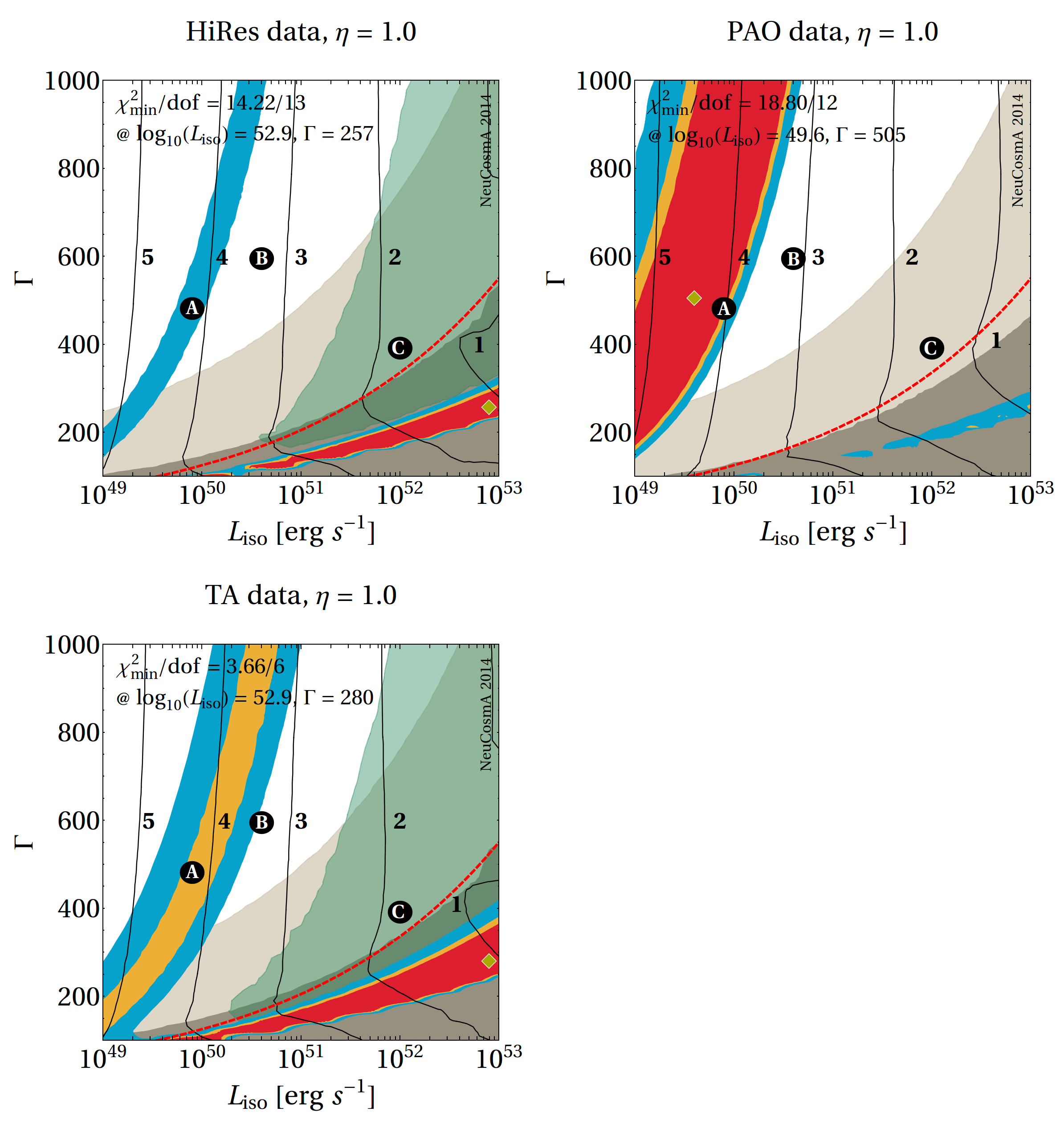}
	\mycaption{\label{fig:contoursHiResAugerTA} Same as \figu{fits}, middle left panel (direct escape with $\eta = 1.0$), but comparing the results of using the HiRes~\cite{Abbasi:2007sv}, Pierre Auger Observatory~\cite{Abraham:2010mj}, and Telescope Array~\cite{AbuZayyad:2012ru} UHECR data points. }
\end{figure}

\begin{figure}[tp]
	\centering
	\includegraphics[width=\textwidth,clip=true]{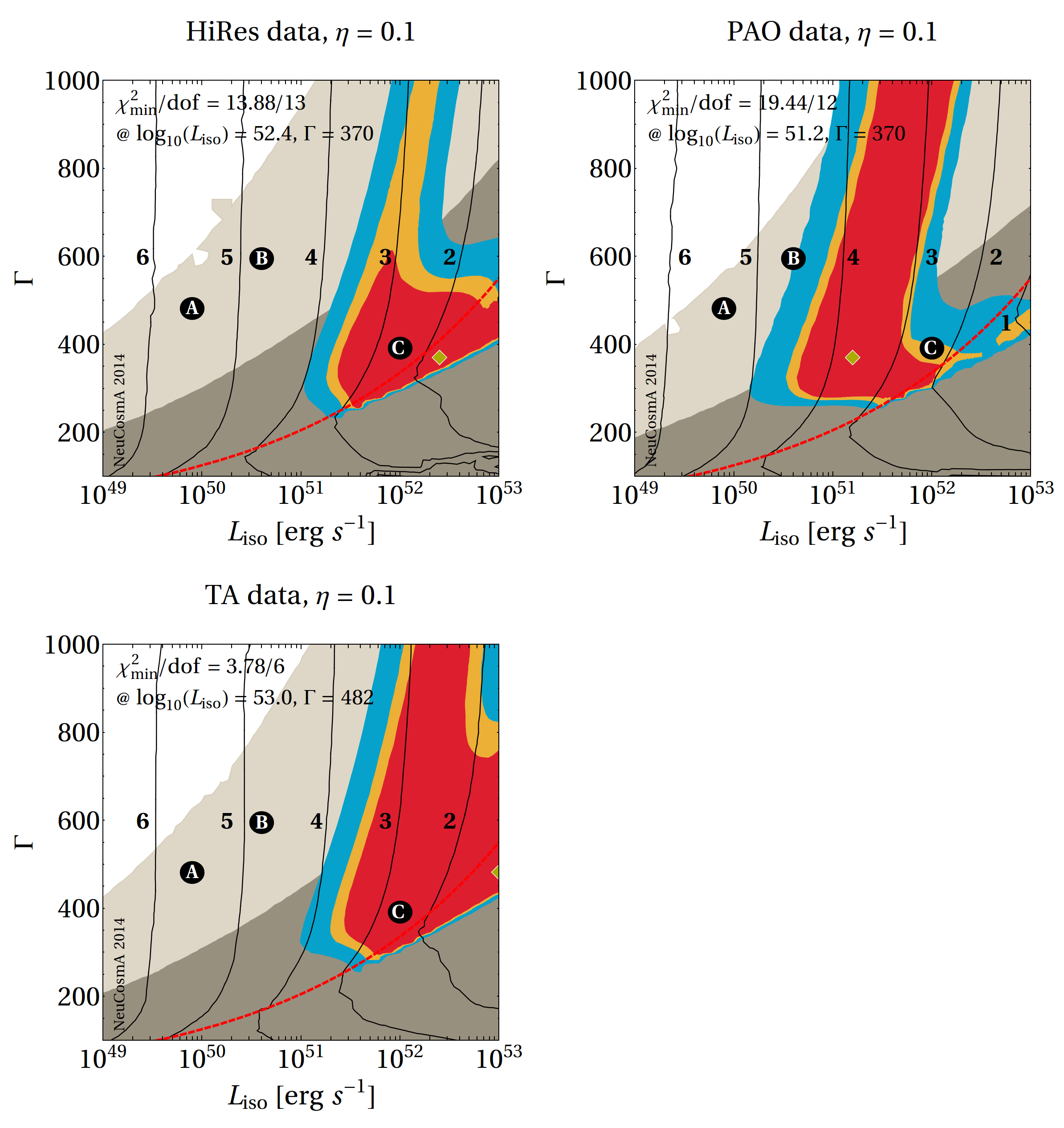}
	\mycaption{\label{fig:contoursHiResAugerTAacc01} Same as \figu{contoursHiResAugerTA}, but for a direct escape-dominated model with $\eta = 0.1$.}
\end{figure}

The results are presented in \figu{contoursHiResAugerTA}. Clearly the confidence regions lie at approximately the same positions in parameter space in the three experiments, but are more extended for the PAO and TA data sets, on account of their larger energy reconstruction uncertainties \cite{Barcikowski:2013wsa}: $\sigma_E = 17\%$ for HiRes, $22\%$ for the PAO, and $21\%$ for the TA. Note that in the PAO case the cosmogenic neutrinos do not reach the 2.44 events (even after 15 years of exposure) needed for a $90\%$ C.L.~exclusion, and so they do not exclude any of the parameter space. For illustration purposes, we also show in \figu{contoursHiResAugerTAacc01} the results of the scans for the three experiments but assuming instead an acceleration efficiency of $\eta = 0.1$; again the size of the regions depend on the energy reconstruction uncertainties.

\begin{figure}[tp]
	\centering
	\includegraphics[trim=.1cm .5cm .5cm .5cm,width=\textwidth,clip=true]{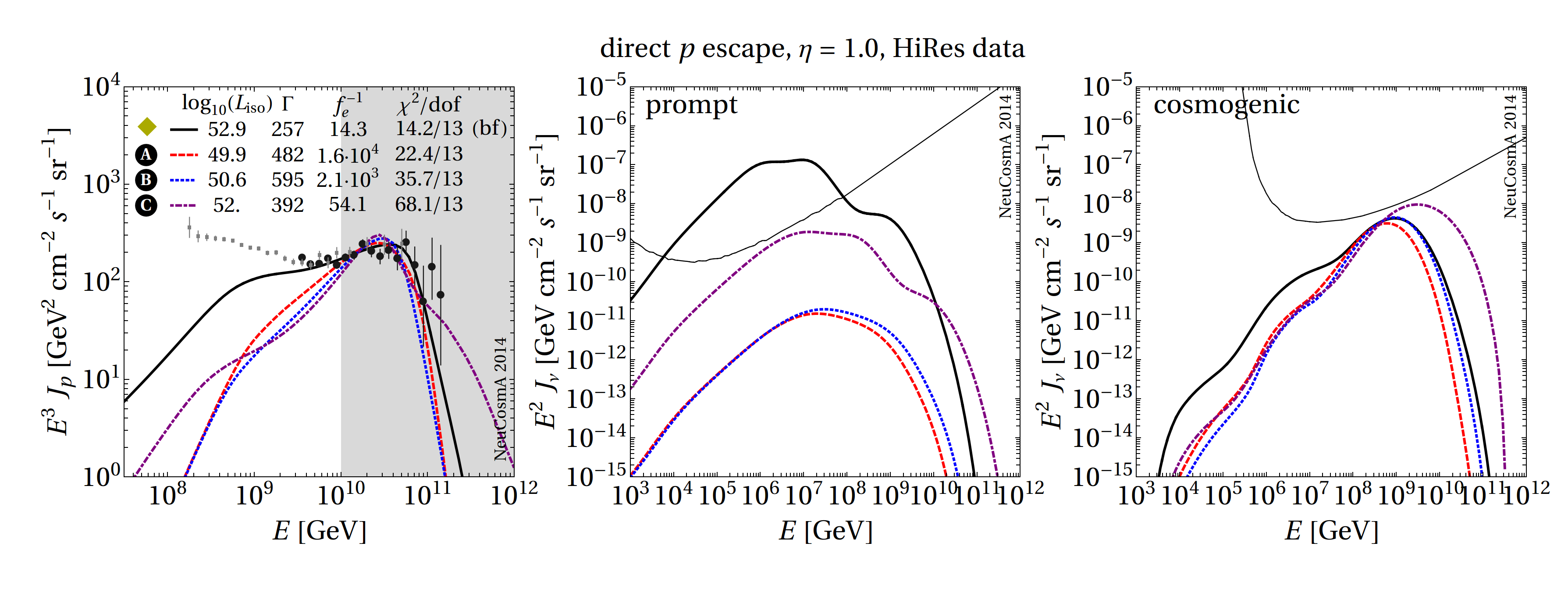}
	\includegraphics[trim=.1cm .5cm .5cm .5cm,width=\textwidth,clip=true]{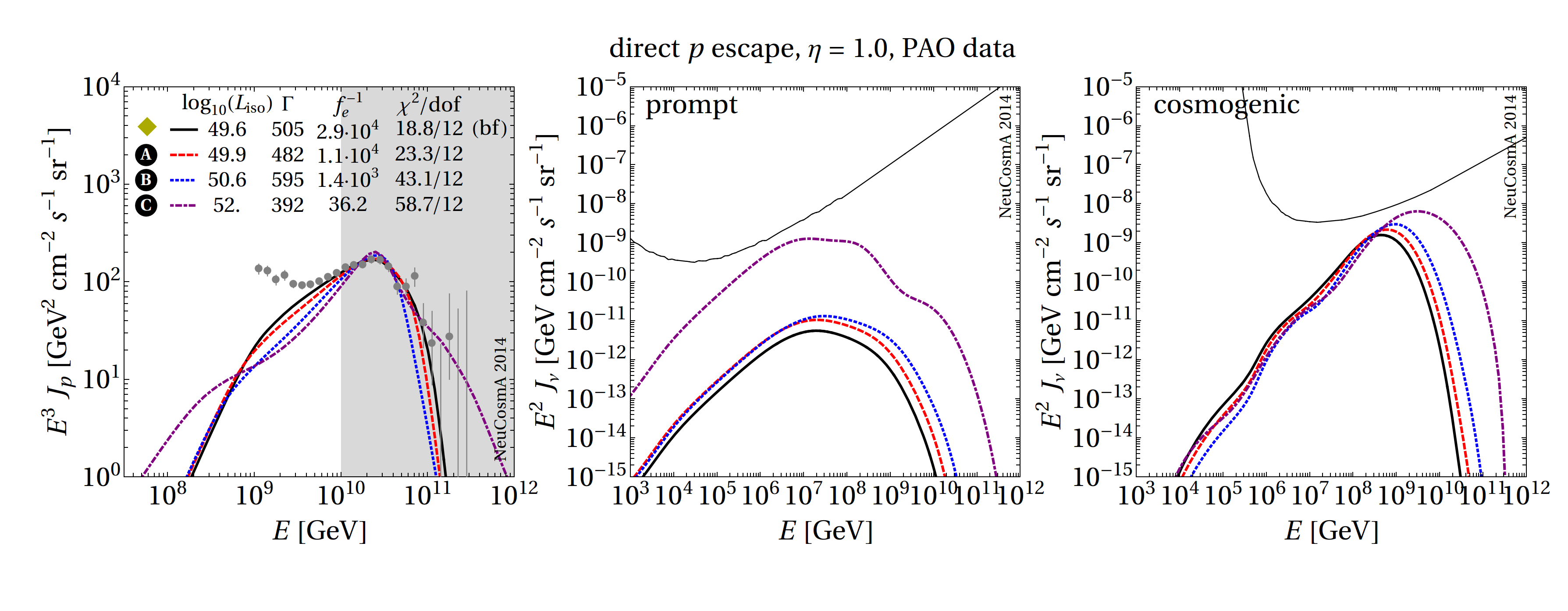}
	\includegraphics[trim=.1cm .5cm .5cm .5cm,width=\textwidth,clip=true]{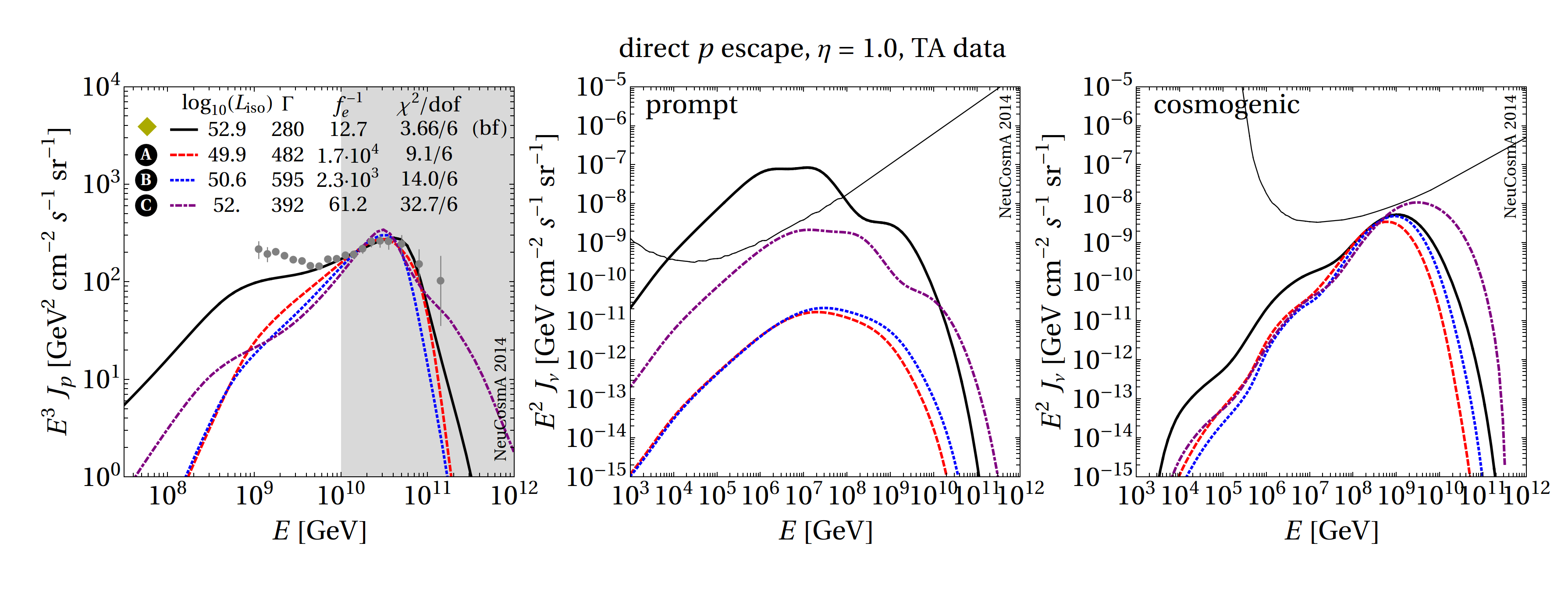}
	\mycaption{\label{fig:spectraHiResAugerTA}Cosmic ray, prompt neutrino, and cosmogenic neutrino spectra (in columns) from GRBs in the range $z=0$ to $6$, for selected points in the parameter space plane $\Gamma$ vs.~$L_\mathrm{iso}$, corresponding to the markers in \figu{contoursHiResAugerTA}. The different rows correspond to the fits to HiRes~\cite{Abbasi:2007sv}, PAO~\cite{Abraham:2010mj}, and TA~\cite{AbuZayyad:2012ru} data. The fit range is gray-shaded.}
\end{figure}

From the spectra plots of the $\eta = 1.0$ case, \figu{spectraHiResAugerTA}, and the values of $f_e^{-1}$ therein, it is evident that the results among the three experiments are broadly comparable and consistent with each other. The only important difference occurs for the PAO data set: the best-fit point lies inside the left fit branch, whereas for HiRes and the TA, it lies inside the right branch. For the PAO, this branch has somewhat disappeared, meaning that the fit there has become poorer. This is an effect of the $\chi^2$ penalty that was implemented to ensure that the fitted spectrum does not exceed the upper ends of the error bars of the data points below $10^{10}\, \mathrm{GeV}$. For the HiRes and TA cases, on the other hand, the penalty does not affect the confidence regions considerably, since the dip in the data below $10^{10}\, \mathrm{GeV}$ is not as pronounced as for the PAO. Overall (for both $\eta$), PAO data seem to prefer higher baryonic loadings that HiRes or TA data.

From \figs~\ref{fig:contoursHiResAugerTA} and \ref{fig:spectraHiResAugerTA}, we can conclude that our choice of the TA data for our main analysis is justified by their being representative of the UHECR observations. Furthermore, we have seen that our conclusions are largely independent of the choice of UHECR experimental data -- which is not surprising, as these can be made compatible by energy scale recalibrations within the uncertainties; see, \eg, \Ref~\cite{Gaisser:2013bla}. Note, however, that using the original data with proper energy scale uncertainties is the correct statistical procedure for our purposes, since re-normalizing the data already implies relative penalties by adjusting some scales more than others.

\section*{Acknowledgements}

We are grateful to Markus Ahlers, John Beacom, Francis Halzen, Kohta Murase, and Nathan Whitehorn for many useful discussions.
PB and WW would also like to thank the WIPAC center in Madison for local hospitality, where this work was initiated.
MB would like to thank the hospitality of CCAPP at the Ohio State University, where parts of this work have been carried out.

WW would like to acknowledge support from DFG grant WI 2639/3-1. MB and PB would like to acknowledge support from the GRK 1147 ``Theoretical Astrophysics and Particle Physics''. PB acknowledges partial support from NASA grant NNX13AH50G. 
This work has also been supported by the FP7 Invisibles network (Marie Curie Actions, PITN-GA-2011-289442), the ``Helmholtz Alliance for Astroparticle Physics HAP'', funded by the Initiative and Networking fund of the Helmholtz association, and DFG grant WI 2639/4-1.


\end{document}